\documentclass[fleqn,10pt]{wlscirep}
\usepackage[utf8]{inputenc}
\usepackage[T1]{fontenc}
\usepackage{lineno}
\usepackage{graphicx}
\usepackage{subcaption} 
\usepackage{courier}
\usepackage{float}
\usepackage{tabularx}
\newcolumntype{Y}{>{\centering\arraybackslash}X}

\usepackage{booktabs} 

\title{A Highly Granular Temporary Migration Dataset Derived From Mobile Phone Data in Senegal}

\author[1,2,*]{Paul Blanchard}
\author[3]{Stefania Rubrichi}
\affil[1]{Institut de Recherche pour le Développement, Paris, 75010, France}
\affil[2]{Department of Economics, Trinity College Dublin, Dublin 2, Ireland}
\affil[3]{Orange Innovation - SENSE, Châtillon, 92320, France}

\affil[*]{corresponding author: Paul Blanchard (paul.blanchard@ird.fr | blanchap@tcd.ie)}

\begin{abstract}
Understanding temporary migration is crucial for addressing various socio-economic and environmental challenges in developing countries. However, traditional surveys often fail to capture such movements effectively, leading to a scarcity of reliable data, particularly in sub-Saharan Africa. This article introduces a detailed and open-access dataset that leverages mobile phone data to capture temporary migration in Senegal with unprecedented spatio-temporal detail. The dataset provides measures of migration flows and stock across 151 locations across the country and for each half-month period from 2013 to 2015, with a specific focus on movements lasting between 20 and 180 days. The article presents a suite of methodological tools that not only include algorithmic methods for the detection of temporary migration events in digital traces, but also addresses key challenges in aggregating individual trajectories into coherent migration statistics. These methodological advancements are not only pivotal for the intrinsic value of the dataset but also adaptable for generating systematic migration statistics from other digital trace datasets in other contexts.
\end{abstract}
\begin{document}

\flushbottom
\maketitle

\thispagestyle{empty}

\section*{Background \& Summary}

The movement of people across space is intricately linked with economic activity and economic development processes. Previous research examining mobility within countries have predominantly focused on the significance of permanent migration in fostering growth and structural transformation \cite{Young2013,Bryan2019}. Such studies mostly delve into the factors influencing and hindering the reallocation of individuals from a less productive rural sector to an urban non-agricultural sector. 

Yet, a growing body of research has highlighted the importance of other forms of short-term mobility in developing countries, such as temporary migration movements. These flows of internal movements have been found to be incredibly common and to largely exceed permanent moves \cite{Baker1995,Coffey2015,Delaunay2016}. They have at first been portrayed as a sign of failure of rural livelihoods \cite{Findley1994,Schareika1997} but have also been described more recently as a structural component within households' livelihood strategies \cite{Bryan2014,Coffey2015,Delaunay2016}.
Despite its proven significance, temporary migration is seldom integrated in national statistical systems in a systematic way. Short-term movements are intrinsically difficult to measure (e.g., due to attrition and recall biases) and require specialized -- and oftentimes costly -- surveys \cite{Lucas1997}. More importantly, the rare surveys measuring temporary migration often adopt standard definitions that do not necessarily allow to capture relatively short trips, which are nonetheless frequent \cite{Coffey2015}. Temporary migration patterns thus remain poorly documented at national scales, and especially so in sub-Saharan Africa.

This article introduces an open access dataset containing highly granular temporary migration estimates derived from mobile phone data in Senegal. The dataset includes measurements of migration flows and stocks across 151 locations spanning the entire country -- these locations encompass the rural areas of 112 districts and 39 cities. Estimates are provided for each half-month period over the 2013-2015 timeframe, considering mobility events lasting from 20 to 180 days. The unique level of granularity offered by this dataset aims to furnish researchers from various disciplines, including economists, demographers, environmental sociologists, and others, with a robust foundation of information to advance our understanding of the characteristics, causes and consequences of temporary movements. Comprehensive data on short-term movements are indeed crucially needed to inform development practitioners and policy makers on various matters and support the design of adequate policy interventions. These interventions include, for instance, responses to the effects of environmental shocks, climate change, epidemics and conflicts on short-term population dynamics.

The development of the proposed dataset arises within a context where digital footprints generated by mobile phone usage have emerged as a promising source of big data for measuring human mobility on broader scales. More importantly, these data exhibit a proven capability to capture subtler human movements with an increased spatio-temporal granularity\cite{Gonzalez2008,Jurdak2015}. In particular, some studies have leveraged mobile phone metadata to quantify seasonal and temporary migration movements in developing contexts \cite{Blumenstock2012,Zufiria2018}. However, none of the corresponding datasets of migration estimates have been made publicly accessible, and the methods usually employed to derive migration measures are subject to certain limitations. For instance, migration events are typically identified as a change in the estimated location between two consecutive time periods -- e.g. calendar months -- calculated as the modal location observed during those time periods \cite{Blumenstock2012,Zufiria2018,Hankaew2019,Lai2019}. The regularization of a user's trajectory at a harmonized but coarser temporal resolution -- i.e. by calculating monthly locations -- necessarily causes some measurement error on the exact start and end dates of migration events as well as on their actual duration. It also implies that relatively short migration events with a duration that is comparable to the time resolution considered are potentially missed.\footnote{For example, an individual can be seen at his home location from March 1 to March 16 of some year, then temporarily migrate for 28 days from March 17 to April 14, and return home from April 15 to the end of the month. Since the majority of days in March and April are spent at the home location, a frequency-based method using monthly locations will assign the user to that location in both months and the migration event will not be detected.} Moreover, those methods provide a limited characterization of the direction of migration flows. Since migration events are simply identified as a location change, it is not possible to distinguish between a departure from and a return to a primary home location. Finally, the production of time-disaggregated temporary migration measures poses a number of methodological issues which have not been clearly addressed yet. Most notably, periods of inactivity necessarily induce some degree of uncertainty in the timing and duration of temporary migration events. This in turn creates situations where, for instance, the assignment of an identified migration departure date to a particular time period (e.g. a week or a month) can be ambiguous if the user is unobserved for some period of time before the departure date.

The dataset is a product of a thorough methodological framework meticulously designed to address a number of these issues. The migration event detection algorithm builds on recent work by Chi \textit{et al.}\cite{Chi2020}, who demonstrated how a clustering approach can enhance accuracy compared to traditional frequency-based methods. An important addition relies on the estimation of a primary residence location prior to detecting temporary migration events, which enables the clear characterization of migration flows' direction by distinguishing departures from and returns to a home location. A set of well-defined algorithmic rules allows to aggregate user-level migration trajectories into consistent migration statistics at a desired spatio-temporal scale. They specifically account for issues related to sampling irregularity while maximizing the retention of information contained in phone-derived trajectories. Furthermore, the validation of migration estimates incorporates systematic methods and supplementary data sources to carefully assess the representativeness of a sample of phone users for generating migration statistics. In addition to the intrinsic value of the dataset, this suite of methodological tools thus constitutes a valuable outcome, as these can be readily adapted and applied to other digital trace datasets for systematically generating migration statistics in other contexts.

\section*{Methods}
\subsection*{Call Detail Records}
We use Call Detail Records (CDR) from the main telecommunication company in Senegal (Sonatel) as the primary input for the construction of temporary migration estimates. CDR are mobile phone metadata collected by telecommunication providers for billing purposes. Each data record corresponds to an instance where a user made or received a call (or a text message), and is associated with a set of attributes that typically include: the phone number of the user, the starting time and date of the call and an identifier of the phone tower that processed the call. 

\begin{figure}[H]
\centering
\begin{subfigure}[b]{.45\textwidth}
  \includegraphics[width=\textwidth]{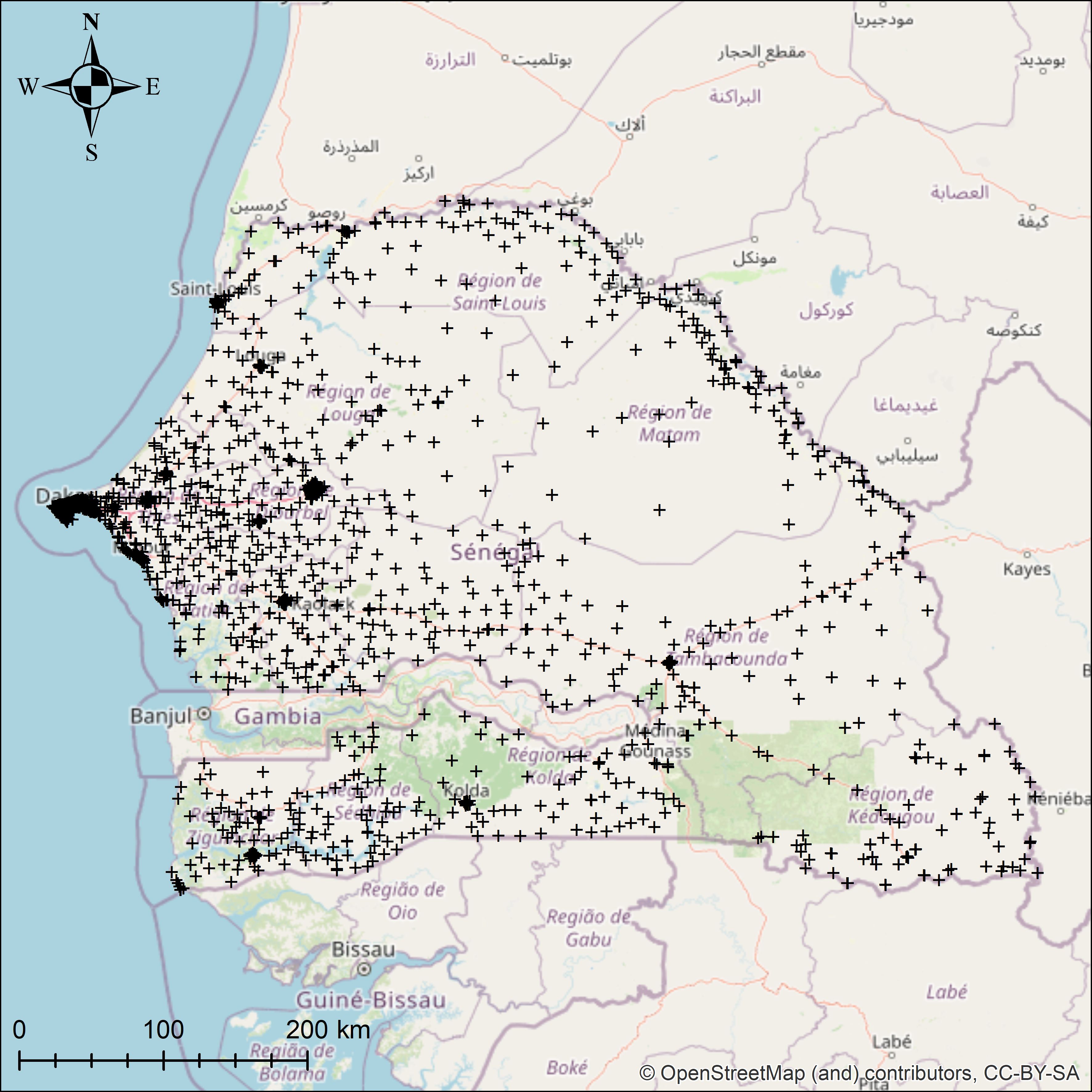}
  \caption{Phone towers}
  \label{fig:bts_map}
\end{subfigure}
\begin{subfigure}[b]{.45\textwidth}
  \includegraphics[width=\textwidth]{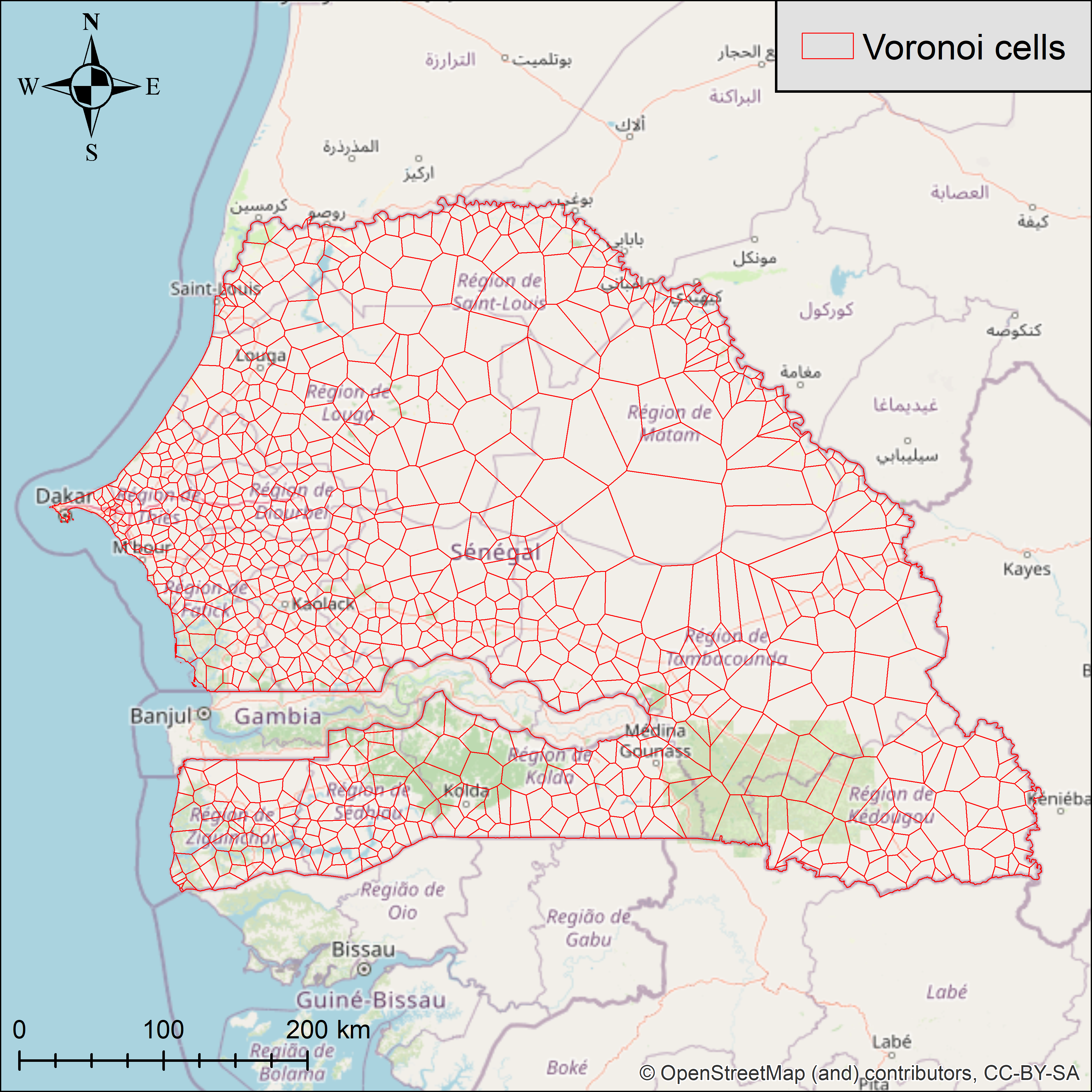}
  \caption{Final voronoi cells}
  \label{fig:voronoi}
\end{subfigure}
\caption{Distribution of phone towers (Figure \ref{fig:bts_map}) and final voronoi cells (Figure \ref{fig:voronoi}) after aggregating cells within cities.}
\label{fig:bts_voronoi}
\end{figure}

A separate dataset provides the point coordinates of each phone tower. For the study period (2013-2015), the Sonatel network was comprised of 2,071 phone towers (Figure \ref{fig:bts_map}). The set of phone tower coordinates is converted into a set of contiguous cells via a voronoi tesselation. Each voronoi cell coincides with the smallest area containing the point location of a device connecting to the corresponding phone tower or, equivalently, it is the approximate area covered by the phone tower. Phone towers are distributed unevenly across the country and their density typically increases with population density. Cells belonging to a single city are thus merged together for mainly two reasons. Firstly, temporary migration is conceptualized as movements across locations such as villages and cities but exclude intra-urban mobility. Secondly, equalizing the sample of cells in terms of their size helps mitigate systematic measurement errors. City polygons are defined based on the GHS Settlement Model 2015 product (GHS-SMOD)\cite{ghssmod}, which identifies 33 urban settlements in Senegal. Voronoi cells intersecting a city polygon are grouped together to form a city cell. However, some secondary urban areas are not captured by GHS-SMOD, resulting in clusters of small cells. Clusters of phone towers that are within a distance of less than 2km from each other are thus detected and the corresponding voronoi cells are merged. This process yields an additional 6 cities. A final network of 916 cells forms a partition of the country extent (Figure \ref{fig:voronoi}) and is comprised of 39 urban cells and 877 rural cells.

To ensure privacy, CDR were pseudonymized by the mobile phone provider via a procedure that replaces phone numbers with unique identifiers. Distinct pseudonymization procedures were applied for the year 2013 and the period 2014-2015. As a result, the unique identifier assigned to a single phone number differs between the two periods and both datasets are processed separately. The 2013 dataset has 9,386,171 unique identifiers and over 28.3 billion records, while the 2014-2015 dataset is comprised of 12,244,494 unique identifiers for over 67 billion observations. To address concerns on the presence of bots and call centers in the sample, 102,313 (resp. 98,086) identifiers that have over 100 records per day on average are removed from the 2013 (resp. 2014-2015) sample -- they account for a total of over 3.5 billion (resp. 5.4 billion) records.

\subsection*{Filtering procedure}

A filtering method is applied to select users satisfying minimal observational constraints, with the objective of ensuring a high level of accuracy in the migration detection procedure. The primary subset used to construct the temporary migration dataset, denoted as \textit{subset A}, consists of users observed for a period covering at least 330 days, on at least 80\% of those days, and with periods of non-observation not exceeding 15 days. The minimal length and frequency of observation are specifically designed to guarantee a baseline level of accuracy in determining users' home location, detecting temporary migration events and estimating departure and return dates. On the other hand, imposing a maximum period unobserved helps mitigate the risk of measurement biases arising from non-random attrition, i.e. periods of inactivity precisely coinciding with users being in migration.

It is important to note that higher observational constraints come at a cost of a lower statistical power since they decrease sample size. Additionally, excluding users based on sampling characteristics may exacerbate selection biases on the cross-section since phone usage patterns can vary with individual characteristics \cite{Blumenstock2010} that potentially correlate with migration decisions. In this respect, previous studies have applied observational criteria aligned with their measurement objectives\cite{BlanchardGollin2023,Blumenstock2012,Hankaew2019,Lai2019}, but have largely disregarded the impact of those constraints on sample composition. Here, we quantify both the benefits of stricter observational constraints, i.e. reduced measurement errors in migration estimates, and their associated drawbacks, i.e. smaller sample sizes and selection biases. Consequently, the selection of filtering parameters defining \textit{subset A} reflects a deliberate trade-off between these factors. Quantitative analyses supporting the choice of these parameters are detailed in the Technical Validation section. 

That being said, while the selection of these parameters involves careful consideration and balancing of the aforementioned benefits and costs, it doesn't follow a formal optimization process and there is inevitably a degree of subjectivity involved. For that reason, we also define a secondary subset with lower observational requirements, denoted as \textit{subset B}, associated with a lower (but still reasonably high, see Technical Validation section) level of accuracy in the migration detection procedure, but a larger sample size and a lower degree of selection induced by the filtering procedure. Migration estimates derived from \textit{subset B} are intended to facilitate robustness checks and enable researchers to test the sensitivity of their results to variations in filtering parameters. Specifically, this subset includes users observed over a period of at least 250 days, with at least 50\% of days observed, and a maximum period unobserved of 25 days. The number of unique identifiers and total number of records for both \textit{subset A} and \textit{subset B} are summarized in Table \ref{tab:sampleSize}. \textit{Subset A} has 1,990,754 unique identifiers in 2013, 2,041,566 for the period 2014-2015, amounting to a total of 47.9 billion records. By contrast, \textit{subset B} has 3,377,994 unique identifiers in 2013, 3,746,640 for the period 2014-2015, and a total of 61.6 billion records.

\subsection*{Migration event detection \label{test}}

The migration event detection algorithm is structured around a conceptualization of human mobility on three distinct scales. First, short-term mobility events such as daily commutes, short trips to cities or weekend getaways are characterized by a short duration, typically a few days. They correspond to movements at a \textit{micro}-scale. Second, temporary migration events correspond to an individual moving from a primary home location to a host area for a period of time going from a couple of weeks to several months before returning to his home location. Those are movements at a \textit{meso}-scale. Third, permanent migration moves imply a long-term change in the usual place of residence and are defined as movements at a \textit{macro}-scale. The time intervals associated with these mobility events are called micro-, meso-, and macro-segments, respectively. For any given individual observed over some period of time, the sets of micro-, meso- and macro-segments constitute three layers of mobility that define the micro-, meso- and macro-location of that individual at any point in time. Note that, in this framework, the macro-location is thus considered as the usual place of residence (i.e. the home location). Given the length of observation and the frequency with which phone users are observed, a raw CDR trajectory generally allows to capture movements at all three scales. As a result, one of the main challenges of identifying segments at a higher scale (e.g., at a meso-scale) is to develop algorithmic methods that smooth out noisy patterns created by movements at lower scales (e.g., at the micro-scale).

\begin{figure}[H]
\captionsetup{width=0.8\textwidth}
\centering
  \includegraphics[width=0.8\textwidth]{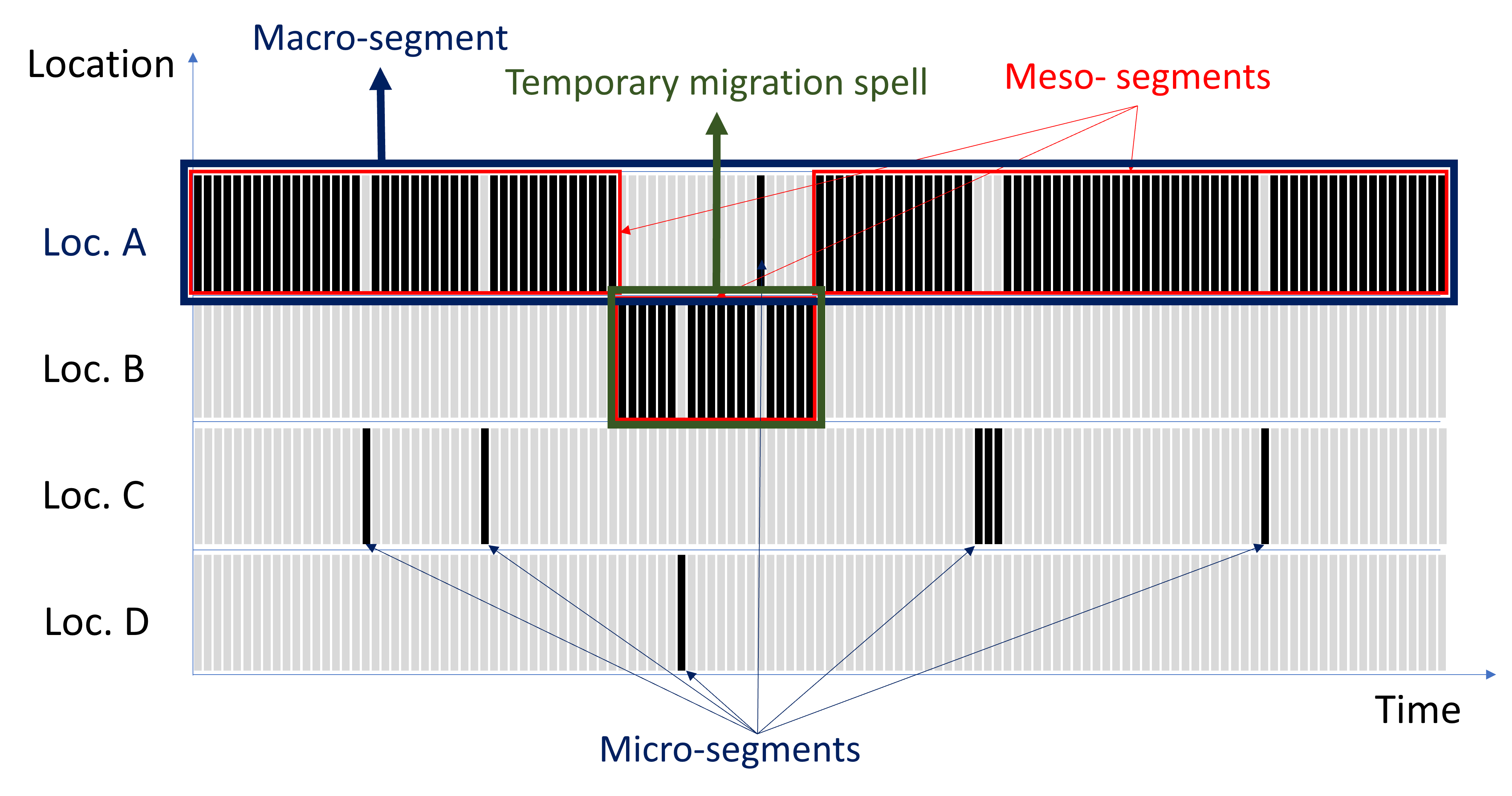}
\caption{Illustration of steps 2 to 4 of the migration detection procedure. Black bars represent the illustrative trajectory of a hypothetical user across four locations: A, B, C and D. The dark thick frame indicates the detection of a macro-segment covering the entire period of observation, which defines the user's home location (A). Red frames describe meso-segments detected with the clustering procedure on daily locations. The green frame designates the only meso-segment detected at a non-home location (B) and thus classified as a temporary migration segment. This CDR trajectory representation is inspired from Figure 1 in Chi \textit{et al.}\cite{Chi2020}.}
\label{fig:migrDetect}
\end{figure}

With these concepts in mind, a four-step methodology is developed to identify temporary migration events in individual CDR trajectories. Firstly, a hierarchical frequency-based procedure is implemented to estimate hourly, daily and monthly locations for each user over his period of observation (\textit{Step 1}). Then, a clustering method is applied to monthly locations to detect macro-segments, which allows to define the usual place(s) of residence over the observed period (\textit{Step 2}). A similar clustering algorithm is applied to daily locations for the detection of meso-segments (\textit{Step 3}). Finally, temporary migration events are identified by overlaying meso- and macro-segments: they correspond to meso-segments at a location which is not the usual place of residence (\textit{Step 4}). Figure \ref{fig:migrDetect} provides a schematic illustration of this migration detection procedure.

To effectively differentiate between long micro-segments and short meso-segments, as well as long meso-segments and short macro-segments, empirical criteria on the duration of mobility events are essential. In this respect, the detection algorithm considers meso-segments with a duration ranging from $\tau_{meso}^{min}=20\;days$ and $\tau_{meso}^{max}=180\;days$. Consequently, macro-segments are naturally defined as periods of at least $\tau_{meso}^{max}$ reflecting the continuous presence of a user at a single location at the macro scale. The relatively low value for $\tau_{meso}^{min}$ allows to capture short migration events, which are more prevalent and often overlooked in survey data compared to longer-term migration spells\cite{Coffey2015}. On the other hand, the choice for the upper-bound duration $\tau_{meso}^{max}$ is mainly constrained by sample characteristics. Specifically, it represents the longest temporary migration events detectable given the observation span of users. As a basic heuristic, a migration event of a certain duration can be detected in a CDR trajectory if the total length of observation is at least twice as long as the migration spell. Indeed, this is the limit over which it is possible to determine that the user spent the majority of his time at a location that can effectively be identified as his primary home location, which then allows to correctly identify the period of time at a distinct location as a temporary migration event. Given that we consider users with a minimum length of observation of approximately a year, we set $\tau_{meso}^{max}=180\;days$. However, parameters $\tau_{meso}^{min}$ and $\tau_{meso}^{max}$ could be flexibly adjusted in future applications based on specific research needs and data constraints.

\subsubsection*{Step 1: hourly, daily, monthly locations}

First, some useful notations and definitions are in order. The studied area is partitioned into contiguous, non-overlapping spatial units that define the full set of potential locations where users can be observed, denoted by $ \mathcal{L}=(\ell_k)_{k\in[1;L]}$, with $L$ the total number of locations. In the present case, $\mathcal{L}$ is the set of voronoi cells introduced above so that $L=916$. The raw CDR trajectory of a user $i$ is denoted by $(x_{t_1}^{i},x_{t_2}^{i},...,x_{t_{T_i}}^{i})$, where each $x_{t}^{i}\in \mathcal{L}$ represents $i$'s observed location at timestamp $t$. $T_i$ is $i$'s total number of CDR.

Consistent with conventional methodologies outlined in previous studies\cite{Blumenstock2012,Zufiria2018,Hankaew2019,Lai2019,Blumenstock2023}, a hierarchical frequency-based method is implemented to determine hourly, daily, and monthly locations. For a user $i$, the hourly location $x_{h,d}^i$ for an hour $h$ of day $d$ is defined as the most frequently visited location during that one-hour time interval, denoted $h_d$:
\begin{equation}
    x_{h,d}^i=mode\bigl\{ \: x_{t}^{i} \: \bigl| \: t\in(t_1,...,t_{T_i}), \:t\in h_d\bigl\} 
\end{equation}

Hourly locations are then aggregated up to daily locations, which are calculated as the most frequent hourly location. As is customary in the literature, night hours between 6pm and 8am are preferred to determine daily locations in order to mitigate the influence of daytime location shifts (e.g. commuting) and maximize the likelihood that the inferred location effectively coincides with the location where the corresponding user spends the night\cite{Blumenstock2012,Vanhoof2018,Hankaew2019}. To limit the loss of information induced by this filtering procedure, we also calculate daily locations based on daytime hourly location between 8am and 6pm and assign those values to user-days without observations at night. The set of night hours for day $d$ is denoted by $\mathcal{N}_d=\bigl\{ \: (h,d) \: \bigl| \: (h,d)\in\{(18,d),...,(23,d)\} \cup \{(0,d+1),...,(7,d+1)\} \bigl\}$ and the set of daytime hours is $\overline{\mathcal{N}_d}$. Then, $\mathcal{D}_i=\{d_1^i,...,d_{D_i}^i\}$ is the set of $D_i$ observed days for user $i$ so that the daily location of user $i$ on any day $d\in\mathcal{D}_i$ is given by:
\begin{equation}
    x_{d}^i=
    \begin{cases}
    mode\bigl\{ \: x_{h,d}^{i} \: \bigl| \: (h,d) \in \mathcal{N}_d \bigl\} , & \text{if}\: \bigl\{ \: x_{h,d}^{i} \: \bigl| \: (h,d) \in \mathcal{N}_d \bigl\}\neq\varnothing\\
    mode\bigl\{ \: x_{h,d}^{i} \: \bigl| \: (h,d) \in \overline{\mathcal{N}_d} \bigl\} , & \text{otherwise}
    \end{cases}
\end{equation} 

Finally, monthly locations are calculated as the modal daily location over a month, with a minimum of 10 days observed imposed in order to guarantee some degree of confidence in the estimated monthly location.

\subsubsection*{Step 2: Macro-segment detection}

Step 2 focuses on the identification of macro-segments, defined as periods of at least $\tau_{meso}^{max}$ during which a user remains consistently present at a single location, while permitting short-term movements (i.e. micro-segments) and temporary migration (i.e. meso-segments) at other locations. The macro-segment detection algorithm uses a clustering procedure on monthly locations, as the frequency-based approach outlined above serves as a simple method to smooth out micro-segments from a raw CDR trajectory. Then, the clustering technique follows the main principles outlined in Chi \textit{et al.}\cite{Chi2020}'s methodology and proceeds in four steps:
\begin{enumerate}[label=(\roman*)]
    \item \underline{Preliminary unique home location estimation}:\\
    A default unique home location $\overline{home_i}$ is estimated for each user $i$. It corresponds to the most frequently observed daily location over $i$'s period of observation:
    \begin{equation}
        \overline{home_i}=mode\bigl\{ \: x_{d}^{i} \: \bigl| \: d \in \mathcal{D}_i \bigl\}
    \end{equation}
    
    \item \underline{Detect contiguous monthly locations}:\\
    Consecutive months at the same location are grouped together, allowing for observation gaps of at most $\epsilon_{gap}^{macro}$ months. $\epsilon_{gap}^{macro}$ is set such that no movement at the macro-scale (i.e. a permanent migration) could occur during unobserved periods. Since macro-segments are defined as periods of at least $\tau_{meso}^{max}$, $\epsilon_{gap}^{macro}$ is set (approximately) equal to $\tau_{meso}^{max}$, $\epsilon_{gap}^{macro}=6\;months\approx180\;days$. Note that, in practice, observation gaps are much shorter given the constraint imposed on the maximum period of non-observation.
    
    \item \underline{Merge monthly location groups}:\\
    Groups of months at a single location are then merged when they are separated by one or more groups accounting for a total duration strictly less than $\tau_{meso}^{max}$. This process essentially groups home stays that may be interspersed with temporary migration spells. 
   
    \item \underline{Resolve overlap}:\\
    Next, the overlap between merged groups that may result from the previous step is resolved. First, merged groups with a duration strictly lower than $\tau_{meso}^{max}$ are removed: as per the definition adopted, they cannot be macro-segments. For two consecutive overlapping groups, overlapping months are assigned to the longest group. Start and end dates of merged groups are updated accordingly and merged groups which now have a duration strictly lower than $\tau_{meso}^{max}$ are removed. To address rare cases of multiple overlaps, this procedure is iterated until no overlapping groups are left. For each user, the final merged groups form his set of detected macro-segments. Given relatively low rates of permanent migration and limitations due to the length of observation relative to $\tau_{meso}^{max}$, the vast majority of users end up with only one macro-segment detected. Those users are assigned the default unique home location determined in the preliminary step, which defines a unique macro-segment for the entire period of observation.
\end{enumerate}

In the illustrative trajectory represented in Figure \ref{fig:migrDetect}, a unique macro-segment corresponding to the dark thicker frame is detected. It defines location $A$ as the usual place of residence over the entire observation period for that hypothetical user.

\subsubsection*{Step 3: Meso-segment detection}

A comparable approach is used to detect meso-segments. The procedure can be decomposed in three steps:
\begin{enumerate}[label=(\roman*)]
    \item \underline{Detect contiguous daily locations}:\\
    Consecutive days at a single location are grouped together, allowing for observation gaps of at most $\epsilon_{gap}^{meso}$. While small values of $\epsilon_{gap}^{meso}$ may fail to smooth out short-mobility events, larger values are associated with significant overlap between groups of days detected. We rely on Chi \textit{et al.}\cite{Chi2020} to determine a reasonable value for $\epsilon_{gap}^{meso}$ and we set it to the optimal value of 7 days they infer from a cross-validation exercise.
    
    \item \underline{Merge daily location groups}:\\
    Groups of daily locations are merged when they are less than $\epsilon_{gap}^{meso}$ days apart. For each user, this results in a set of intermediary meso-segments. Similar to Chi \textit{et al.}\cite{Chi2020}, we filter out meso-segments with a proportion of days at the identified location lower than some parameter $\phi$, that we set to 0.5. This helps to limit cases where a meso-segment might capture frequent movements between multiple locations rather than a temporary migration event at a single location.
    
    \item \underline{Resolve overlap}:\\
    As in the macro-segment detection procedure, merging groups of days at a single location can lead to some overlap between intermediary meso-segments. The overlap between pairs of consecutive segments is resolved by taking the middle of the overlap as the end date of the first segment and the following day as the start date of the second one. This process is iterated until no overlap is left.
\end{enumerate}

The result of this clustering procedure is illustrated in Figure \ref{fig:migrDetect} where three detected meso-segments are represented with red frames.

Three attributes are determined for each meso-segment: a meso-location, a duration, and the macro-location associated with the period covered by the segment. The meso-location is a direct output of the meso-segment detection procedure. Then, for any segment $S_i$ of a user $i$, a lower-bound duration $minDuration(S_i)$ -- referred to as the ``observed duration'' -- is calculated as the time elapsed between the identified start and end dates of the segment. The upper-bound duration $maxDuration(S_i)$ -- referred to as the ``maximum duration'' -- is the time elapsed between the observed day just preceding the segment and the observed day directly following $S_i$. The relative gap between the lower- and upper-bound duration estimates thus represents the uncertainty in the meso-segment duration measure. Finally, the macro-location associated with the meso-segment is straightforward for users with a unique macro-segment, which constitutes the majority of cases. For other users with multiple macro-segments across the period of observation, if a meso-segment is entirely covered by a macro-segment, it is assigned the corresponding macro-location. If a meso-segment overlaps between two macro-segments, it is assigned the macro-location of the macro-segment with the largest overlap.

\subsubsection*{Step 4: Identification of migration events}

Temporary migration events are identified as meso-segments with a duration of at least $\tau_{meso}^{min}$, occuring at a destination that differs from the macro-location -- which defines the home location at the time of the mobility event. For instance, in the illustration provided in Figure \ref{fig:migrDetect}, three meso-segments are detected (highlighted in red frames), all presumed to have an observed duration of at least 20 days. Among them, only the second meso-segment exhibits a meso-location (location $B$) distinct from the macro-location (location $A$) and is therefore identified as a temporary migration event (green frame).

\subsection*{From user-level migration history to migration statistics}
\subsubsection*{Weighting scheme}

In conventional surveys, statistics on a target population are derived from a sample of individuals. The extrapolation from the sample to the population level is permitted by a meticulously defined sampling process. Individuals are selected from a sampling frame, which represents the target population, using a well-defined sampling design. However, mobile phone data simply provide a selected subset of the population, which composition is not governed by a similar sampling procedure. Because phone ownership and usage patterns vary among different demographic groups\cite{Blumenstock2010,Wesolowski2012b,Lai2019}, directly inferring population-level statistics from a sample of mobile phone data is inherently subject to sampling biases. The size of these biases ultimately depend on the magnitude of migration behavior differentials between phone users and non-users, combined with the prominence of non-users within the target population. Moreover, since a statistical bias represents the difference between a sample-based statistic and the true value in a target population, its magnitude is contingent on how this target population is actually defined.

With this in mind, the selection issue in the production of phone-based migration statistics can be addressed in mainly two ways. First, the target population can be simply restricted to a minimal subset that the data effectively represent. Second, some degree of extrapolation to a larger target population can be achieved by using observable characteristics of users to implement correction methods. Both approaches are considered in two distinct sets of migration estimates.

\begin{figure}[H]
\captionsetup{width=0.8\textwidth}
\centering
  \includegraphics[width=0.8\textwidth]{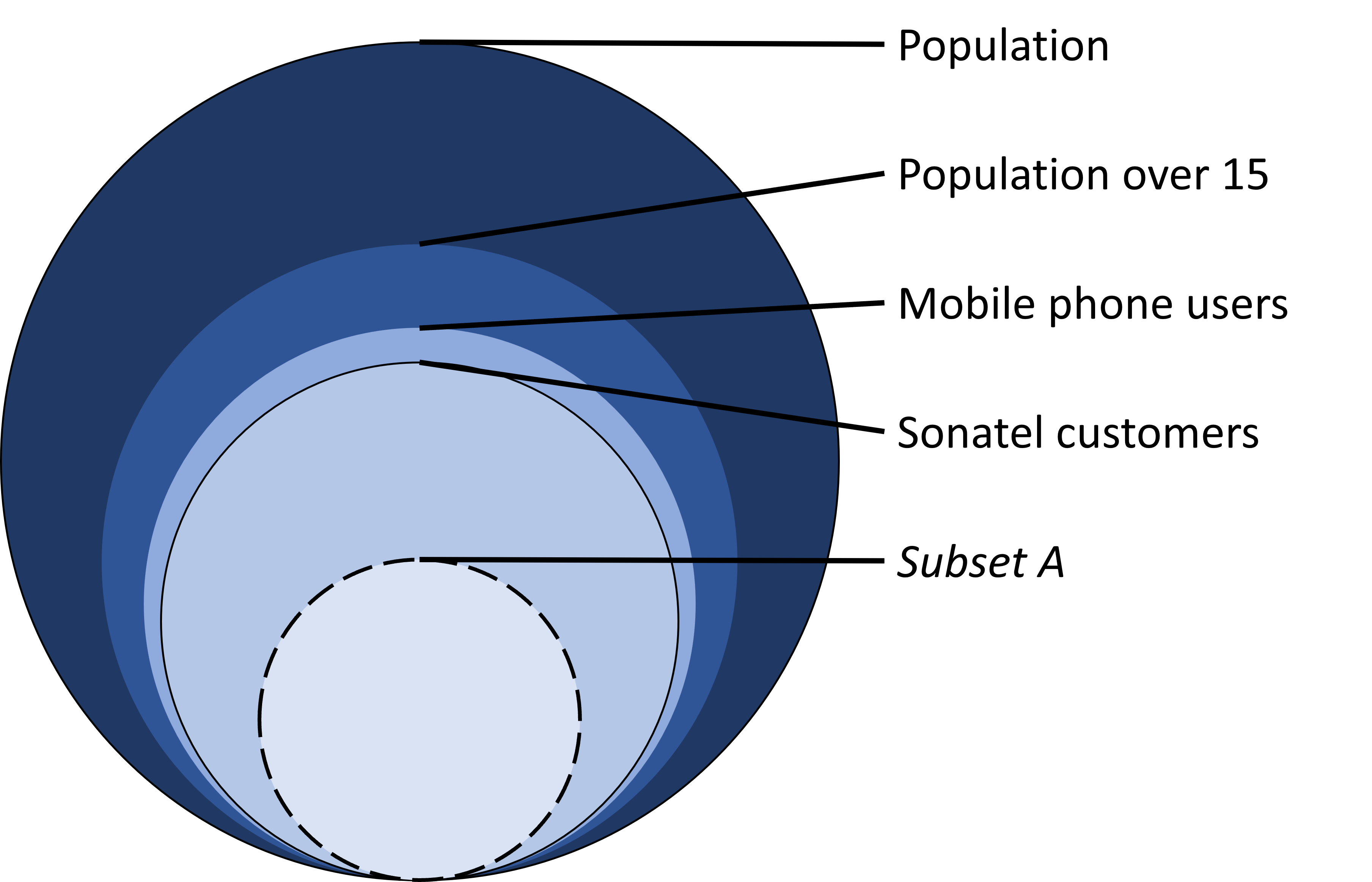}
\caption{Relative size of various subsets of the population in Senegal. The diagram is at scale, which means that the area of each disk represents the size of the corresponding subset relative to the total population.}
\label{fig:vennDiagram}
\end{figure}

The first one is comprised of statistics directly derived from a given subset (i.e., \textit{subset A} or \textit{subset B}). They are referred to as the \textit{unweighted} estimates. Operating under the minimal assumption that the migration outcomes of users in the subset are comparable to those of the overall population of phone users, the sample of users is considered as representative of that population. Evidence supporting this assumption is provided in the Technical Validation section. As a result, the target population associated with the \textit{unweighted} estimates is confined to the subset of mobile phone users, which constitutes a sizable portion of the adult population (see Figure \ref{fig:vennDiagram}). According to the 2014 \textit{Listening to Senegal} survey\cite{L2S2014}, mobile phone users comprised 72\% of the population over 18, thus constituting at least 37\% of the entire population.

A second set of migration estimates, called the \textit{weighted} estimates, is produced with a correction method that allows to consider a broader target population, extending to the entire adult population (i.e. the population over 15). This choice is motivated by the fact that mobile phone ownership among individuals below 15 is indeed considered as negligible. This segment of the population is entirely absent from the mobile phone dataset, and their movements are unlikely to be captured. In fact, estimates of mobile phone ownership by age derived from the 2017 Demographic and Health Survey (DHS)\cite{dhs} effectively reveal a notable decrease among individuals aged between 20 and 15, from over 75\% to 23\%. Moreover, it is assumed that local differences in migration outcomes between users in the sample and individuals in the target population are small, which we refer to as the \textit{local representativeness} assumption. Specifically, we define 39 urban strata -- coinciding with the 39 identified cities in Senegal -- and 185 rural strata, which correspond to the rural areas of third-level administrative units further segmented into areas with low population density (i.e., below the rural median) and high population density (i.e., above the rural median). Differences in migration outcomes between users and the target population are then presumed limited within each individual stratum. In the Technical Validation section, we provide evidence supporting the notion that differences at a local level between phone users and the target population are effectively limited. Then, a correction method is implemented to neutralize imbalances in a key characteristic that is easily observable in CDR data: their home location. Within each stratum, users are assigned a weight equal to the ratio of the stratum-level target population over the total number of observed users identified as residing in that stratum. Moreover, we allow for weights to vary over time to accommodate the fluctuating number of users actually observed across time units. For any location $\ell$ and time period $t$, the value of the weight $w_{\ell t}$ is then:
\begin{equation}
    w_{\ell t}=\frac{pop_{\ell}}{N_{\ell t}^{users}}
\end{equation}
Where $pop_{\ell}$ is the size of the target population in location $\ell$ and $N_{\ell t}^{users}$ is the total number of users residing in $\ell$ who are effectively observed during time period $t$.

Consequently, for any given time period, the sum of weights across users is equal to the total population over 15. In short, the weighting scheme corrects for disparities in the population-to-users ratio across strata, which are primarily caused by variations in mobile phone ownership and usage. For instance, urban areas are generally over-represented in the sample: in \textit{subset A}, 80\% of users live in cities whereas those only account for 54\% of the population over 15. As a result, under the \textit{local representativeness} assumption, the \textit{weighted} migration estimates are unbiased estimators for the true migration outcomes of the population over 15. 

Note that the weighting scheme is designed as if the sample had been randomly drawn at the stratum-level, with the fraction of individuals selected from the target population varying across strata. However, in practice, other forces drive the underlying selection mechanism, and a limitation of the rectification method is its failure to account for these biases. For instance, the sample of users often disproportionately represents men, even within strata. Future research could enhance the proposed weighting scheme by incorporating socio-demographic information -- either made available by the data provider or inferred from usage patterns\cite{Wesolowski2013,Felbo2015} -- in order to address these local sampling biases. Despite the acknowledged and documented issue of selection in CDR samples, the literature has generally overlooked concrete correction methods to address it for constructing representative mobility measures. Hence, we argue that the proposed weighting scheme and its underlying logic represent a significant improvement for producing near-representative migration statistics from a non-random sample of digital traces.

\subsubsection*{Regularizing unbalanced user-level trajectories}

The migration detection model furnishes the location history of each individual user in the form of successive meso-segments. Migration statistics are derived by aggregating these heterogeneous trajectories at a specific spatio-temporal resolution. For any given time unit $t$ and pair of locations $o$ and $d$, it is possible to calculate migrations flows during $t$, i.e. the number of migration departures from $o$ to $d$ and returns from $d$ back to $o$, as well as the migration stock, which corresponds to the number of users residing in $o$ being in migration at destination $d$ during $t$.

\begin{figure}[H]
\centering
\begin{subfigure}[b]{\textwidth}
  \includegraphics[width=\textwidth]{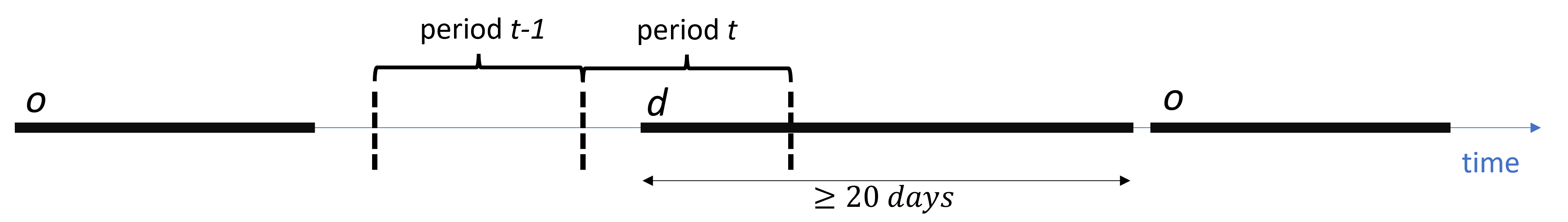}
  \caption{Uncertain departure date}
  \label{fig:departure_uncertainty_a}
\end{subfigure}

\begin{subfigure}[b]{\textwidth}
  \includegraphics[width=\textwidth]{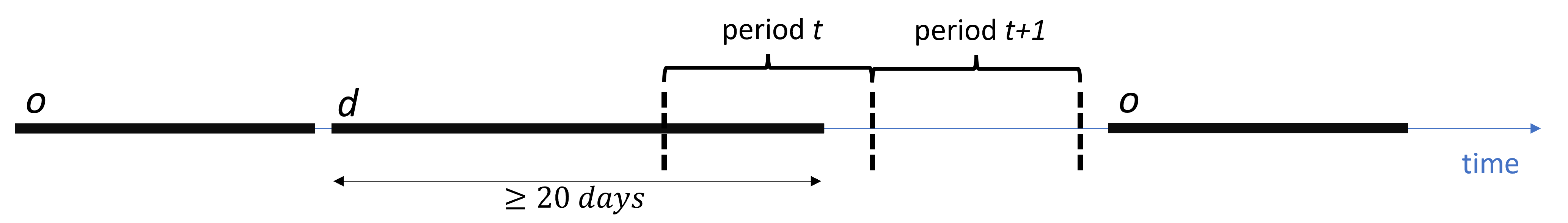}
  \caption{Uncertain return date}
  \label{fig:return_uncertainty_a}
\end{subfigure}

\begin{subfigure}[b]{\textwidth}
  \includegraphics[width=\textwidth]{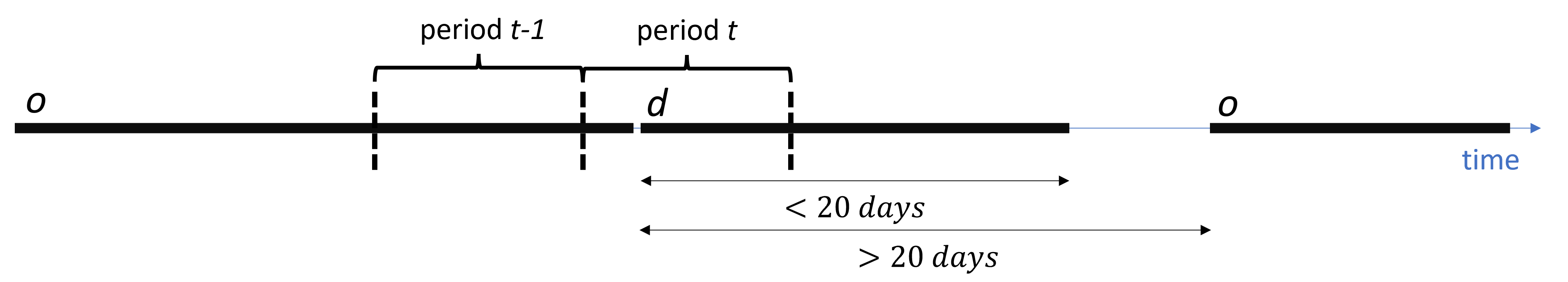}
  \caption{Unambiguous departure date, uncertain duration}
  \label{fig:departure_uncertainty_b}
\end{subfigure}

\begin{subfigure}[b]{\textwidth}
  \includegraphics[width=\textwidth]{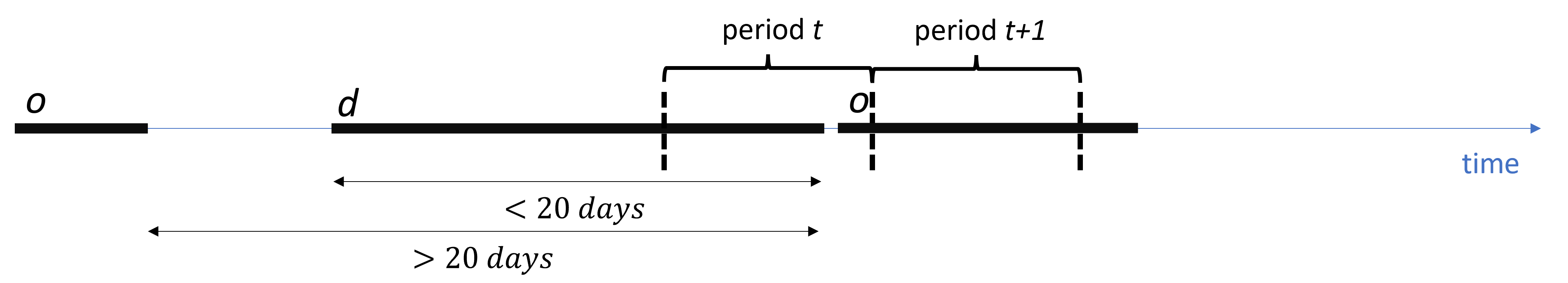}
  \caption{Unambiguous return date, uncertain duration}
  \label{fig:return_uncertainty_b}
\end{subfigure}

\caption{Uncertainty in the calculation of migration flows by time unit.}
\label{fig:flow_uncertainty}
\end{figure}

A user $i$ residing in $o$ is considered to have departed for migration to destination $d$ at time $t$ if he has a migration meso-segment at $d$ that started during $t$. Similarly, user $i$ returned from $d$ to his home location $o$ at time $t$ if a migration segment at $d$ ended during $t$. However, observational gaps imply some degree of uncertainty regarding the actual start and end dates of meso-segments, thereby complicating the computation of migration flows in practice. Illustrative examples are shown in Figure \ref{fig:flow_uncertainty} considering a minimum duration of $\tau_{meso}^{min}=20\;days$ to define temporary migration events. In Figure \ref{fig:departure_uncertainty_a}, user $i$ residing in $o$ has a migration segment at destination $d$ with an observed departure date within period $t$. However, $i$ is unobserved in period $t-1$, rendering it uncertain as to the specific period when the migration departure actually occurred (i.e. $t$, $t-1$, or $t-2$). Similarly, in Figure \ref{fig:return_uncertainty_a}, a migration segment ends within period $t$ but the observational gap that follows raises the possibility for user $i$ to have returned home in period $t+1$ or $t+2$. These ambiguities are partly resolved by introducing a tolerance parameter $\epsilon^{tol}$. In situations analogous to that of Figure \ref{fig:departure_uncertainty_a}, $\epsilon^{tol}$ sets the maximum acceptable time unobserved before the start of period $t$ to still consider that user $i$ departed for migration at $d$ during period $t$. Likewise, for migration returns, $\epsilon^{tol}$ sets the maximum acceptable time unobserved after the period when the user is seen returning home in order to consider that the user effectively returned during that time period. The migration statistics disaggregated by half-month provided in the dataset are produced with $\epsilon^{tol}$ equal to 7 days. 

Then, uncertainty on the start and end dates of meso-segments naturally leads to some level of uncertainty on the actual duration of meso-segments, which gives raise to a second category of ambiguous cases. For example, in Figure \ref{fig:departure_uncertainty_b}, the start date of the segment at destination $d$ unequivocally falls within period $t$, but the observed duration is lower than 20 days and the segment is not classified as a migration segment. Yet, the observational gap following the segment indicates that its actual duration may possibly be greater than 20 days, in which case user $i$ should be regarded as having departed for migration at time $t$. Figure \ref{fig:return_uncertainty_b} shows a similar situation where the return date is unambiguously assigned to period $t$ but the uncertain duration induced by the observational gap preceding the segment complicates its classification as a migration segment. The migration estimates provided in the dataset are simply based on meso-segments with an observed duration ($minDuration$) greater than $\tau_{meso}^{min}$, which are referred to as \textit{high-confidence} estimates. ``Lower-confidence'' estimates of migration departures and returns were also produced considering meso-segments with an observed duration lower than $\tau_{meso}^{min}$ but a maximum duration ($maxDuration$) greater than $\tau_{meso}^{min}$, similar to scenarios depicted in Figure \ref{fig:departure_uncertainty_b}-\ref{fig:return_uncertainty_b}. Many such configurations giving rise to ambiguous cases are possible, and an exhaustive set of algorithmic rules were implemented to address each one of them. In practice, due to the rather strict observational constraints imposed (as outlined in the Filtering procedure section), uncertainty on the actual duration of meso-segments remains minimal. As a result, these lower-confidence estimates show negligible difference with the primary high-confidence estimations, and the dataset therefore exclusively contains high-confidence estimates. Nonetheless, a comprehensive description of the algorithm's treatment of various configurations, along with illustrative diagrams, is furnished in the Supplementary Material. This empowers researchers to apply the methodology to alternative digital trace datasets not necessarily exhibiting comparably high sampling frequencies, while also aiding in the understanding of the code.

\begin{figure}[H]
\centering
\begin{subfigure}[b]{\textwidth}
  \includegraphics[width=\textwidth]{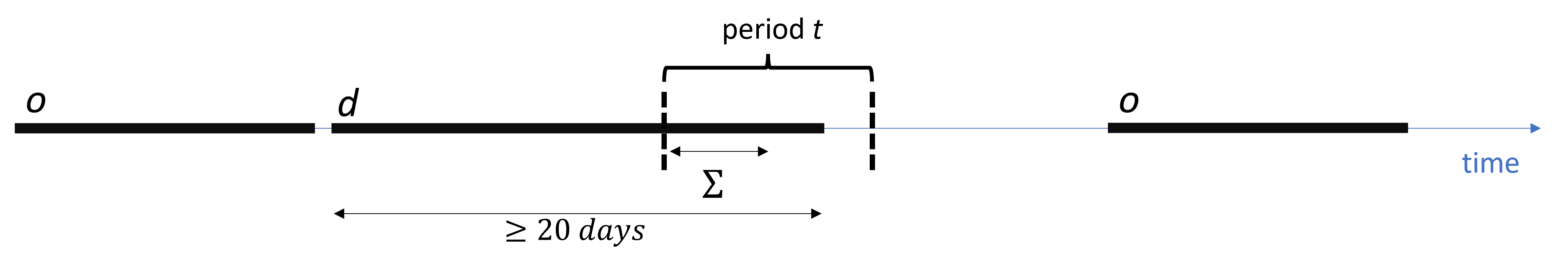}
  \caption{A user in migration in time period $t$}
  \label{fig:stock_certain}
\end{subfigure}

\begin{subfigure}[b]{\textwidth}
  \includegraphics[width=\textwidth]{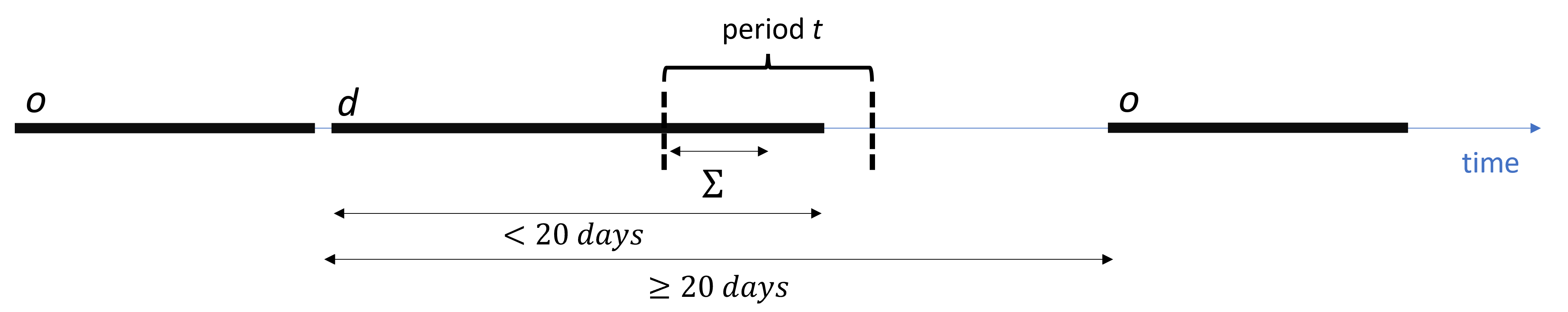}
  \caption{Ambiguous migration status due to uncertainty in segment duration}
  \label{fig:stock_uncertainty_a}
\end{subfigure}
\caption{Determining the migration status of a user in period $t$.}
\label{fig:stock_uncertainty}
\end{figure}

Then, the migration stock from $o$ to $d$ during a time period $t$ is calculated by aggregating the migration status of users, i.e. whether a user is in migration or not at time $t$. A user $i$ is defined as being in migration at time $t$ if that user exhibits a migration segment that overlaps time period $t$ on at least $\Sigma$ days (see Figure \ref{fig:stock_certain}). Migration stock estimates provided in the dataset are generated with a value of $\Sigma$ equal to 8 days. With this value, we simply impose that the overlap represents at least half the time unit since half-months have a duration of at most 16 days.

Determining the migration status of user $i$ for a time period $t$ can also be subject to some ambiguities, arising primarily from uncertainty in the duration of meso-segments. For instance, in Figure \ref{fig:stock_uncertainty_a}, user $i$ may or may not be in migration at destination $d$ in period $t$, depending on his actual location during the following observation gap. The possibility exists that the segment has an actual duration greater than 20 days, in which case $i$ should be considered as being in migration at time $t$. High-confidence estimates of migration stocks are derived exclusively from meso-segments with an observed duration greater than $\tau_{meso}^{min}$. Lower-confidence estimates were also calculated considering meso-segments with an observed duration below $\tau_{meso}^{min}$ but a maximum duration greater than $\tau_{meso}^{min}$. Again, these two sets of estimates show little difference in practice and only high-confidence estimates are included in the dataset, but a comprehensive description of algorithmic rules along with illustrative diagrams are provided in the Supplementary Material.

Finally, estimating time-disaggregated migration rates requires some measures of the actual number of users observed at any given time period $t$, serving as the denominator for such rates. Similar to the computation of migration flows and stock, the presence of observational gaps and attrition introduces variations in the number of users observed over time in each location. In essence, a user $i$ is classified as ``observed at time $t$'' for a specific migration measure (i.e. departures, returns, or stock) if their trajectory allows to unambiguously determine his migration status at time $t$ for that migration measure -- e.g. $i$ departed for migration during $t$ or did not depart for migration during $t$. For example, in the scenario depicted in Figure \ref{fig:departure_uncertainty_a} and assuming that the tolerance parameter is exceeded, the user would be deemed unobserved for the calculation of the departure rates at time period $t$. Again, all possible configurations in the trajectory of users are analyzed and all cases where users are considered as unobserved for measures of migration departures, returns and stock, are identified. Details on each of those cases along with illustrative diagrams are provided in the Supplementary Material, facilitating the understanding of the rules implemented as well as the corresponding code. It is important to highlight that the conditions defining the observational status of a user for a time period $t$ depend on the migration measure considered, as well as the minimum migration duration threshold $\tau_{meso}^{min}$, the tolerance parameter $\epsilon^{tol}$ and the parameter $\Sigma$. Additionally, these numbers are also employed in the calculation of weights used to produce the \textit{weighted} estimates.

\section*{Data Records}

Migration estimates are provided at the (origin*destination*time)-level. The origin and destination locations considered are comprised of 39 cities and 112 rural areas of third-level administrative units (i.e., districts), defining the spatial resolution of the dataset. Note that all 39 cities are considered as individual spatial units. Therefore, the dataset contains city-level migration estimates rather than estimations at the level of urban areas for each district. Time units coincide with ``half-months'', defined as the periods going from the 1\textsuperscript{st} to the 15\textsuperscript{th}, and from the 16\textsuperscript{th} to the end of each month. Each year is thus comprised of 24 half-months and the dataset covers the period 2013-2015. 

The full dataset is organized in 12 datasets. Each dataset provides either \textit{weighted} or \textit{unweighted} estimates and is derived from either \textit{subset A} or \textit{subset B}. In addition, for each of these four combinations, separate datasets provide estimates considering only migration events with a duration of at least 20, 30, or 60 days. A standard file name \texttt{type\_X\_DDdays.csv.gz} is used to uniquely identify each dataset, with \texttt{type} being either \texttt{weighted} or \texttt{unweighted}, \texttt{X}$\in\{$\texttt{A},\texttt{B}$\}$ denotes the subset from which migration estimates are derived, and \texttt{DD} is either \texttt{20}, \texttt{30}, or \texttt{60}. For instance, the file \texttt{weighted\_A\_20days.csv.gz} contains weighted estimates derived from \textit{subset A} considering temporary migration events of at least 20 days.

Each dataset contains time series of migration departures, migration returns, and migration stock, both in absolute terms and as a fraction of the total number of users observed at origin. These metrics are provided for each origin-destination pair over the period from 2013 to 2015, with time units defined as half-month intervals. Note that half-month periods that coincide with the start and end of each CDR dataset (i.e., the 2013 and 2014-2015 datasets) are excluded from the final estimates due to increased uncertainty at the boundaries of the observation periods. For example, considering migration events of at least 20 days, all users seen at a non-home location from January 1, 2013 to at least January 8, 2013 and returning to their home location before January 20, 2013 are not classified as migrants during the first half of January. However, if they had departed to the non-home location before December 31, 2012, they should indeed be identified as migrants. The extent of these high-uncertainty periods at the edges of CDR datasets depends on the minimum duration used to define temporary migration events. Therefore, for migration estimates based on events of at least 20 days, we only exclude estimates for the first and last half-months of 2013, the first half-month of 2014 and the last half-month of 2015. For events of at least 30 days, estimates for the first and last two half-months of 2013, the first two half-months of 2014 and the last two half-months of 2015 are excluded. For events of at least 60 days, we exclude estimates for the first and last four half-months of 2013, the first four half-months of 2014, and the last two half-months of 2015. All variable names along with detailed descriptions are provided in Table \ref{tab:var_desc}. 

\section*{Technical Validation}
\subsection*{Migration event detection accuracy}

Observational requirements for the measure of human mobility necessarily depend on the type of movements one aims to capture. Broadly speaking, measuring long-term changes in the place of residence requires extended periods of observation (e.g. several years) with modest sampling frequencies, while capturing commuting movements asks for high sampling frequencies (e.g. multiple observations per day) over potentially shorter observation periods. Minimal sampling characteristics for the measure of temporary migration movements is qualitatively somewhere in between. The proposed migration detection algorithm essentially requires that users are seen often enough during a sufficiently long period of time in order to be able to (i) confidently identify a home location and (ii) detect the temporary changes in the usual location observed -- that we have called ``meso-movements''. We investigate this issue in quantitative terms by conducting a sensitivity analysis of the proposed migration detection algorithm with respect to users' observational characteristics. More specifically, we evaluate the impact of the length of time a user is observed and the fraction of days with observations (i.e. the frequency of observation) on the level of accuracy associated with both the prediction of home locations and the detection of temporary migration events. This allows us to provide estimates of the detection accuracy associated with both \textit{subset A} and \textit{subset B} upon which the dataset is constructed, and thus validate the choice of filtering parameters applied to define these subsets.

To do this, we consider a benchmark subset of users in the 2013 dataset that meet stringent observational constraints: they are observed for at least 360 days and on at least 95\% of days (195,070 such users satisfy those constraints). A random subset of 10,000 users is selected among those with a unique home location identified and at least one migration event of at least 20 days detected. The strict observational constraints imposed on this subset allow to consider the migration detection outputs as reflecting (i) the actual home locations of users and (ii) their actual temporary migration moves. Then, we consider a set of observational constraints that we apply to each user by selecting a random subset of CDR of that user that satisfy those constraints. We re-apply the detection algorithm to these subsets of CDR and compare the outputs with those obtained with the full trajectories in order to evaluate the accuracy of the model for the set of observational constraints considered. We reproduce this procedure for various sets of observational constraints to appreciate the overall sensitivity of the detection model to observational characteristics.

\begin{figure}[H]
\centering
\begin{subfigure}[b]{.48\textwidth}
  \includegraphics[width=\textwidth]{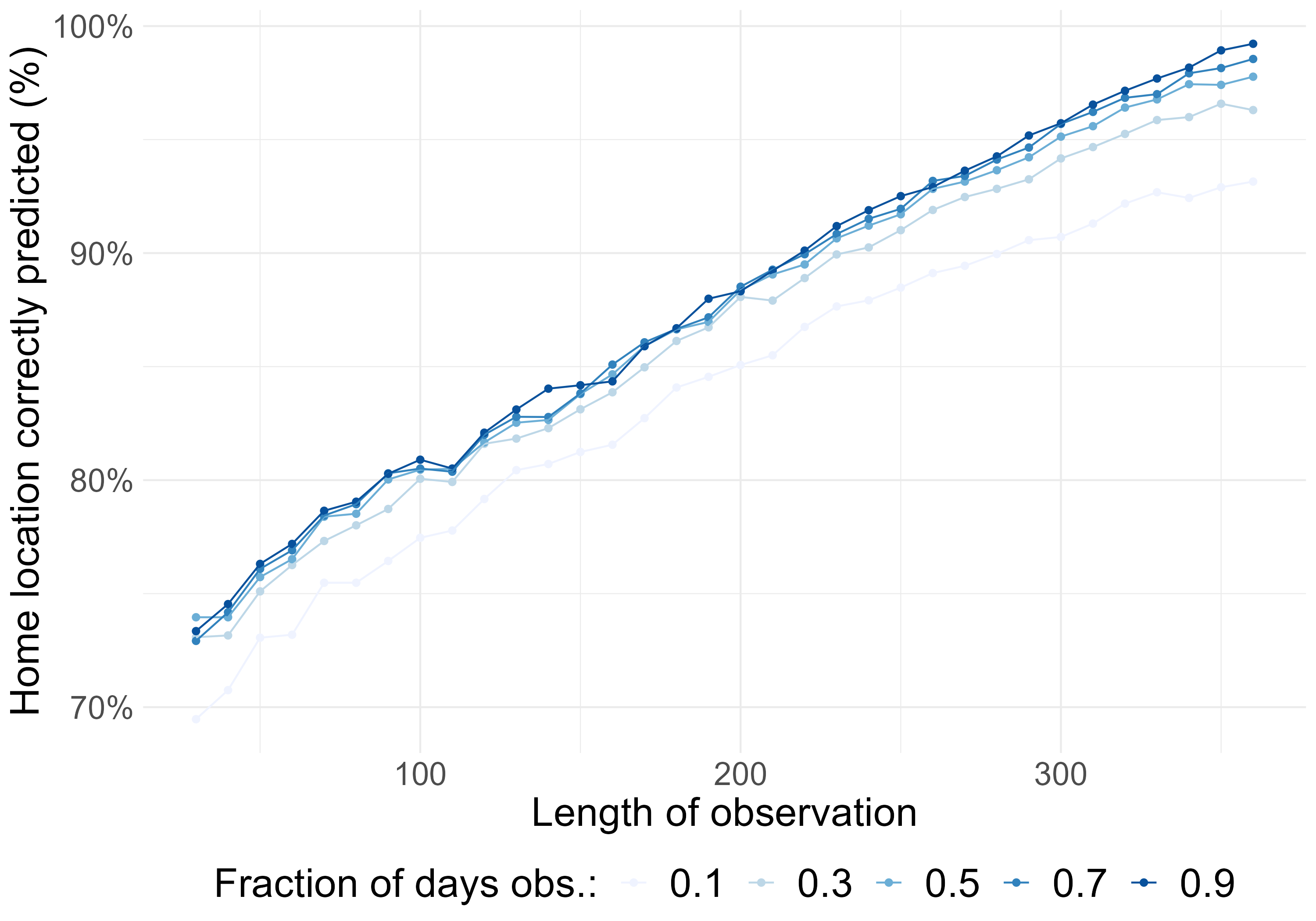}
  \caption{Home estimation accuracy}
  \label{fig:homeAccuracy}
\end{subfigure}
\begin{subfigure}[b]{.48\textwidth}
  \includegraphics[width=\textwidth]{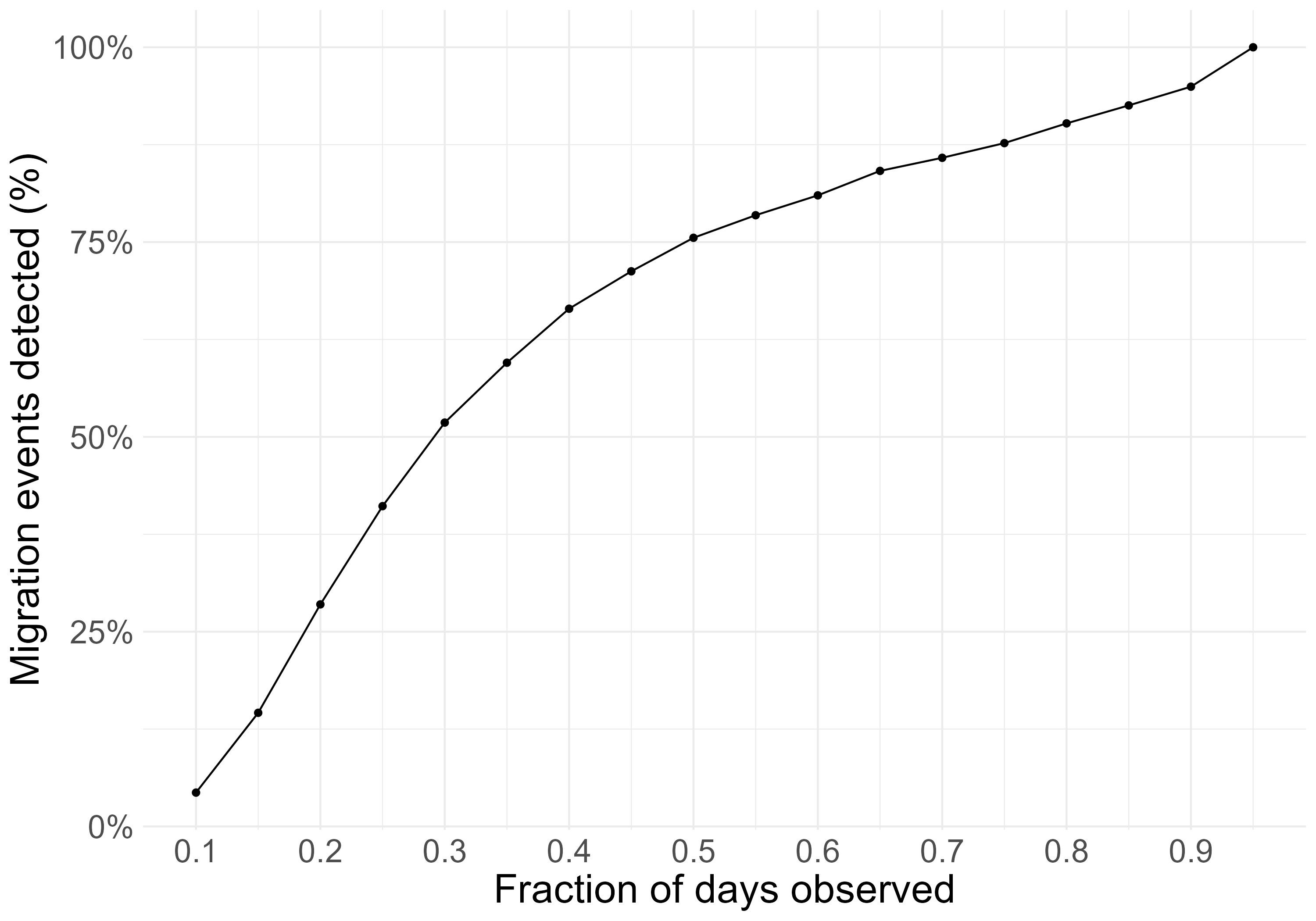}
  \caption{Migration detection accuracy}
  \label{fig:migrationAccuracy}
\end{subfigure}
\caption{Model accuracy for home location predictions as a function of the length and frequency of observation (panel (a)) and accuracy of the migration detection algorithm as a function of the fraction of days observed (panel (b)).}
\label{fig:accuracy}
\end{figure}

First, we evaluate the impact of the length of observation $\Delta$ and frequency of observation $\Omega$ (henceforth also referred to as the ``density'' of the trajectory) on the accuracy of home location predictions. For each set of parameters $(\Delta,\Omega)$, the model accuracy is simply defined as the fraction of users with a correctly predicted home location. Figure \ref{fig:homeAccuracy} shows estimates of the model accuracy for lengths of observation ranging between 30 and 360 days and for different values of $\Omega$. It is clear that the density of trajectories $\Omega$ has little incidence on the accuracy of home location predictions. For instance, even with only 10\% of days observed, the level of accuracy continues to exceed 90\% for lengths of observation of at least 290 days. More generally, for any given length of observation, the level of accuracy only varies by a few percentage points with values of $\Omega$ ranging from 0.1 to 0.9. On the other hand, accuracy seems to increase linearly with the length of observation. For $\Omega=0.9$, it increases from 73\% to 99\% when the length of observation imposed increases from 30 to 360 days. 

Second, we focus on the impact of the frequency of observation $\Omega$ on the accuracy of the migration detection model, holding $\Delta$ fixed.\footnote{Looking at the impact of $\Delta$ on the ability of the algorithm to detect migration events is not particularly relevant. Shorter lengths of observation simply imply missing migration events occurring during the period unobserved.} Here, the model accuracy for any given value of $\Omega$ is defined as the fraction of real migration segments (i.e. those detected in the benchmark subset) that are effectively identified in subsets of CDR of density $\Omega$. Removing observations from a full trajectory can lead to migration events being still detected although with slightly different start and end dates. Therefore, a real migration segment is considered as identified in a subset of density $\Omega$ if a migration segment overlapping at least half the real migration segment is detected in this subset of CDR. Results for $\Delta$ set to 360 days and $\Omega$ varying from 0.1 to 0.95 are provided in Figure \ref{fig:migrationAccuracy}. Unsurprisingly, the frequency of observation has a significant impact on the accuracy of the migration detection model. From 95\% of migration events detected with a density of 0.9, it decreases to as low as 4\% when the fraction of days observed is equal to 0.1. The convex shape of the relationship indicates that the level of accuracy starts to deteriorates sharply when $\Omega$ falls below approximately 0.5, and drops below 50\% for values of $\Omega$ that are less than 0.3. On the other hand, densities greater than 0.8 allow to sustain a high level of accuracy beyond 90\%. 

The accuracy of both home location estimation and migration event detection is then assessed for the observational constraints associated with \textit{subset A} and \textit{subset B} respectively. Results are summarized in Table \ref{tab:accuracy_BySubset}. Home detection accuracy is high for both subsets: 92\% for \textit{subset A} and 98\% for \textit{subset B}. We estimate that at least 90\% of migration events are effectively detected by the algorithm in \textit{subset A}. As expected, this figure is lower for \textit{subset B} but still relatively high at 76\%.

\subsection*{Validity of unweighted estimates: Are Sonatel users representative of the population of phone users?}

Under the assumption that users in \textit{subset A} and \textit{subset B} have temporary migration outcomes that are comparable to those of the overall population of mobile phone users, \textit{unweighted} estimates are considered as representative of that population. 

Given the absence of CDR data for the entire population of mobile phone users, conducting a direct comparison of migration outcomes between our subsets of users and the broader population of mobile phone users is unfeasible. Likewise, without personal information available in CDR data, we cannot compare the characteristics of users in \textit{subset A} and \textit{subset B} with those of the population phone users. However, leveraging ICT Access Surveys\cite{ict} data, we perform statistical tests to at least evaluate whether Sonatel users differ from the population of phone users across various observable characteristics, including gender, age, education, zone of residence and wealth. Results are showed in Table \ref{tab:sonatelVSusers}. Overall, the findings suggest that Sonatel users are generally comparable to the broader population of phone users, particularly in terms of characteristics potentially associated with temporary migration determinants (e.g. assets ownership, wealth, gender). One notable difference is that Sonatel users tend to be slightly more urban than users of other operators (57.4\% against 50.5\%). Nevertheless, the existence of potential differences between Sonatel users and users from other operators remains a minor concern in the context of Senegal. Indeed, as illustrated in Figure \ref{fig:vennDiagram}, the market was largely dominated by Sonatel during the study period: over 88\% of mobile phone users have a Sonatel SIM card and 77\% report Sonatel as their main provider\cite{L2S2014}. However, this issue may assume greater significance in settings where the telephony market is more fragmented.

\subsection*{Validity of weighted estimates: local representativeness assumption and selection on home locations}

The \textit{local representativeness} assumption crucially underpins the production of the \textit{weighted} estimates, allowing to extend the target population to the population over 15. This assumption posits that differences in temporary migration outcomes between users in the sample and the target population remain limited. Although this cannot be directly verified in practice due to the absence of data on temporary movements, we adopt a second-best approach and conduct two distinct exercises that help evaluate the validity of the \textit{local representativeness} assumption. First, we use secondary survey data to compare mobile phone users with the adult population at large at a local level along a number of observable characteristics. Second, we compare the specific phone users in the selected subsets with the overall population with respect to the only characteristic that is readily observable with CDR data, i.e. the home location. 

In the first validation exercise, we leverage data from the 2017 Senegal Demographic and Health Survey (DHS)\cite{dhs}, which focuses on individuals aged 15 and above. The survey provides information on individuals' mobile phone ownership along with various characteristics such as wealth, occupation, financial inclusion, and education level. To assess the \textit{local representativeness} assumption, we employ statistical tests (t-tests) to examine differences between phone users and the overall population across these dimensions. These tests are conducted separately for each zone (rural/urban) within the 14 regions of Senegal to ensure consistency with the representativeness level of the DHS. Results are illustrated for the region of Kaolack in Table \ref{tab:dhs_ttest_kaolack} while results for the 13 other regions are left in the Supplementary Material. The most notable observation is that very few coefficients exhibit statistical significance, a pattern that we consistently observe across all regions and zones. In particular, phone users are generally found to be slightly wealthier compared to the overall population, but those differences are never significant, even at a 10\% level. Similarly, across specific regions and zones, phone users generally tend to have better access to amenities like drinking water, sanitation, and electricity, higher levels of education, lower unemployment rates, and greater participation in the agricultural sector. Again, these disparities are minor and statistically insignificant. These results tend to confirm that, at a local level, phone users and the broader adult population are statistically indistinguishable along numerous key dimensions. Assuming some degree of correlation between those characteristics and mobility choices, this supports the notion that, at a local level, temporary migration outcomes of phone users do not significantly differ from those of the adult population as a whole. Two exceptions are worth highlighting. Unsurprisingly, phone users are older than the overall population over 15 given lower ownership rates in the 15-20 age category. Most notably, a clear gender divide in mobile phone ownership exists and phone users are disproportionately more male, especially in rural areas. In rural Kaolack, 56\% of phone users are male against 48\% in the population over 15 and this difference is statistically significant at a 5\% level. Comparable differences are found in 11 of the 14 rural regions of Senegal. Therefore, a plausible constraint on the validity of the \textit{local representativeness} assumption is that mobile phone data under-represent women and younger individuals.

\begin{figure}[H]
\centering
\begin{subfigure}[b]{.41\textwidth}
  \includegraphics[width=\textwidth]{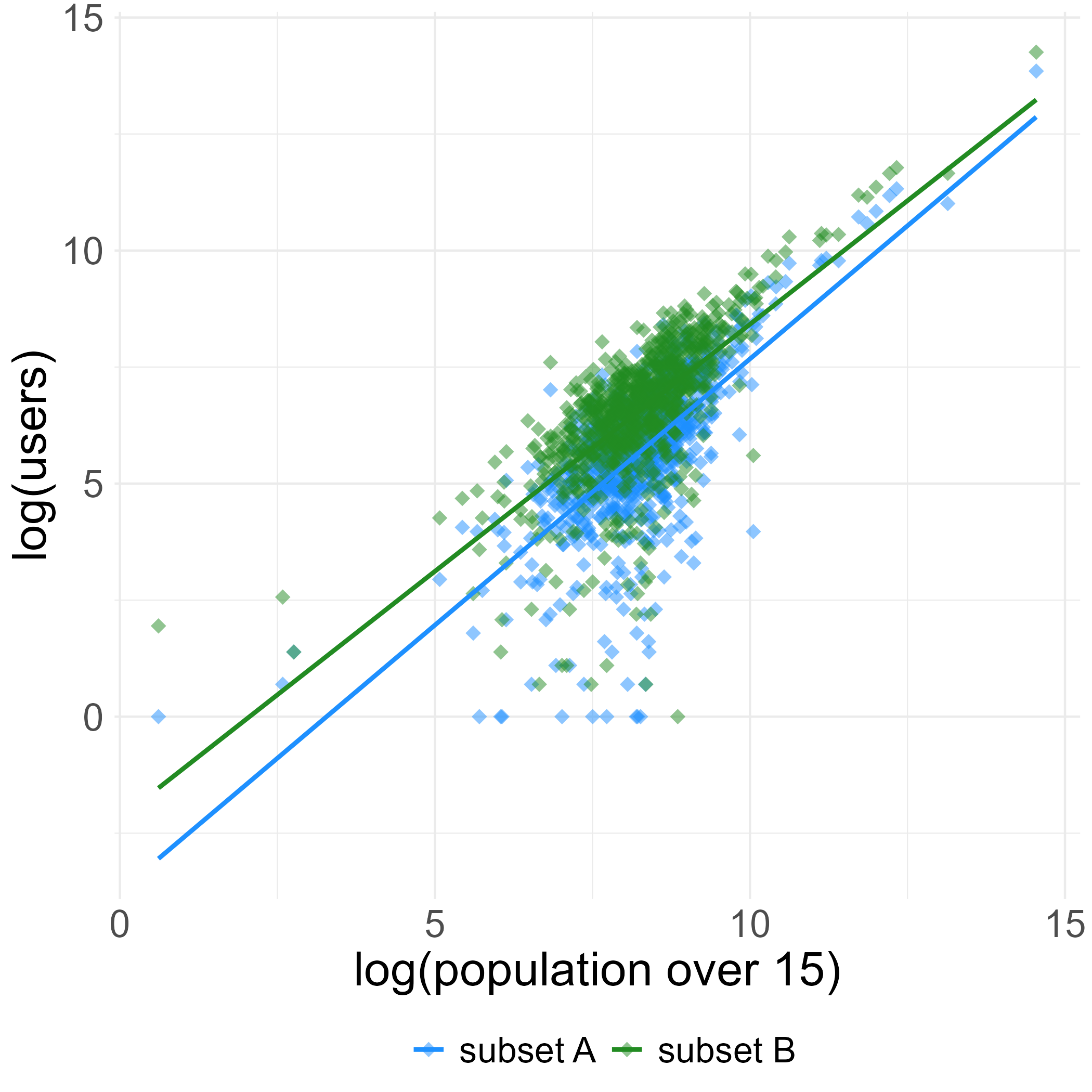}
  \caption{Users versus adult population}
  \label{fig:usersVSpop_a}
\end{subfigure}
\begin{subfigure}[b]{.55\textwidth}
\raisebox{5mm}{
  \includegraphics[width=\textwidth]{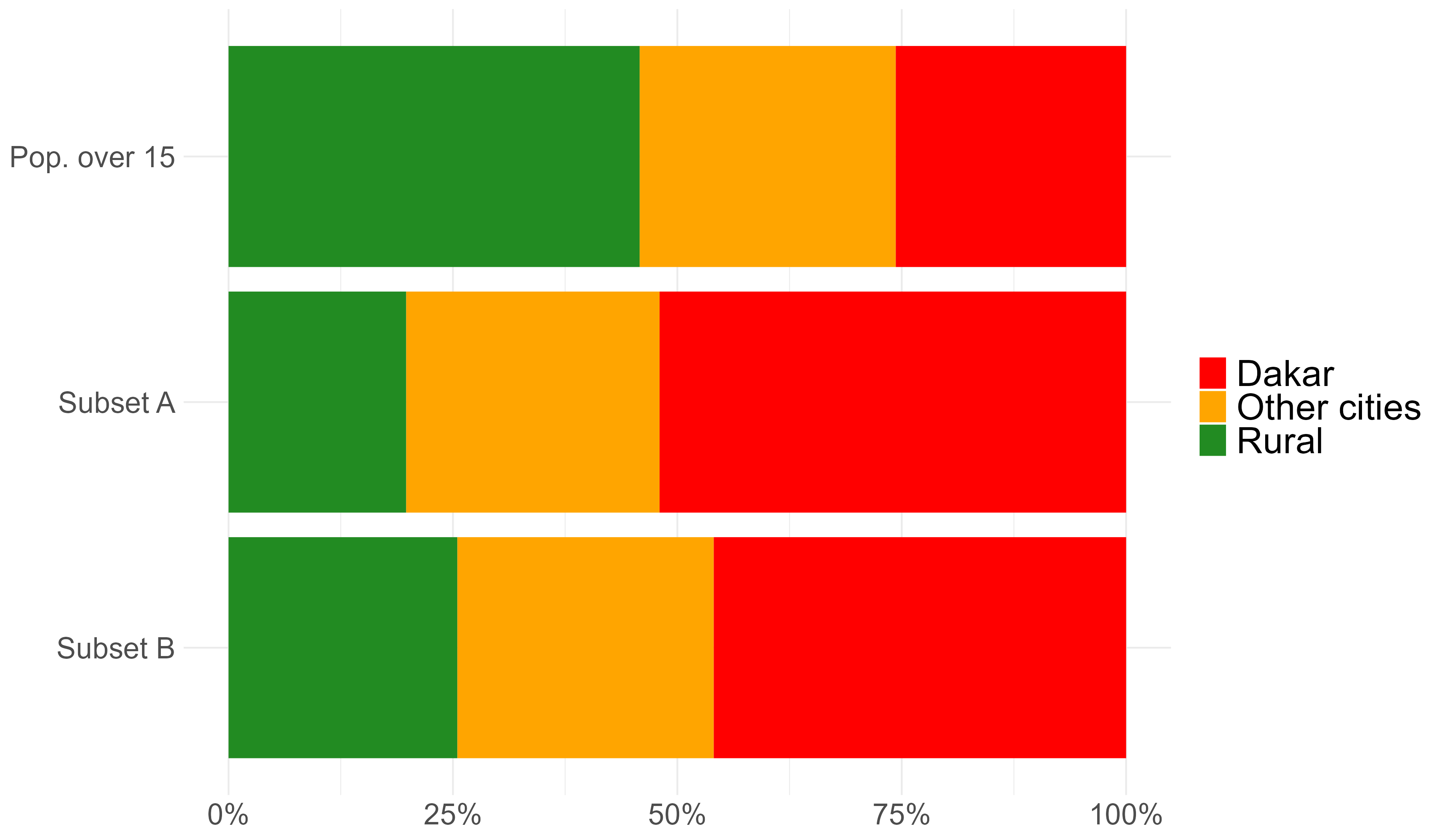}}
  \caption{Distribution of users across three categories.}
  \label{fig:usersVSpop_b}
\end{subfigure}
\caption{Comparison of the distribution of users across locations with the distribution of the population over 15. Panel (a) shows the logged number of users by voronoi cell in the 2013 dataset against the population over 15. The voronoi-level population over 15 is estimated by combining census-based department-level estimations of the fraction of the population over 15\cite{IPUMS} with voronoi-level total population estimates obtained by overlaying voronoi polygons with the 2017 100m-resolution gridded population product from the WorldPop Research Group\cite{Qader2022}. Panel (b) shows the distribution of users in \textit{subset A} and \textit{subset B} in the 2013 dataset across three categories of voronoi cells: Dakar, other cells classified as urban, and cells classified as rural. The figure also provides the distribution of the target population over 15 across these categories for comparison.}
\label{fig:usersVSpop}
\end{figure}

Secondly, we directly compare the phone users of our sample with the adult population based on a characteristic that is easily observable from CDR data: their residence location. Figure \ref{fig:usersVSpop_a} shows the number of users by voronoi cell against the population over 15, revealing a positive but imperfect correlation for both \textit{subset A} and \textit{subset B}. In Figure \ref{fig:usersVSpop_b}, we calculate the distribution of phone users in \textit{subset A} and \textit{subset B} across three categories of locations: Dakar, other urban locations, and rural locations. Comparing this with the distribution of the adult population, we see that a disproportionately high fraction of users are in Dakar, while a lower fraction are in rural areas compared to the overall adult population.

\begin{figure}[H]
\centering
\begin{subfigure}[b]{.41\textwidth}
  \includegraphics[width=\textwidth]{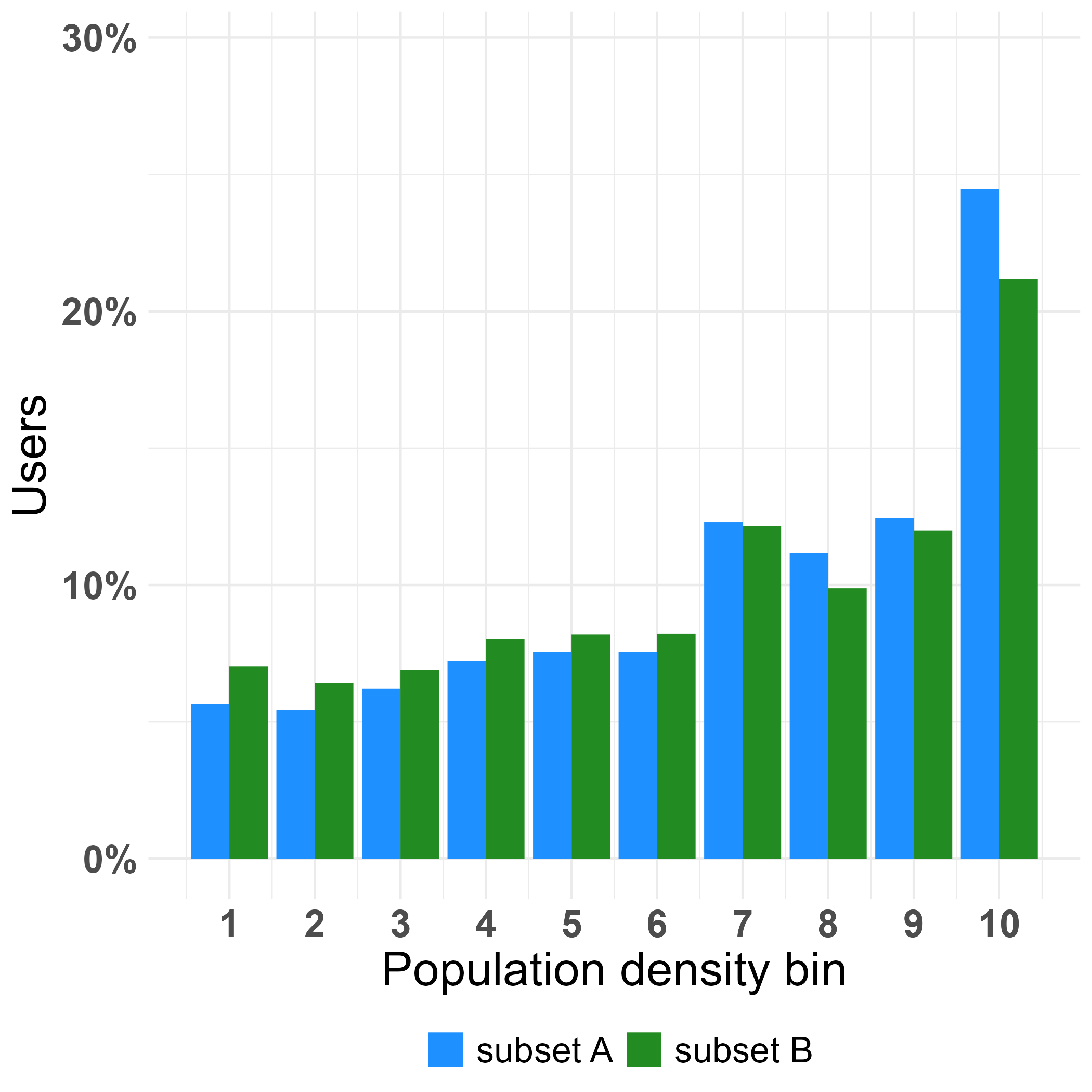}
  \caption{Distribution of users across density bins.}
  \label{fig:homeLocSelection_a}
\end{subfigure}
\begin{subfigure}[b]{.55\textwidth}
\raisebox{5mm}{
  \includegraphics[width=\textwidth]{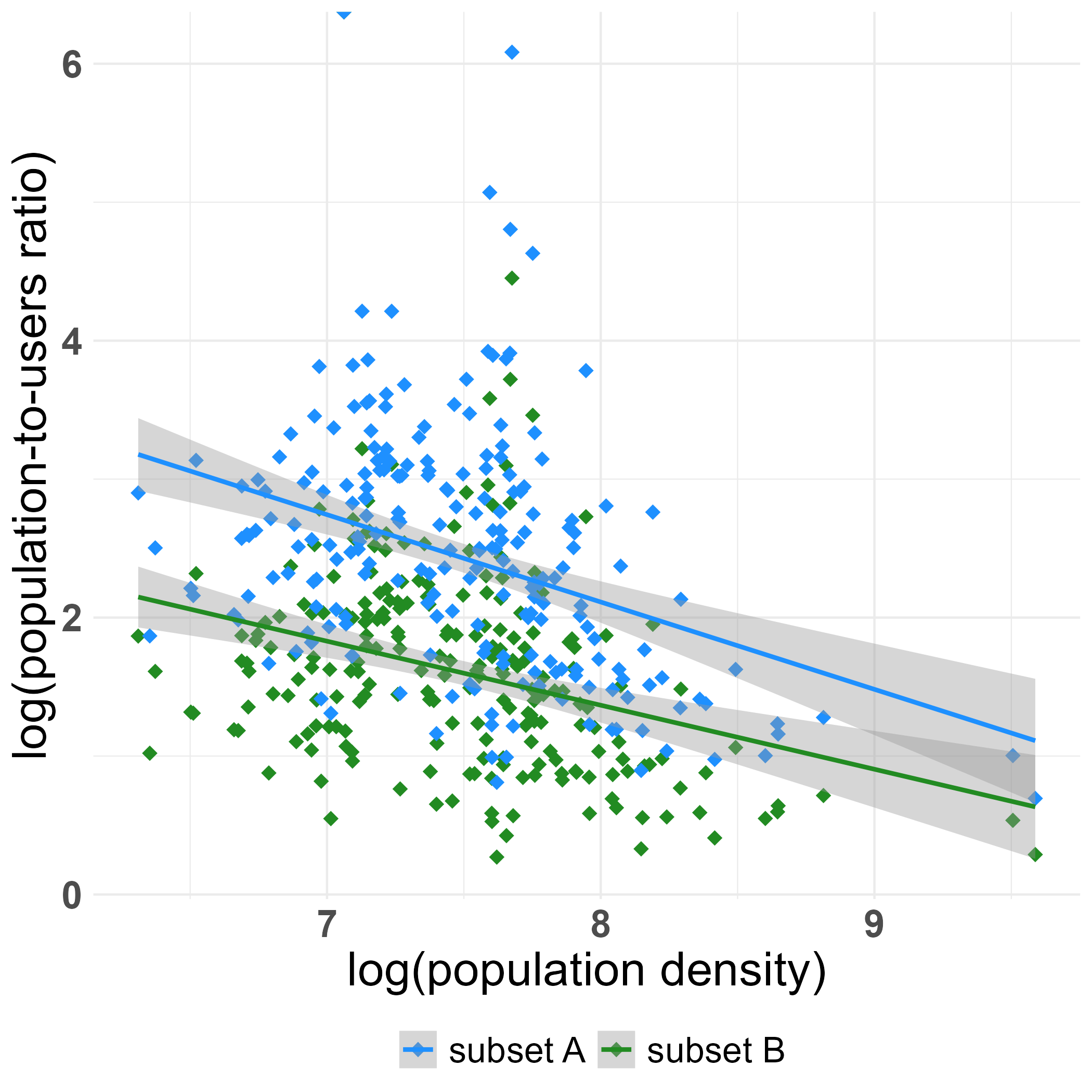}}
  \caption{Population-to-users ratio versus population density at the stratum-level.}
  \label{fig:homeLocSelection_b}
\end{subfigure}
\caption{Systematic bias of home locations toward denser areas. Panel (a) represents the distribution of phone users in the 2013 dataset across groups of cells defined based on population density deciles. Dakar is excluded from this analysis as it accounts for over 20\% of the population and would cover the top two density bins. In any case, the selection towards Dakar is already clear in Figure \ref{fig:usersVSpop_b}. Panel (b) shows the ratio of the population over 15 over the number of users at the stratum-level in the 2013 dataset against population density.}
\label{fig:homeLocSelection}
\end{figure}

To go further, we explicitly investigate the relationship between the distribution of users and population density. Voronoi cells are ordered by population density and grouped into ten bins each accounting for 10\% of the population over 15. The degree of selection of home locations with respect to population density is then assessed by calculating the distribution of users' home locations across these density bins. Note that in the absence of selection, the fraction of users found in each bin should match the share of the population it hosts (i.e. 10\%). Results are showed in Figure \ref{fig:homeLocSelection_a} and a clear pattern of selection emerges where the fraction of users increases with population density, for both \textit{subset A} and \textit{subset B}. In short, although phone users are broadly similar to the overall population over 15 locally (\textit{local representativeness}), our samples tend to over-represent individuals residing in denser areas. This analysis highlights the significance of this selection pattern and underscores the relevance of the weighting scheme described in the Methods section, which precisely addresses these imbalances in the mobile phone data sample composition. An alternative way to see this is provided in Figure \ref{fig:homeLocSelection_b} that represents the population-to-users ratio against population density, calculated at the level of strata used as weighting units and defined in the Methods section. The graph reveals a negative correlation indicating that denser areas are associated with higher numbers of users relative to the local population (i.e. lower population-to-users ratios). The proposed weighting scheme allows to neutralize this systematic bias by making the ratio of the adult population over the (weighted) number of users constant and equal to 1 across strata.

\subsection*{Validation of filtering parameters: their impact on sample size and selection on home location}

The subsets used to compute temporary migration statistics (i.e. \textit{subset A} and \textit{subset B}) are constructed via a filtering procedure detailed in the Methods section. This involves imposing minimal constraints on users' frequency and length of observation, as well as the maximum time non-observed, primarily to ensure accuracy in migration detection outcomes. However, these higher observational constraints also result in smaller sample sizes and may exacerbate selection biases on the cross-section. In the first sub-section of the Technical Validation, we perform a quantitative analysis to examine the relationship between migration detection model accuracy and observational constraints. This analysis supports the notion that the filtering parameters used to construct \textit{subset A} and \textit{subset B} allow to maintain high levels of accuracy. In this sub-section, we validate the choice of filtering parameters by quantifying the costs associated with higher observational constraints, specifically in terms of reduced sample size and increased selection bias with respect to the distribution of home locations across space.

\begin{figure}[H]
\centering
\begin{subfigure}[b]{.48\textwidth}
  \includegraphics[width=\textwidth]{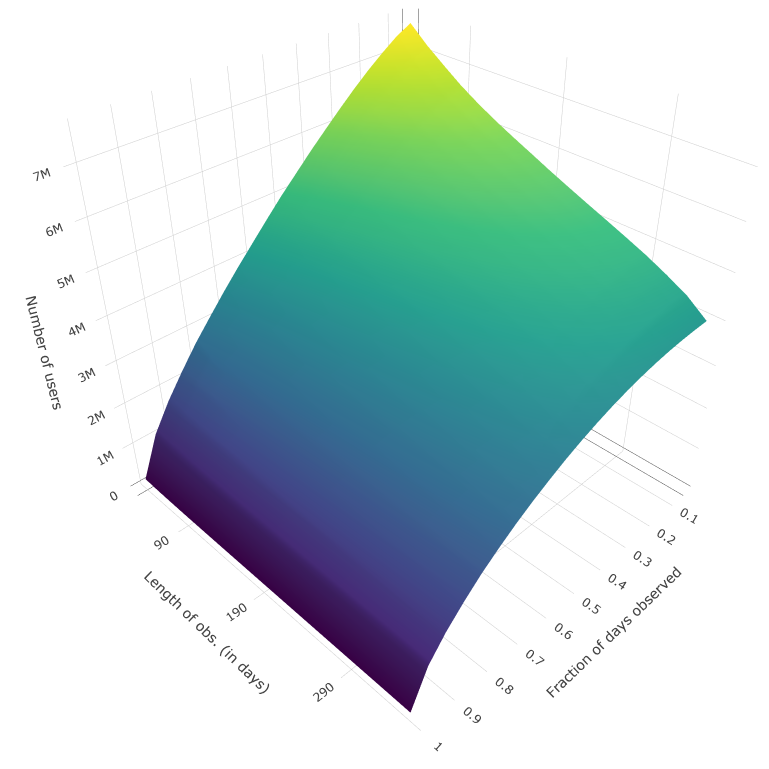}
  \caption{Length and frequency of observation}
\label{fig:filtering_sampleSize_a}
\end{subfigure}
\begin{subfigure}[b]{.48\textwidth}
  \includegraphics[width=\textwidth]{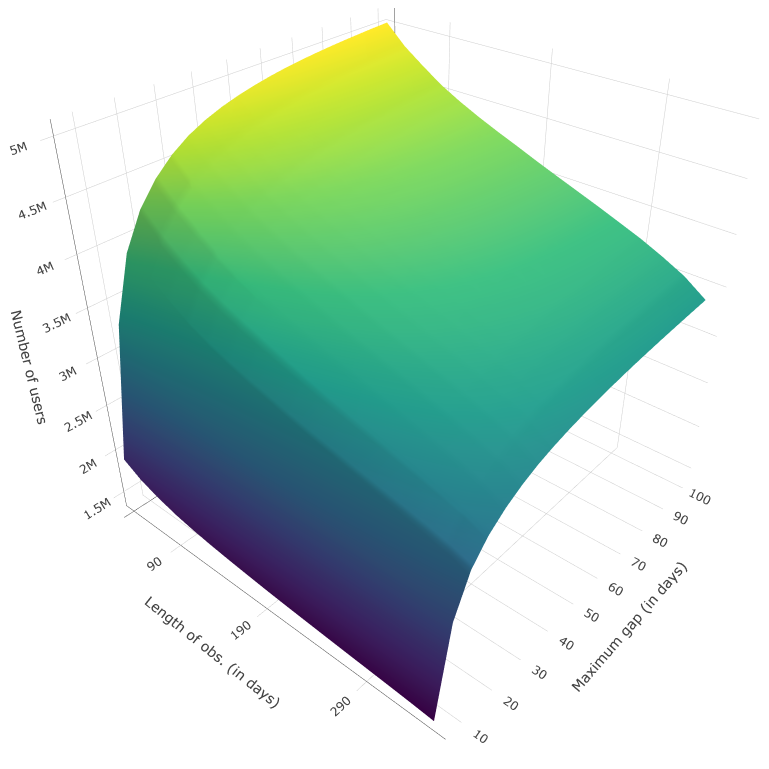}
  \caption{Length of observation and maximum time non-observed}
\label{fig:filtering_sampleSize_b}
\end{subfigure}
\begin{subfigure}[b]{.48\textwidth}
  \includegraphics[width=\textwidth]{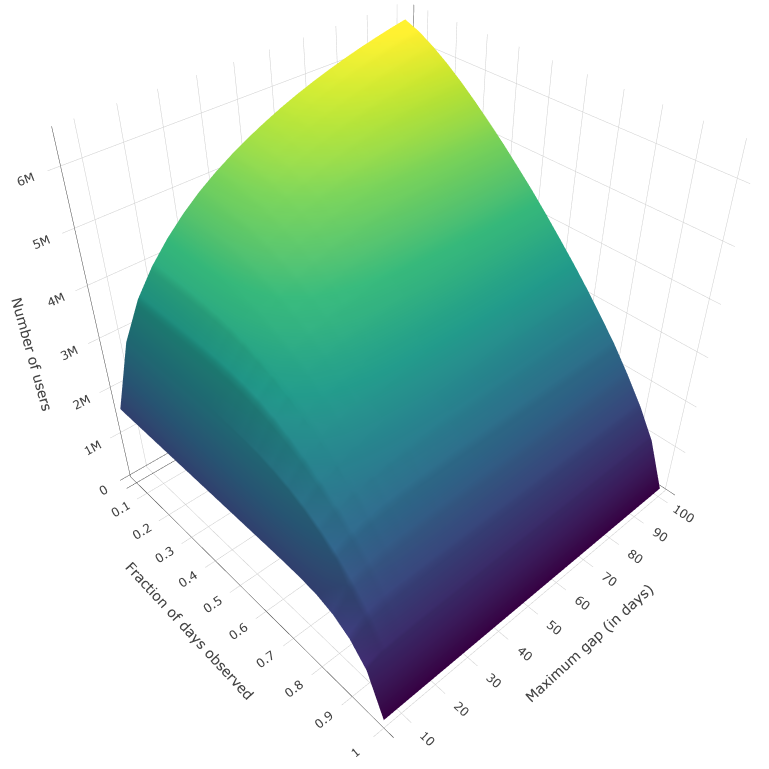}
  \caption{Frequency of observation and maximum time non-observed}
\label{fig:filtering_sampleSize_c}
\end{subfigure}
\caption{Impact of filtering parameters on the number of users left in a subset. Panel (a) shows three-dimensional surface representing the number of users in the 2013 dataset as a function of the minimal length and frequency of observation imposed, with the maximum time non-observed set to 100 days. Panel (b) represents the number of users as a function of the length of observation and the maximum time non-observed, setting the fraction of days with observations to 0.5. Panel (c) shows the number of users as a function of the fraction of days with observations and the maximum time non-observed, with the minimal length of observation set to 110 days.}
\label{fig:filtering_sampleSize}
\end{figure}

We first examine the impact of our three main observational constraints (frequency of observation, length of observation, maximum time non-observed) on sample size. To facilitate the visualization of the results, we evaluate the joint impact of any pair of constraints while holding the third fixed. Figure \ref{fig:filtering_sampleSize_a} shows a three-dimensional surface representing the number of users remaining in a subset as a function of the minimal frequency ($\Omega$) and length of observation ($\Delta$) imposed. While increasing the constraint on $\Delta$ has a negative but limited impact on sample size, the frequency of observation has a much larger impact on sample size. Then, Figure \ref{fig:filtering_sampleSize_b} represents the sample size as a function of the minimal length of observation and the maximum observational gap allowed. Consistent with Figure \ref{fig:filtering_sampleSize_a}, the constraint on the length of observation has a relatively minor impact on sample size. However, a clear non-linear impact of the maximum observational gap allowed is observed, with sample size decreasing sharply for values below 10-15 days. Finally, Figure \ref{fig:filtering_sampleSize_c} provides results consistent with those of Figures \ref{fig:filtering_sampleSize_a}-\ref{fig:filtering_sampleSize_b}, where the minimum fraction of days observed has a significant marginal impact on sample size while imposing a maximum observational gap lower than 10-15 days causes significant losses in sample size. 

\begin{figure}[H]
\centering
\begin{subfigure}[b]{.48\textwidth}
  \includegraphics[width=\textwidth]{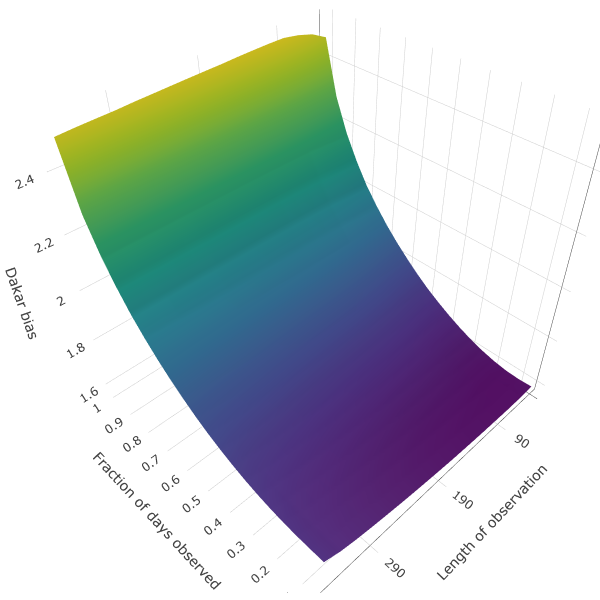}
  \caption{Length and frequency of observation}
\label{fig:DakarBias_a}
\end{subfigure}
\begin{subfigure}[b]{.48\textwidth}
  \includegraphics[width=\textwidth]{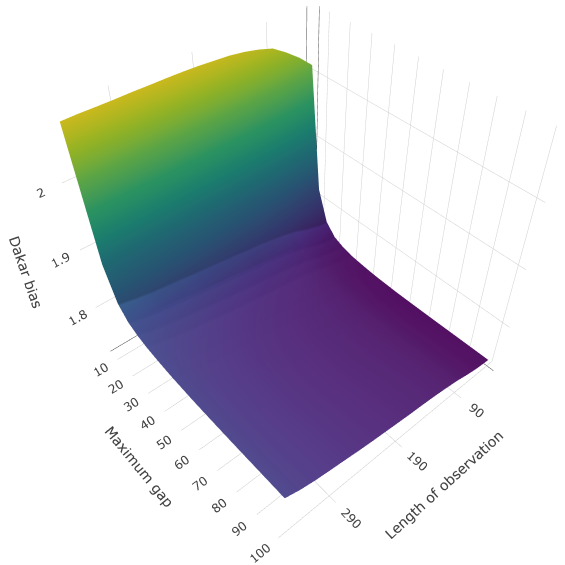}
  \caption{Length of observation and maximum time non-observed}
\label{fig:DakarBias_b}
\end{subfigure}
\begin{subfigure}[b]{.48\textwidth}
  \includegraphics[width=\textwidth]{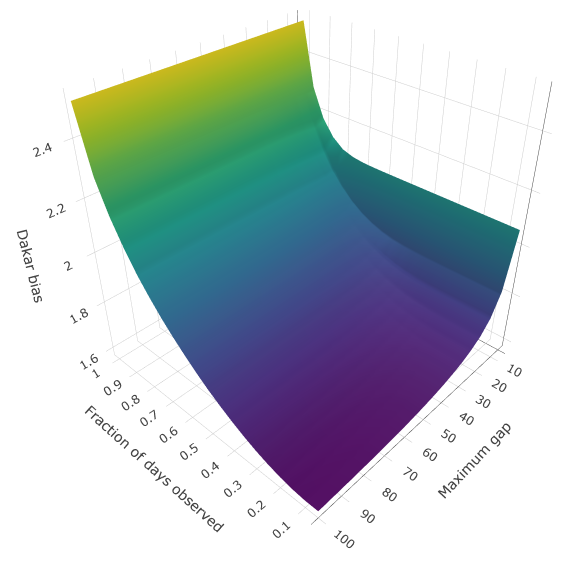}
  \caption{Frequency of observation and maximum time non-observed}
\label{fig:DakarBias_c}
\end{subfigure}
\caption{Impact of filtering parameters on the bias of home locations toward Dakar. ``Dakar bias'' is defined for any given subset of users as the ratio between the fraction of users in the subset residing in Dakar and the fraction of individuals effectively living in Dakar in the target population. Panel (a) shows a three-dimensional surface representing Dakar bias in subsets of the 2013 dataset as a function of the minimum length and frequency of observation imposed, with the maximum time non-observed set to 100 days. Panel (b) represents Dakar bias as a function of the length of observation and the maximum time non-observed, setting the fraction of days with observations to 0.5. Panel (c) shows Dakar bias as a function of the fraction of days with observations and the maximum time non-observed, with the minimal length of observation set to 110 days.}
\label{fig:DakarBias}
\end{figure}

\begin{figure}[H]
\centering
\begin{subfigure}[b]{.48\textwidth}
  \includegraphics[width=\textwidth]{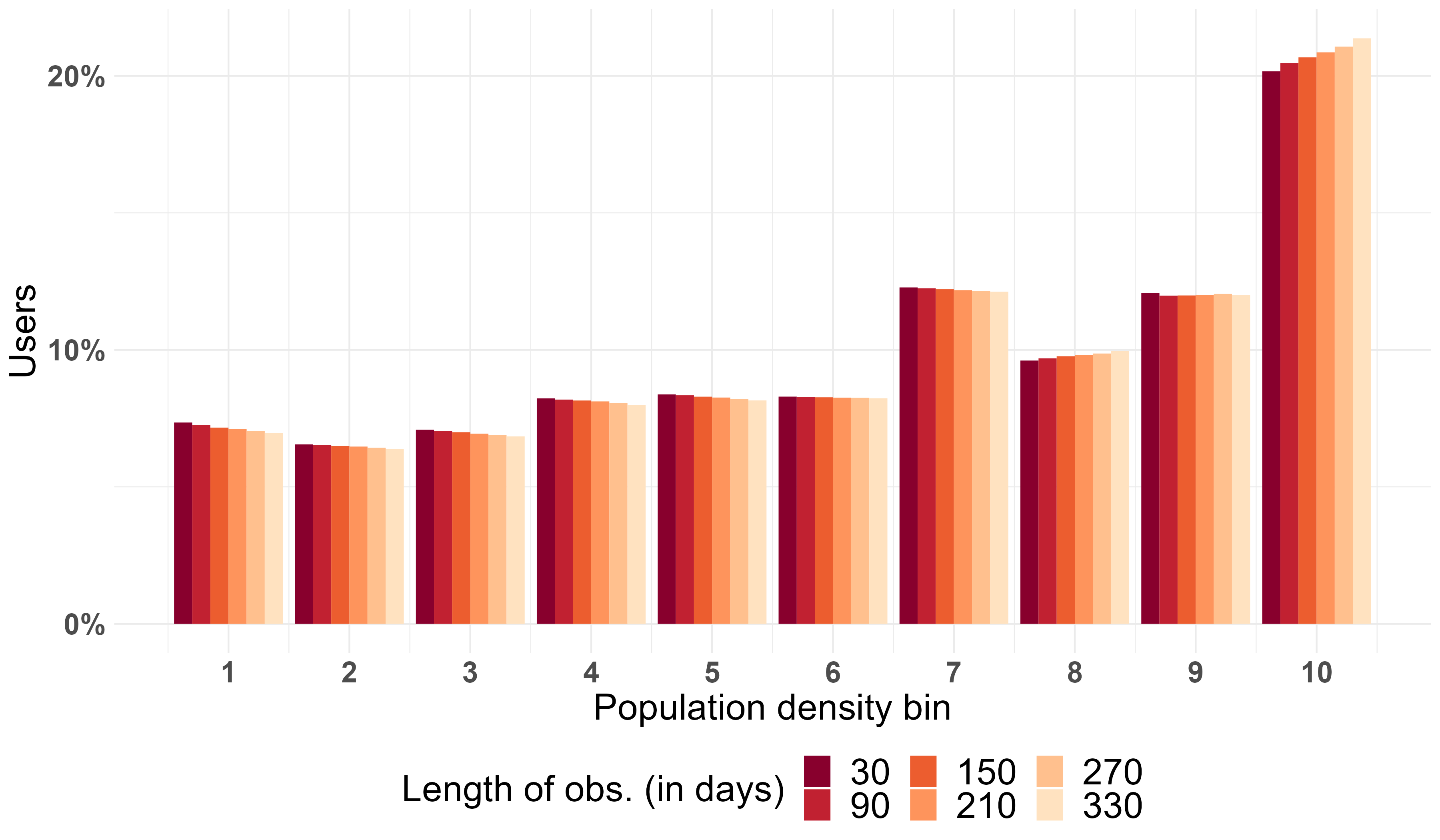}
  \caption{Home location bias versus the minimum length of observation}
\label{fig:SelectionNoDakar_a}
\end{subfigure}
\begin{subfigure}[b]{.48\textwidth}
  \includegraphics[width=\textwidth]{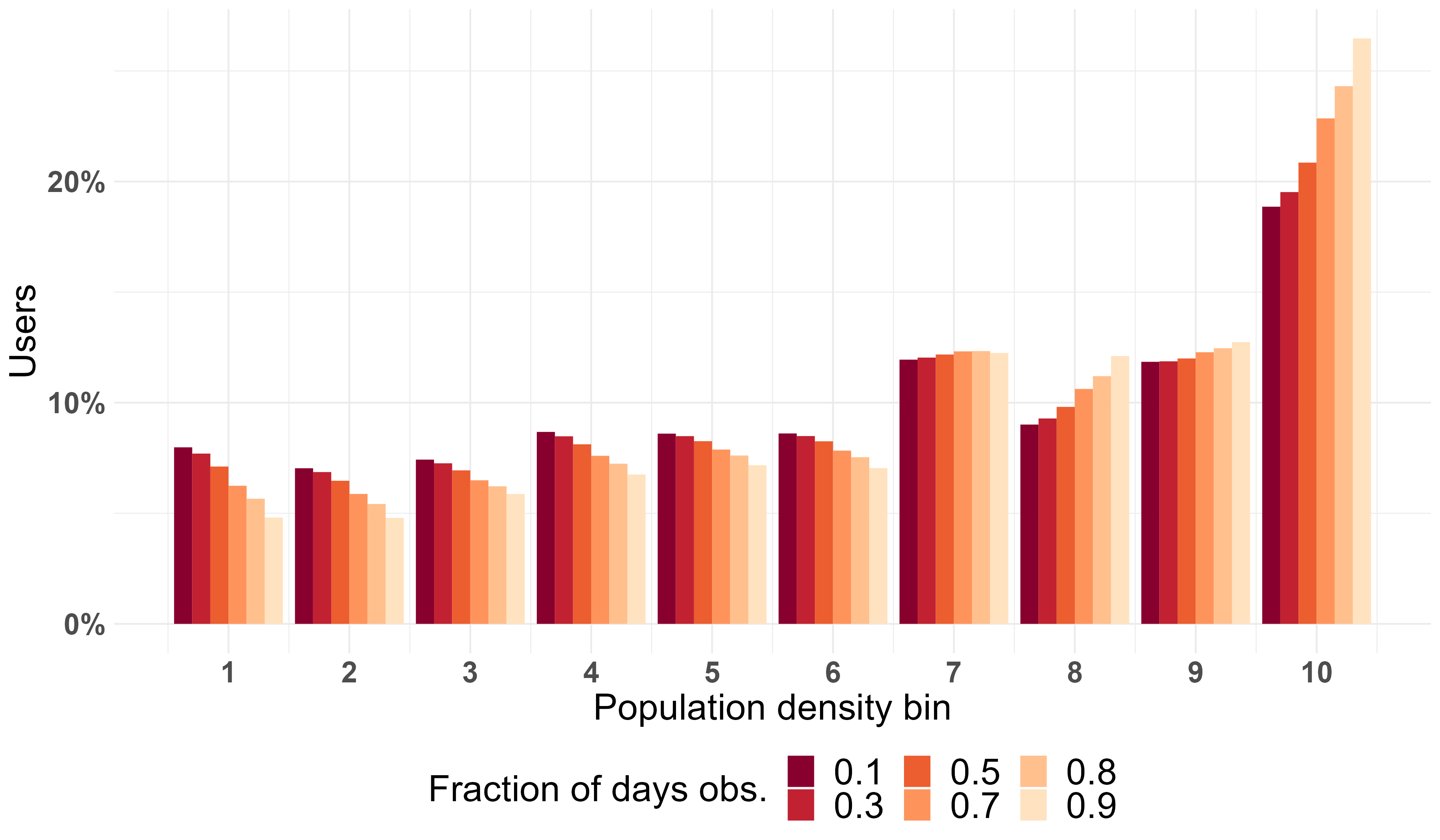}
  \caption{Home location bias versus the frequency of observation}
\label{fig:SelectionNoDakar_b}
\end{subfigure}
\begin{subfigure}[b]{.48\textwidth}
  \includegraphics[width=\textwidth]{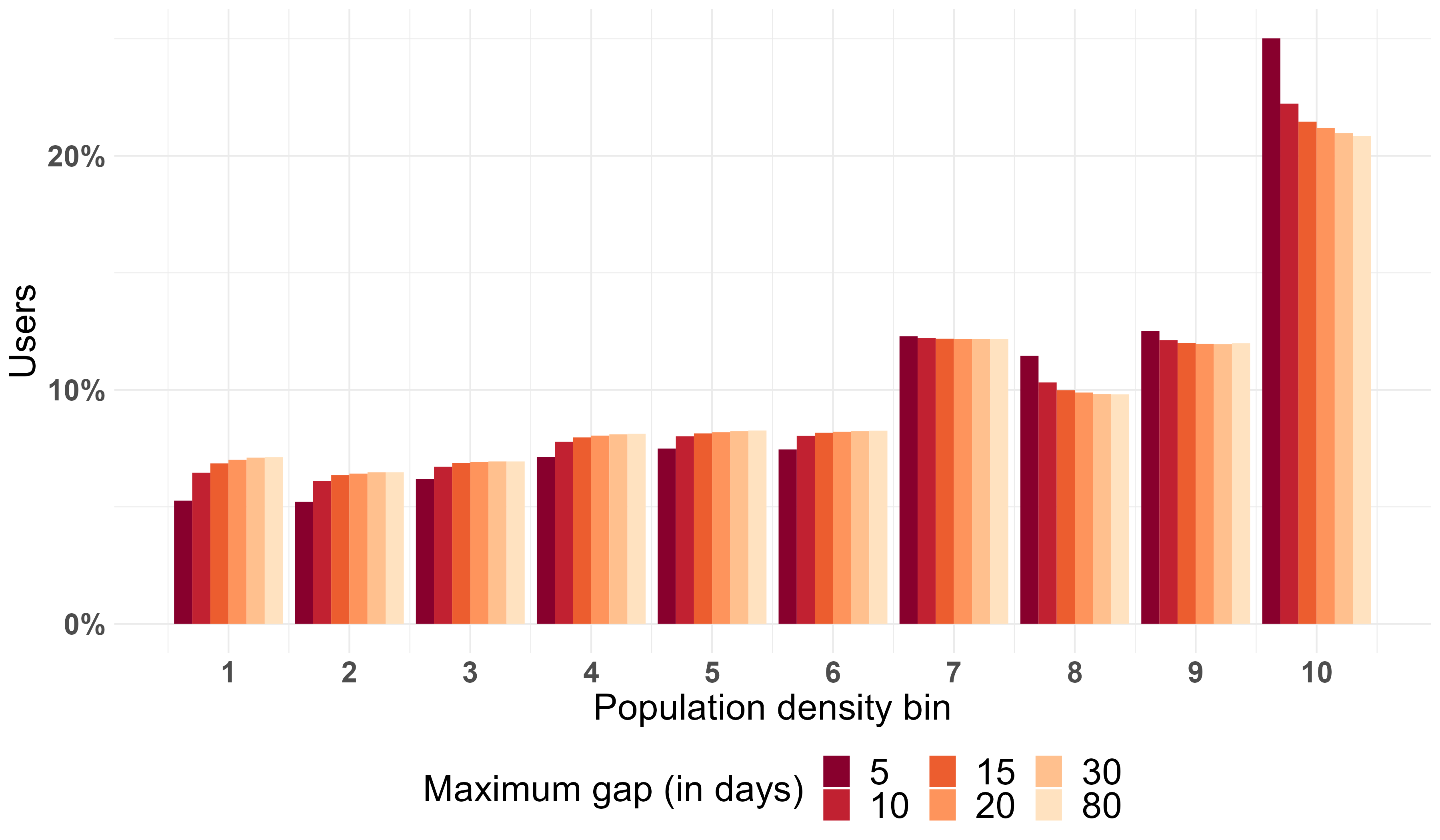}
  \caption{Home location bias versus the maximum time non-observed}
\label{fig:SelectionNoDakar_c}
\end{subfigure}
\caption{Impact of filtering parameters on the bias of home locations toward denser areas. Panel (a) represents the distribution of non-Dakar phone users across population density bins, for subsets of the 2013 dataset associated with different values of the minimum length of observation, and a minimum fraction of days observed and maximum time non-observed fixed and set equal to 0.5 and 100 days respectively. As in Figure \ref{fig:homeLocSelection_a}, each bin is a group of cells with similar population density that account for 10\% of the non-Dakar population over 15. Similarly, panel (b) shows the distribution of non-Dakar phone users across population density bins, for subsets of the 2013 dataset associated with different values of the minimum fraction of days observed, and a length of observation and maximum time non-observed equal to 210 days and 100 days respectively. Panel (c) shows the distribution of non-Dakar phone users across population density bins, for subsets of the 2013 dataset associated with different values of the maximum time non-observed, and a length of observation and a fraction of days observed equal to 210 days and 0.5 respectively.}
\label{fig:SelectionNoDakar}
\end{figure}

Then, we evaluate the impact of filtering parameters on the pattern of selection toward denser areas documented above. Specifically, we first estimate the impact of filtering parameters on the bias toward Dakar in the sample composition. We define this bias for any given subset as the ratio of the fraction of users with an inferred home location in Dakar over the fraction of the population over 15 effectively residing in Dakar. For example, a value of 2 would indicate that the sample contains twice as many users in Dakar than there would be if the sample had been randomly drawn from the target population. In the spirit of Figure \ref{fig:filtering_sampleSize}, we represent in Figure \ref{fig:DakarBias} the value of this bias as a function of any pair of constraints, holding the third parameter fixed. Figures \ref{fig:DakarBias_a} and \ref{fig:DakarBias_b} clearly show that selecting users with a higher length of observation has practically no impact on the bias toward Dakar. In contrast, Figures \ref{fig:DakarBias_a} and \ref{fig:DakarBias_c} reveal that augmenting the minimum frequency of observation exacerbates the bias. For instance, regardess of the minimum length of observation, the bias increases from about 1.6 for a minimum fraction of days of 0.1 to 2-2.2 when this constraint is raised to 0.9 (Figure \ref{fig:DakarBias_a}). Similar to the impact of filtering parameters on sample size (Figure \ref{fig:filtering_sampleSize}), reducing the maximum time unobserved increases the bias only for values below a threshold of about 10-15 days (Figures \ref{fig:DakarBias_b} and \ref{fig:DakarBias_c}). Notably, this impact diminishes significantly when higher values for the minimal frequency of observation are considered, specifically for a minimum fraction of days observed around 0.8 and above (Figure \ref{fig:DakarBias_c}). Secondly, we assess the degree of selection induced by filtering parameters on the composition of the rest of the sample. To do this, we calculate the distribution of phone users across groups of cells defined based on population density deciles (as in Figure \ref{fig:homeLocSelection_a}), for different values of filtering parameters. Figures \ref{fig:SelectionNoDakar_a},\ref{fig:SelectionNoDakar_b}, and \ref{fig:SelectionNoDakar_c} show how these distributions vary with the minimum length of observation, the minimum frequency of observation, and the maximum time non-observed, respectively. Once again, increasing the minimum length of observation has minimal impact on the distribution of users across density bins (Figure \ref{fig:SelectionNoDakar_a}), and reducing the maximum time non-observed allowed slightly exacerbates the bias toward denser areas only for values around 10 days and below. Finally, the frequency of observation is the main parameter influencing the bias in the distribution of users across density bins. As illustrated in Figure \ref{fig:SelectionNoDakar_b}, increasing the minimum fraction of days observed to select a subset magnifies the tilt toward categories of denser cells. For instance, increasing the fraction of days observed from 0.1 to 0.9 decreases the fraction of users in the category of least dense cells (bin 1) from 8\% to 5\% and increases the share of users in the densest areas (bin 10) from 19\% to 26\%. 

In summary, the cost of the filtering procedure in terms of reduced sample size and increased selection is primarily driven by the frequency of observation parameter, which is also the main determinant of the migration detection model accuracy. Moreover, it is worth noting that the impact on selection is clear but largely contained. Imposing a high minimum fraction of days with observations induces further distortions toward Dakar and denser areas, but the resulting subsets still provide wide coverage with consistent fractions of users found in the most remote locations. Consequently, the minimum fraction of days observed of 0.8 used to construct \textit{subset A} is viewed as a credible tradeoff allowing to achieve a high level of accuracy while avoiding a significant reduction in sample size and an unreasonable distortion in sample composition. The value used for the construction of \textit{subset B} (0.5) is thought of as resulting from a tradeoff assigning relatively more weight to the cost on selection and less to the accuracy of the migration detection model. On the other hand, the constraint on the length of observation remains quite stringent for both subsets since it has only a limited impact on sample size and selection. Finally, we do not consider values below 15 days for the maximum time non-observed given the potentially large impact implied on sample size and selection. Also, with temporary migration events being defined with a minimum duration of 20 days, constraining observational gaps to a maximum of 15-25 days is in fact sufficient to avoid cases of non-random attrition where users would be non-observed precisely while in migration.

It is essential to note that the distortion in sample composition caused by the filtering procedure is more concerning for unweighted estimates. The observed pattern of selection leads to a small over-representation of individuals in cities and denser areas and therefore tends to move the sample away from the target population of mobile phone users. On the other hand, weighted estimates systematically address discrepancies between the distribution of users' home locations in a given subset and the distribution of the target population. 

Furthermore, it is crucial to acknowledge that without information on users' socio-economic characteristics, we cannot fully assess the impact of filtering parameters on the validity of the \textit{local representativeness} assumption. Future research could delve into this aspect. Nevertheless, given the modest selection patterns observed in the distribution of home locations, there are reasons to believe that selection at a local level induced by the filtering procedure remains limited.

\section*{Usage Notes}

The dataset is available for download at a public \textit{figshare} repository ([repository name and link will be provided here]). The complete dataset is broken down into multiple files as described in the Data records section. Each file is associated with a type of estimates (\textit{weighted} or \textit{unweighted}), a specific subset (A or B), and a minimum migration event duration threshold.

A shapefile delineating the boundaries of all spatial units used in the dataset (i.e., origin and destination locations) is provided in the \textit{figshare} repository. This allows users to map temporary migration estimates and combine the dataset with other spatial data, such as climate or land use information. The spatial units are constructed from the voronoi cells showed in Figure \ref{fig:voronoi}. Cells classified as urban form individual spatial units. On the other hand, rural cells are assigned to a unique district based on a maximum population criterion and cells that belong to a unique district are grouped. This process results in 112 rural locations and 39 urban locations (i.e. cities) for a total of 151 distinct locations. The shapefile is provided as a GeoPackage file (\texttt{spatial\_units\_SciData.gpkg}) with a single layer. The attribute table has three columns: i) \textit{id} is a unique identifier for each spatial unit, ii) \textit{name} gives the name of the corresponding spatial unit, which is either the city name or a name of the form \textit{NAME-rural} where \textit{NAME} is the name of the corresponding district, and iii) \textit{zone\_category} indicates whether the spatial unit is classified as rural or urban. 

The temporary migration estimates in the dataset can be aggregated to coarser spatial and temporal resolutions, but users should be aware of certain limitations. Absolute migration flows (e.g., the number of departures and returns) can be summed without restriction. For instance, to find the total number of migration departures in Senegal for 2013, one can simply add up the departures across all origin-destination pairs and half-months for that year. However, computing a departure rate is not feasible since the denominator -- i.e. the number of users observed -- cannot be directly derived from the dataset. This calculation would require knowing the number of unique users observed throughout the entire year 2013, specifically those with CDR trajectories allowing to detect any migration movements if they effectively occurred. Then, migration stock estimates can be aggregated spatially without restriction for any given half-month period. For example, the total stock of temporary migrants to Dakar in the first half of August 2013 is obtained by adding up the stock of migrants to Dakar across all origin locations for that particular half-month. Nonetheless, aggregating migration stocks over time periods longer than the minimum duration defined for temporary migration events may not always be meaningful. Alternative metrics could be considered to provide temporary migration measures over extended periods of time. For instance, one could calculate the number of unique users with at least one migration event of at least 20 days having occurred during 2013. Such metrics may be included in future versions of the dataset.

\section*{Code availability}

The full code allowing to process raw mobile phone data and produce a final temporary migration dataset at the desired level of granularity is made available on GitHub at \href{https://github.com/blanchap/TempMigration_SciData}{https://github.com/blanchap/TempMigration\_SciData}. A README file containing additional information on the content of each script and how to properly execute them is provided in the GitHub repository.

\bibliography{arxiv}

\section*{Acknowledgements}
This study was partly conducted within the OPAL project. Mobile phone data were provided by Sonatel (Senegal) as part of the D4D Challenge and the OPAL project.

\section*{Author contributions statement}
P.B. and S.R. designed the research. P.B. and S.R. analysed the data. P.B. performed the statistical analysis. P.B. wrote the Article.  All authors edited and approved the final version of the Article.

\section*{Competing interests}
The authors declare no competing interests.

\newpage

\section*{Figures \& Tables}

\begin{table}[H]
    \centering
    \begin{tabular}{cccc}
    \toprule
         \textbf{Subset} & \textbf{Unique identifiers, 2013} & \textbf{Unique identifiers, 2014-2015} & \textbf{Total records}\\\addlinespace
        \textit{subset A} & 1,990,754 & 2,041,566 &  47,857,866,128\\\addlinespace
        \textit{subset B} & 3,377,994 & 3,746,640 & 61,566,733,246\\
        \bottomrule
    \end{tabular}
    \caption{Number of unique identifiers and total number of records in \textit{subset A} and \textit{subset B}.}
    \label{tab:sampleSize}
\end{table}

\begin{figure}[H]
\captionsetup{width=0.8\textwidth}
\centering
  \includegraphics[width=0.8\textwidth]{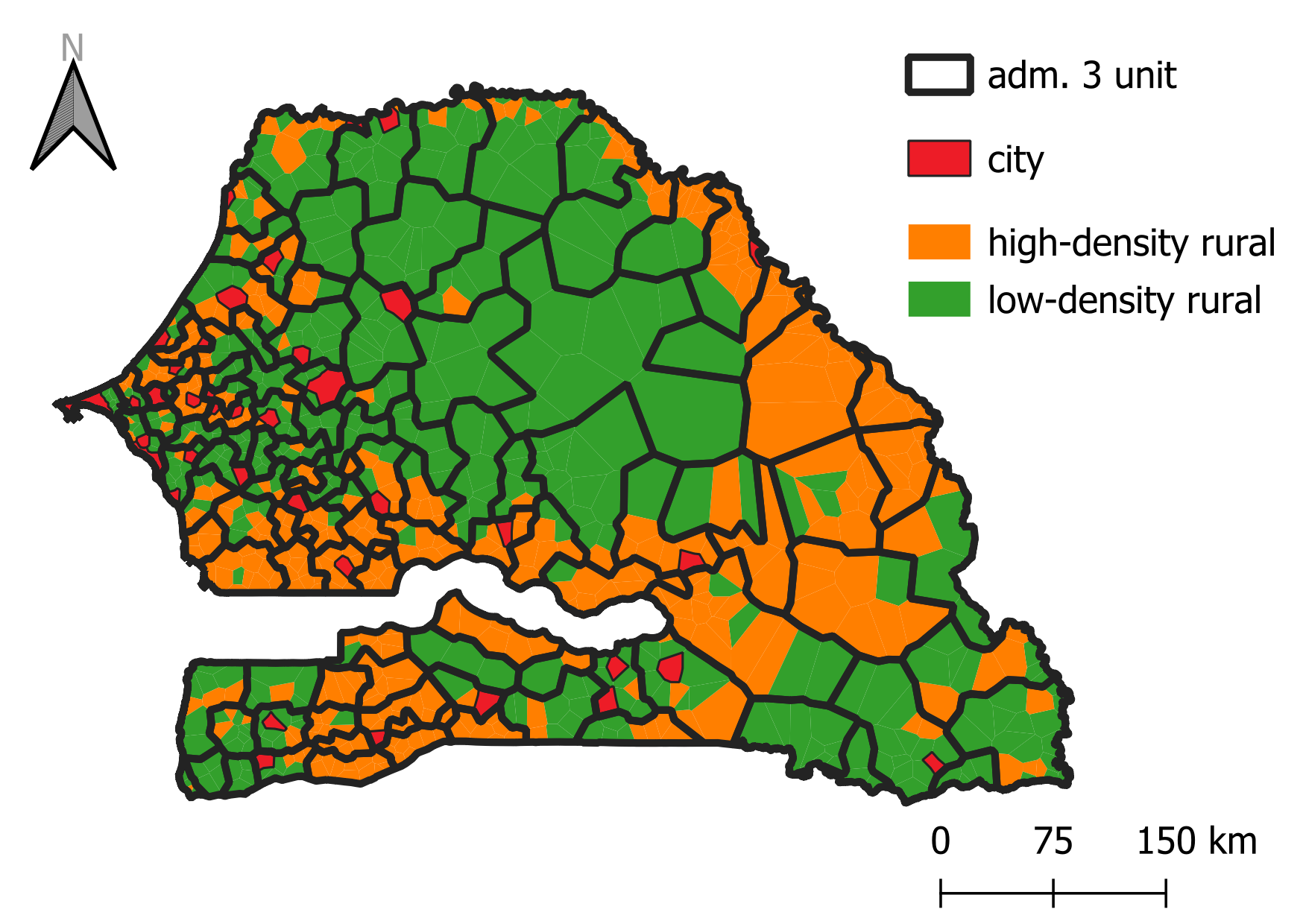}
\caption{Strata used to design the weighting scheme employed in the \textit{weighted} migration estimates. Each of the 39 individual cities constitutes an urban stratum (in red). Black lines delineate groups of voronoi cells belonging to a single third-level administrative unit, where voronoi cells are assigned an administrative unit based on a maximum population criterion. The median population density across rural voronoi cells (1550 inh./km$^2$) is used to define low- and high-density rural areas, represented in green and orange respectively. Each (administrative unit,rural density category) couple constitutes a rural stratum. Note that voronoi-level population estimates are obtained by overlaying voronoi polygons with the 2017 100m-resolution gridded population product from the WorldPop Research Group\cite{Qader2022}.}
\label{fig:strata}
\end{figure}

\begin{table}[H]
    \centering
    \begin{tabular}{p{7cm} p{9cm}}
    \toprule
       \multicolumn{1}{c}{\textbf{Variable name}} &  \multicolumn{1}{c}{\textbf{Description}} \\ \addlinespace
       \textit{N\_depart} &  Number of migration departures\\ \addlinespace
       \textit{N\_return}  & Number of migration returns\\\addlinespace
       \textit{N\_migrants}  & Number of individuals in migration, i.e. the stock of migrants\\\addlinespace
       \textit{N\_users\_observed\_depart}  & Total number of users residing in origin $o$ observed at time $t$, for departure counts\\\addlinespace
       \textit{N\_users\_observed\_return} & Total number of users residing in origin $o$ observed at time $t$, for return counts\\\addlinespace
       \textit{N\_users\_observed\_stock} & Total number of users residing in origin $o$ observed at time $t$, for migration stock\\\addlinespace
       \textit{rate\_depart} & Rate of departures calculated as $\frac{N\_depart}{N\_users\_observed\_depart}$ \\\addlinespace
       \textit{rate\_return} & Rate of migration returns calculated as $\frac{N\_return}{N\_users\_observed\_return}$ \\\addlinespace
       \textit{rate\_migrants} & Migration stock rate calculated as $\frac{N\_migrants}{N\_users\_observed\_stock}$\\\addlinespace
       \bottomrule
    \end{tabular}
    \caption{Description of variables provided in the dataset. Each row of the dataset provides measures of temporary migration from an origin $o$, to a destination $d$, for a half-month denoted by $t$. In datasets providing \textit{weighted} estimates, variable names include the suffix “\textit{\_adj}” to indicate that variables were adjusted with the weighting scheme outlined in the Methods section (e.g. \textit{N\_depart\_adj} instead of \textit{N\_depart}). Note that, by construction, the ``adjusted'' number of users observed provided in weighted estimates datasets coincides with the target population for the corresponding origin location.}
    \label{tab:var_desc}
\end{table}

\begin{table}[H]
    \centering
    \begin{tabular}{ccc}
    \toprule
         & \textbf{\textit{subset A}} & \textbf{\textit{subset B}}\\\addlinespace
        \textit{Home detection accuracy} & 98\% & 92\% \\\addlinespace
        \textit{Migration event detection accuracy} & 90\% & 76\% \\
        \bottomrule
    \end{tabular}
    \caption{Estimated accuracy of home location estimation and temporary migration event detection in \textit{subset A} and \textit{subset B}.}
    \label{tab:accuracy_BySubset}
\end{table}

\begin{table}[H]
\centering
\setlength{\aboverulesep}{0pt}
\setlength{\belowrulesep}{0pt}
\begin{tabularx}{\textwidth}{l *{3}{Y}}
  \toprule
 & \textbf{Sonatel users} & \textbf{All users} & \textbf{Diff.} \\ 
    & (1) & (2) & (1)-(2) \\ 
  \midrule
Male dummy & 0.559 & 0.554 & 0.005$^{}$ \\ 
  Age & 37.237 & 36.975 & 0.262$^{}$ \\ 
  Years of education & 6.785 & 6.101 & 0.685$^{***}$ \\ 
  Urban dummy & 0.574 & 0.505 & 0.07$^{***}$ \\ 
  Has electricity & 0.910 & 0.897 & 0.013$^{}$ \\ 
  Has piped water & 0.869 & 0.847 & 0.022$^{}$ \\ 
  Has a fridge & 0.447 & 0.415 & 0.033$^{**}$ \\ 
  Has a radio & 0.710 & 0.726 & -0.016$^{}$ \\ 
  Has a TV & 0.761 & 0.743 & 0.018$^{}$ \\ 
  Richest quintile dummy & 0.170 & 0.170 & 0$^{}$ \\ 
  Poorest quintile dummy & 0.178 & 0.196 & -0.019$^{}$ \\ 
   \bottomrule
\end{tabularx}
\caption{Comparison of Sonatel users with the overall population of phone users. Statistics were derived from the individual-level Access Survey dataset for Senegal conducted by Research ICT Africa\cite{ict}.}
\label{tab:sonatelVSusers}
\end{table}

\begin{table}[H]
\centering
\setlength{\aboverulesep}{0pt}
\setlength{\belowrulesep}{0pt}
\resizebox{\textwidth}{!}{
\begin{tabular}{|l|>{\centering}m{0.7in}|>{\centering}m{0.7in}|>{\centering}m{0.7in}|>{\centering}m{0.7in}|>{\centering}m{0.7in}|>{\centering\arraybackslash}m{0.7in}|}
  \toprule 
  & \multicolumn{3}{c|}{\textbf{\underline{Urban}}} & \multicolumn{3}{c|}{\textbf{\underline{Rural}}} \\ 
 & \textbf{Phone users} & \textbf{All} & \textbf{Diff.} & \textbf{Phone users} & \textbf{All} & \textbf{Diff.} \\ 
   & (1) & (2) & (1)-(2) & (3) & (4) & (3)-(4) \\ 
 \midrule
wealth group: richest & 0.095 & 0.085 & 0.01$^{}$ & 0.096 & 0.087 & 0.01$^{}$ \\ 
  wealth group: richer & 0.200 & 0.184 & 0.016$^{}$ & 0.263 & 0.223 & 0.04$^{}$ \\ 
  wealth group: middle & 0.189 & 0.192 & -0.003$^{}$ & 0.290 & 0.291 & -0.001$^{}$ \\ 
  wealth group: poorer & 0.255 & 0.262 & -0.007$^{}$ & 0.280 & 0.295 & -0.015$^{}$ \\ 
  wealth group: poorest & 0.261 & 0.277 & -0.016$^{}$ & 0.071 & 0.104 & -0.033$^{}$ \\ 
  Years of education & 5.439 & 5.272 & 0.167$^{}$ & 3.074 & 2.812 & 0.262$^{}$ \\ 
  Age & 30.371 & 28.446 & 1.925$^{**}$ & 31.408 & 29.337 & 2.071$^{*}$ \\ 
  Male & 0.468 & 0.451 & 0.017$^{}$ & 0.564 & 0.476 & 0.088$^{**}$ \\ 
  Married & 0.525 & 0.461 & 0.064$^{*}$ & 0.693 & 0.621 & 0.072$^{}$ \\ 
  Has a bank account & 0.226 & 0.179 & 0.047$^{}$ & 0.150 & 0.090 & 0.06$^{}$ \\ 
  occupation: not working & 0.302 & 0.343 & -0.042$^{}$ & 0.200 & 0.213 & -0.013$^{}$ \\ 
  occupation: agriculture & 0.037 & 0.036 & 0.001$^{}$ & 0.530 & 0.548 & -0.019$^{}$ \\ 
  occupation: sales & 0.253 & 0.237 & 0.016$^{}$ & 0.087 & 0.071 & 0.016$^{}$ \\ 
  occupation: household/domestic & 0.033 & 0.044 & -0.01$^{}$ & 0.013 & 0.012 & 0.001$^{}$ \\ 
  occupation: unskilled & 0.138 & 0.148 & -0.01$^{}$ & 0.099 & 0.097 & 0.002$^{}$ \\ 
  Household size & 10.462 & 10.474 & -0.012$^{}$ & 14.009 & 13.704 & 0.305$^{}$ \\ 
  Water access & 0.697 & 0.688 & 0.009$^{}$ & 0.518 & 0.489 & 0.03$^{}$ \\ 
  Access to sanitation & 0.845 & 0.825 & 0.02$^{}$ & 0.364 & 0.323 & 0.042$^{}$ \\ 
  Electricity & 0.844 & 0.836 & 0.008$^{}$ & 0.320 & 0.282 & 0.037$^{}$ \\ 
   \bottomrule
\end{tabular}}
\caption{Differences in characteristics between phone users and the population, Kaolack. Statistics were derived from the Senegal 2017 DHS men and women individual datasets. Results for all other regions are left in the Supplementary Material.} 
\label{tab:dhs_ttest_kaolack}
\end{table}

\newpage

{\large Blanchard and Rubrichi}
\hfill
{\large Supplementary Material}\\

\begin{center}
    {\LARGE \textbf{A Highly Granular Temporary Migration Dataset Derived From Mobile Phone Data in Senegal}}
\end{center}
\vspace{0.75cm}

\section{Algorithmic rules to aggregate user-level trajectories}

The following diagrams illustrate the algorithmic rules used to identify migration departures, migration returns and the status of migration for a given time unit, for both high- and low-confidence estimates. An explanatory note below each diagram provides a description of the corresponding configuration and the criteria applied. Thick segments along the time arrow reflect meso-segments for a hypothetical user while empty spaces correspond to observational gaps. Segment locations are indicated at the top left of segments, using some common notations throughout all diagrams. ``H'' denotes the home location, ``!H'' any non-home location and ``not !H'' simply means ``any location that is not !H''. When no location is specified, it is assumed that the corresponding segment could be at any location.\\

It is important to note that some configurations are somehow redundant from a logical perspective, but need to be treated separately in the algorithm. All cases are still presented in this appendix with the intent to facilitate the understanding of the code for anyone wishing to reproduce or simply use it.

\subsection{Identifying migration departures: high-confidence}

\begin{figure}[H]
\renewcommand{\figurename}{Diagram}
\caption{High-confidence migration departure: case 1}
\label{fig:migrationDiagram_depart_high_1}
\centering
  \includegraphics[width=\textwidth]{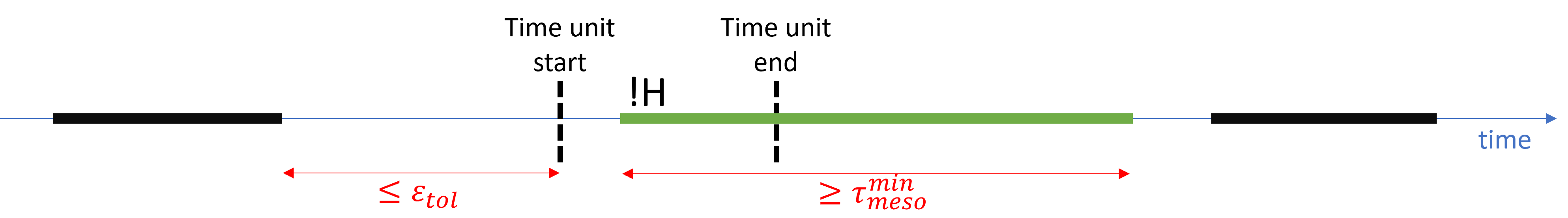}
\caption*{\small{\textit{Note}: The green segment is a migration segment with a start date within the time unit. The observation gap between the time unit start date and the observation preceding the green segment is lower than the tolerance parameter $\epsilon^{tol}$. As a result, the start date of the green segment is counted as a migration departure in the high-confidence estimation.}}
\end{figure}

\begin{figure}[H]
\renewcommand{\figurename}{Diagram}
\caption{High-confidence migration departure: case 2}
\label{fig:migrationDiagram_depart_high_2}
\centering
  \includegraphics[width=\textwidth]{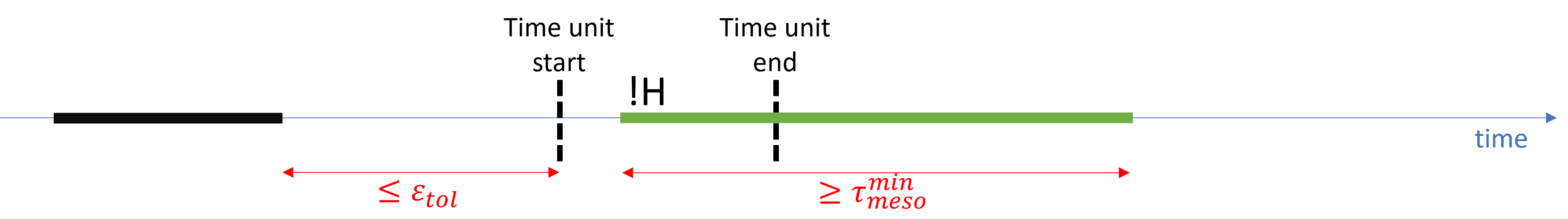}
\caption*{\small{\textit{Note}: This is the same configuration as in diagram \ref{fig:migrationDiagram_depart_high_1}, but the user exits the sample at the end of the green segment.}}
\end{figure}

\subsection{Identifying migration departures: low-confidence}

\begin{figure}[H]
\renewcommand{\figurename}{Diagram}
\caption{Low-confidence migration departure: case 1}
\label{fig:migrationDiagram_depart_low_1}
\centering
  \includegraphics[width=\textwidth]{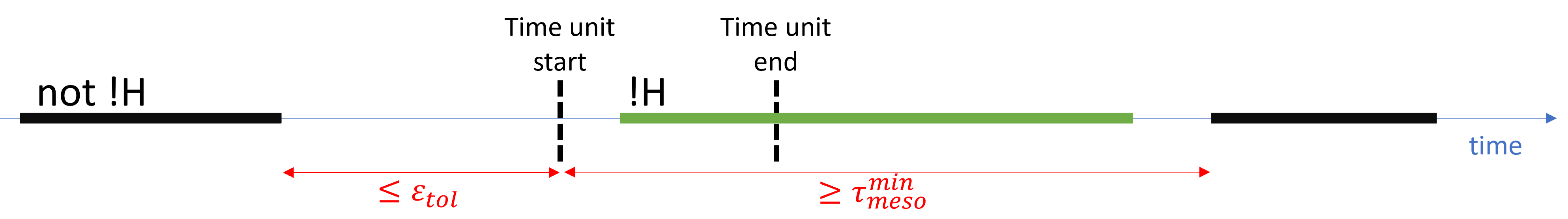}
\caption*{\small{\textit{Note}: The green segment at !H does not necessarily have an observed duration greater than $\tau^{temp}$. The maximum duration possible to consider the segment started during the time unit is the time elapsed between the time unit start date and the day preceding the start of the following segment. When this is greater than $\tau^{temp}$ and the tolerance criterion is not exceeded, this configuration results in one additional migration departure in the low-confidence estimate.}}
\end{figure}

\begin{figure}[H]
\renewcommand{\figurename}{Diagram}
\caption{Low-confidence migration departure: case 2}
\label{fig:migrationDiagram_depart_low_2}
\centering
  \includegraphics[width=\textwidth]{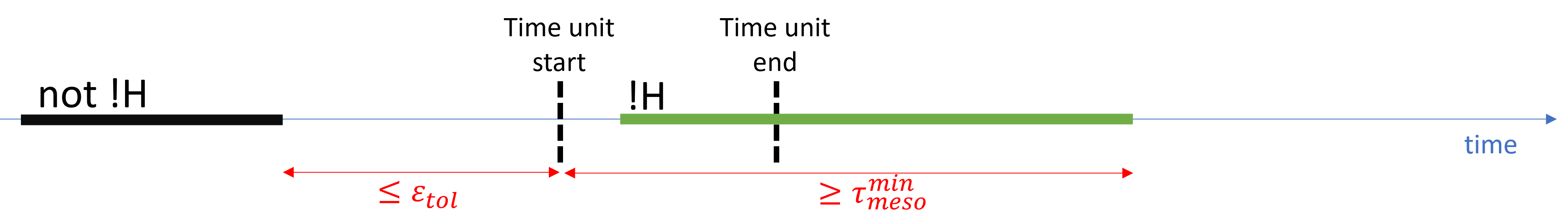}
\caption*{\small{\textit{Note}: This is the same configuration as in diagram \ref{fig:migrationDiagram_depart_low_1}, although the user exits the sample at the end of the green segment. The maximum duration possible to consider the segment started during the time unit is the time elapsed between the time unit start date and the observed end date of the segment. If this is greater than $\tau^{temp}$ and the tolerance criterion is not exceeded, this configuration results in one additional migration departure in the low-confidence estimate.}}
\end{figure}

\begin{figure}[H]
\renewcommand{\figurename}{Diagram}
\caption{Low-confidence migration departure: case 3}
\label{fig:migrationDiagram_depart_low_3}
\centering
  \includegraphics[width=\textwidth]{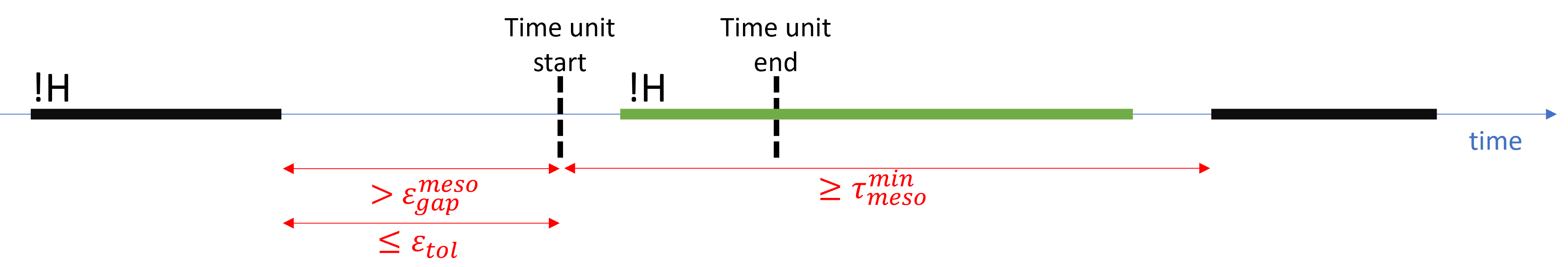}
\caption*{\small{\textit{Note}: This is the same configuration as in diagram \ref{fig:migrationDiagram_depart_low_1}, although the segment preceding the green segment is at the same location !H. In this case, the green segment cannot have started less than $\epsilon_{gap}^{meso}+1$ days after that segment. The diagram presents a situation where the gap between the end of the preceding segment and the first day of the time unit is strictly larger than $\epsilon_{gap}^{meso}$, so that the green segment may have started on the first day of the time unit. The situation is then equivalent to that described in diagram \ref{fig:migrationDiagram_depart_low_1}.}}
\end{figure}

\begin{figure}[H]
\renewcommand{\figurename}{Diagram}
\caption{Low-confidence migration departure: case 4}
\label{fig:migrationDiagram_depart_low_4}
\centering
  \includegraphics[width=\textwidth]{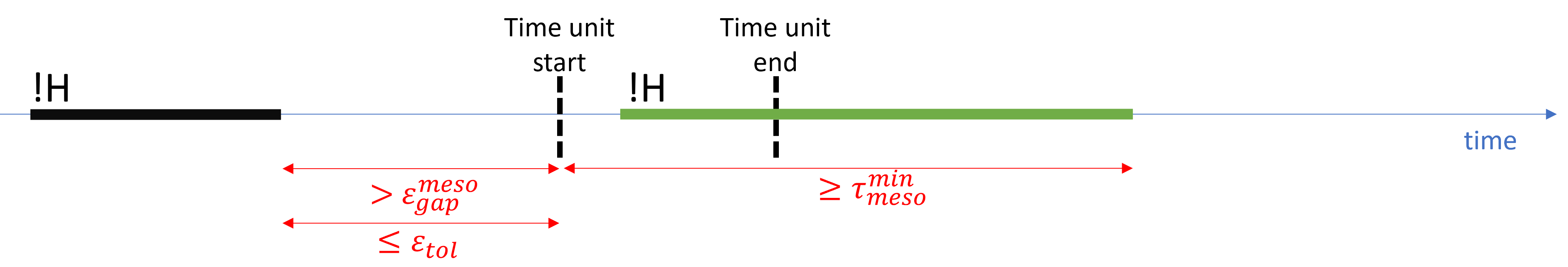}
\caption*{\small{\textit{Note}: This is the same configuration as in diagram \ref{fig:migrationDiagram_depart_low_3}, but the user exits the sample at the end of the green segment. The maximum duration is therefore lower as the maximum end date coincides with the observed end date.}}
\end{figure}

\begin{figure}[H]
\renewcommand{\figurename}{Diagram}
\caption{Low-confidence migration departure: case 5}
\label{fig:migrationDiagram_depart_low_5}
\centering
  \includegraphics[width=\textwidth]{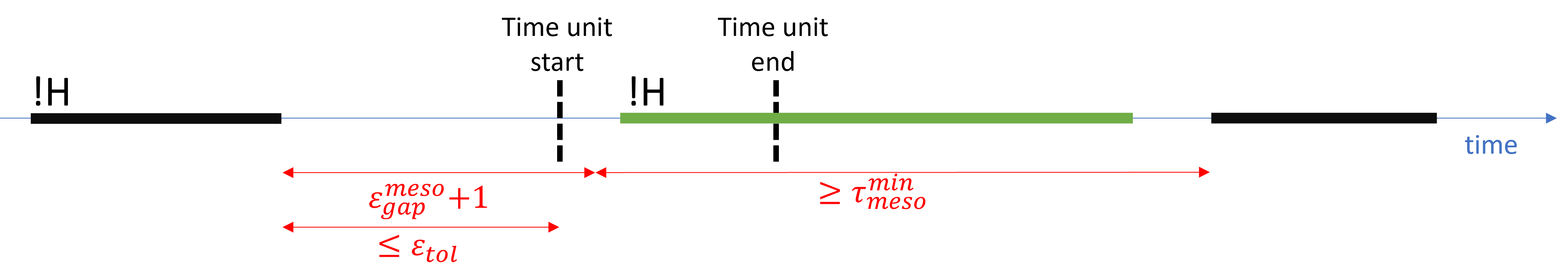}
\caption*{\small{\textit{Note}: This is the same configuration as in diagram \ref{fig:migrationDiagram_depart_low_3}, although the date corresponding to $\epsilon_{gap}^{meso}+1$ days after the end of the preceding segment falls within the time unit. The maximum duration of the green segment is therefore slightly lower because the minimum start date possible for the green segment is greater than the first day of the time unit.}}
\end{figure}

\begin{figure}[H]
\renewcommand{\figurename}{Diagram}
\caption{Low-confidence migration departure: case 6}
\label{fig:migrationDiagram_depart_low_6}
\centering
  \includegraphics[width=\textwidth]{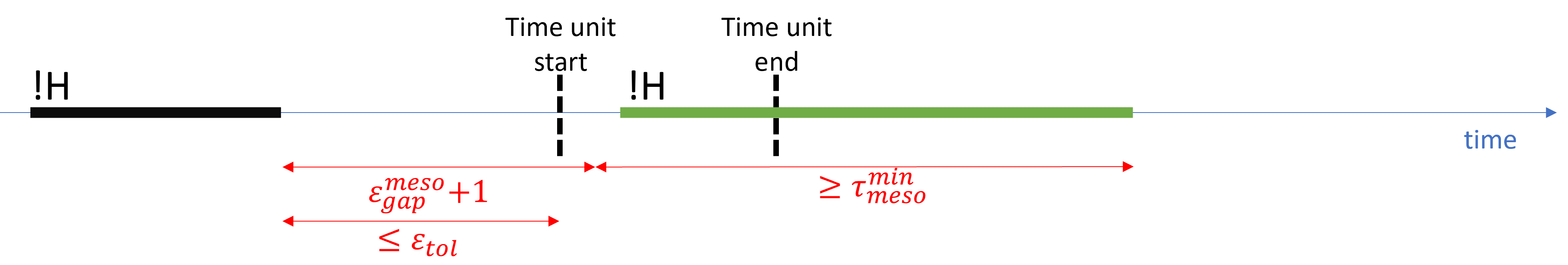}
\caption*{\small{\textit{Note}: This is the same configuration as in diagram \ref{fig:migrationDiagram_depart_low_5}, but the user exits the sample at the end of the green segment. The maximum duration is therefore lower as the maximum end date coincides with the observed end date.}}
\end{figure}

\subsection{Identifying migration returns: high-confidence}

\begin{figure}[H]
\renewcommand{\figurename}{Diagram}
\caption{High-confidence migration return: case 1}
\label{fig:migrationDiagram_return_high_1}
\centering
  \includegraphics[width=\textwidth]{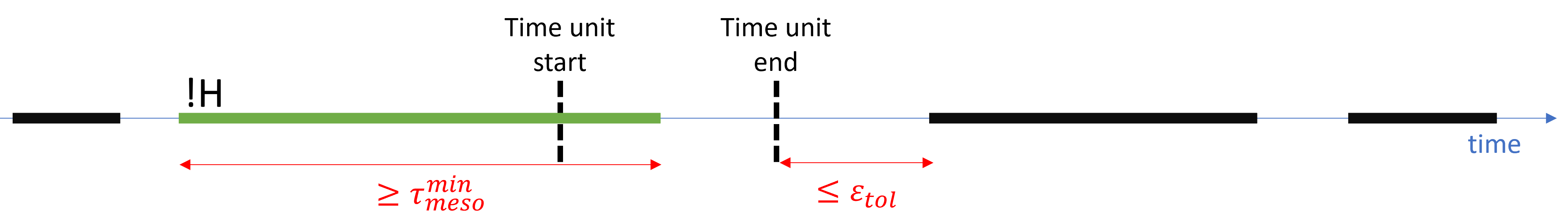}
\caption*{\small{\textit{Note}: The green segment is at a non-home location and has a duration greater than $\tau^{temp}$: it is a migration segment. The observed end date falls within the time unit but the observation gap following the segment indicates that the user may have actually returned after the time unit. Since the time elapsed between the end of the time unit and the day preceding the following segment is less than the tolerance criterion $\epsilon^{tol}$, the user is considered to have returned during the time unit in the high-confidence estimate.}}
\end{figure}

\begin{figure}[H]
\renewcommand{\figurename}{Diagram}
\caption{High-confidence migration return: case 2}
\label{fig:migrationDiagram_return_high_2}
\centering
  \includegraphics[width=\textwidth]{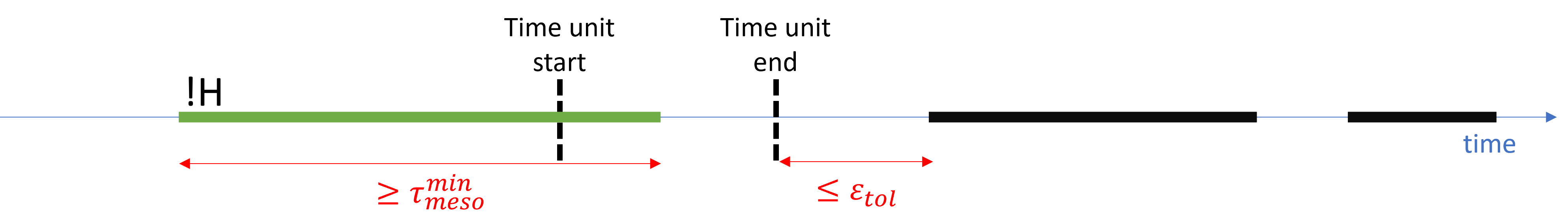}
\caption*{\small{\textit{Note}: This is the same configuration as in diagram \ref{fig:migrationDiagram_return_high_1}, but the user is never observed before the green segment.}}
\end{figure}

\subsection{Identifying migration returns: low-confidence}

\begin{figure}[H]
\renewcommand{\figurename}{Diagram}
\caption{Low-confidence migration return: case 1}
\label{fig:migrationDiagram_return_low_1}
\centering
  \includegraphics[width=\textwidth]{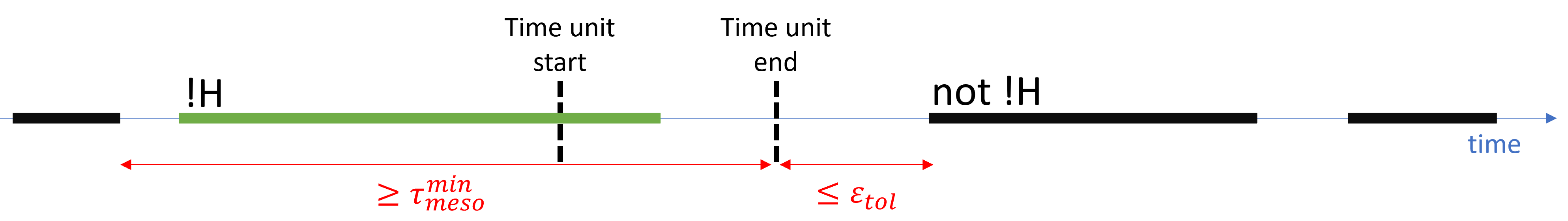}
\caption*{\small{\textit{Note}: The green segment at !H does not necessarily have an observed duration greater than $\tau^{temp}$. The maximum duration possible to consider the segment ended during the time unit is the time elapsed between day following the end date of the preceding segment and the time unit end date. When this is greater than $\tau^{temp}$ and the tolerance criterion is not exceeded, this configuration results in one additional migration return in the low-confidence estimate.}}
\end{figure}

\begin{figure}[H]
\renewcommand{\figurename}{Diagram}
\caption{Low-confidence migration return: case 2}
\label{fig:migrationDiagram_return_low_2}
\centering
  \includegraphics[width=\textwidth]{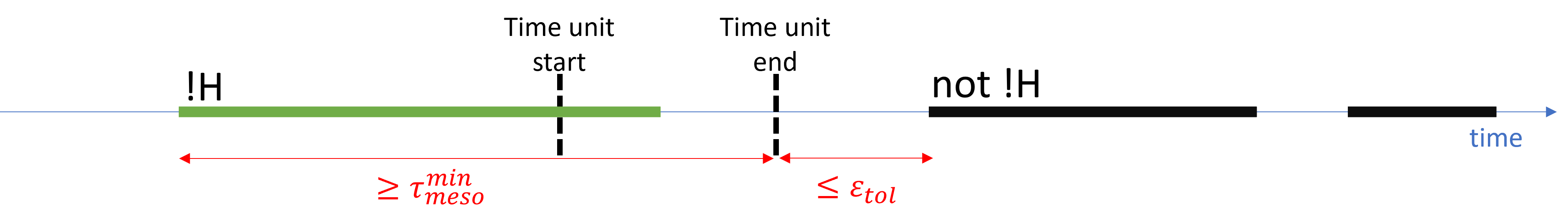}
\caption*{\small{\textit{Note}: This is the same configuration as in diagram \ref{fig:migrationDiagram_return_low_1}, although the user is never observed before the green segment. The maximum duration possible to consider the segment ended during the time unit is the time elapsed between the observed start date of the green segment and the time unit end date. If this is greater than $\tau^temp$ and the tolerance criterion is not exceeded, this configuration results in one additional migration return in the low-confidence estimate.}}
\end{figure}

\begin{figure}[H]
\renewcommand{\figurename}{Diagram}
\caption{Low-confidence migration return: case 3}
\label{fig:migrationDiagram_return_low_3}
\centering
  \includegraphics[width=\textwidth]{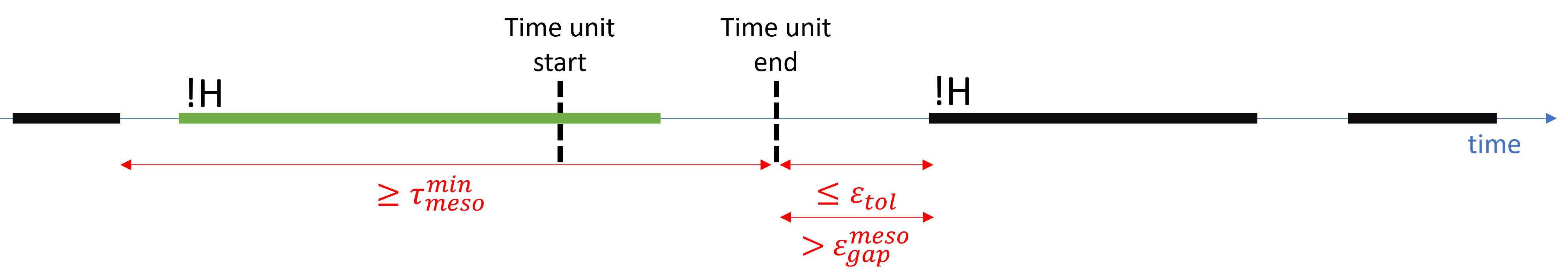}
\caption*{\small{\textit{Note}: This is the same configuration as in diagram \ref{fig:migrationDiagram_return_low_1}, although the segment following the green segment is at the same location !H. In this case, the green segment cannot have ended less than $\epsilon_{gap}^{meso}+1$ days before that segment. The diagram presents a situation where the gap between the start of the following segment and the last day of the time unit is strictly larger than $\epsilon_{gap}^{meso}$, so that the green segment may have lasted up until the very last day of the time unit. The situation is then equivalent to that described in diagram \ref{fig:migrationDiagram_return_low_1}.}}
\end{figure}

\begin{figure}[H]
\renewcommand{\figurename}{Diagram}
\caption{Low-confidence migration return: case 4}
\label{fig:migrationDiagram_return_low_4}
\centering
  \includegraphics[width=\textwidth]{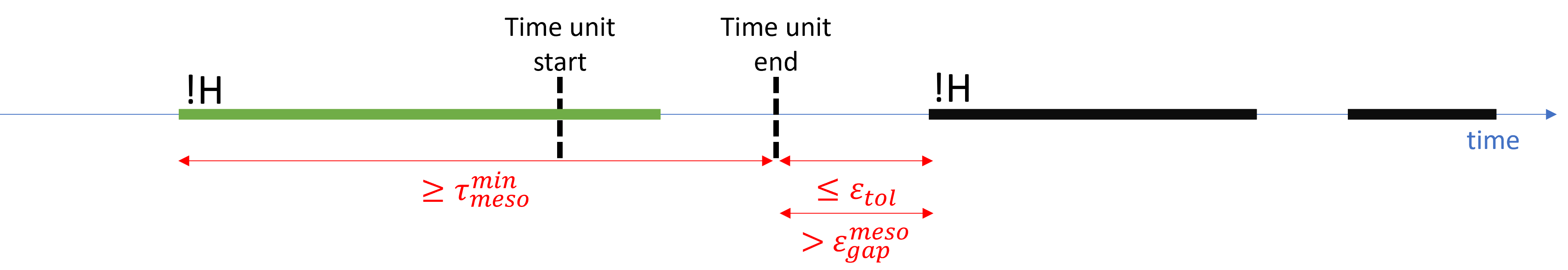}
\caption*{\small{\textit{Note}: This is the same configuration as in diagram \ref{fig:migrationDiagram_return_low_3}, but the user is never observed before the green segment. The maximum duration is therefore lower as the minimum start date coincides with the observed start date. The situation is otherwise equivalent to that of diagram \ref{fig:migrationDiagram_return_low_3}.}}
\end{figure}

\begin{figure}[H]
\renewcommand{\figurename}{Diagram}
\caption{Low-confidence migration return: case 5}
\label{fig:migrationDiagram_return_low_5}
\centering
  \includegraphics[width=\textwidth]{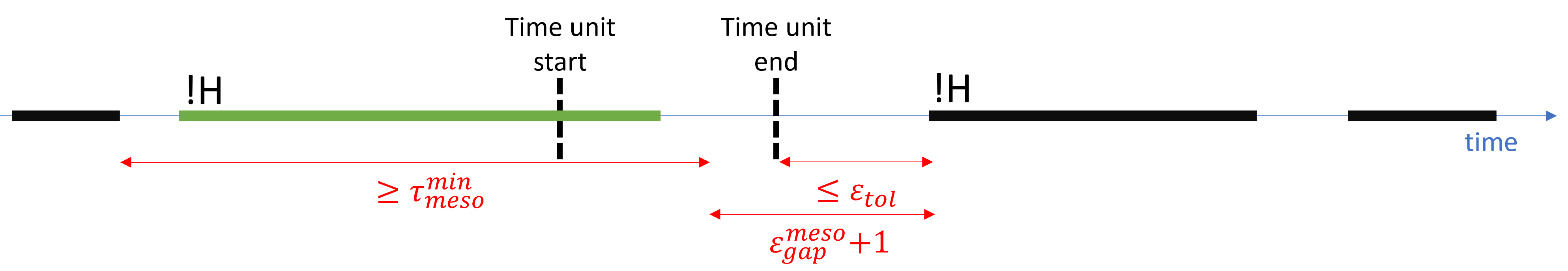}
\caption*{\small{\textit{Note}: This is the same configuration as in diagram \ref{fig:migrationDiagram_return_low_3}, although the date corresponding to $\epsilon_{gap}^{meso}+1$ days before the start of the following segment falls within the time unit. The maximum duration of the green segment is therefore slightly lower because the maximum end date possible for the green segment precedes the last day of the time unit. The situation is otherwise equivalent to that of diagram \ref{fig:migrationDiagram_return_low_3}.}}
\end{figure}

\begin{figure}[H]
\renewcommand{\figurename}{Diagram}
\caption{Low-confidence migration return: case 6}
\label{fig:migrationDiagram_return_low_6}
\centering
  \includegraphics[width=\textwidth]{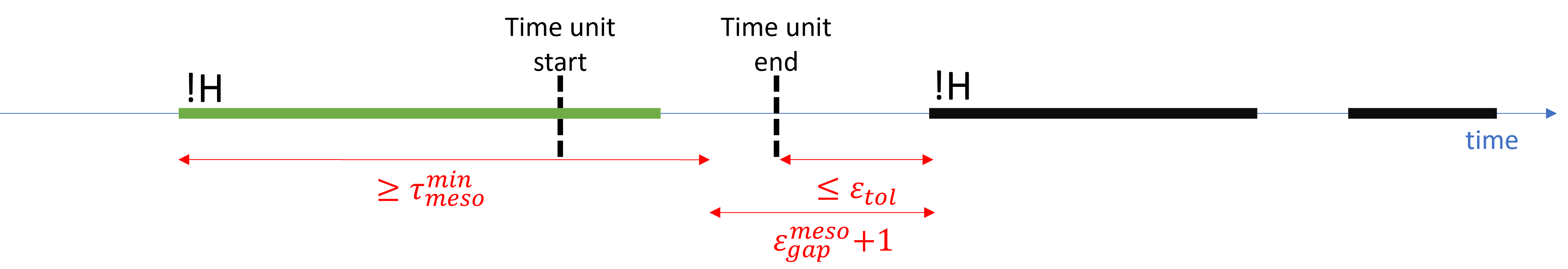}
\caption*{\small{\textit{Note}: This is the same configuration as in diagram \ref{fig:migrationDiagram_return_low_5}, but the user is never observed before the green segment. The maximum duration is slightly lower because the minimum start date coincides with the observed start date. The situation is otherwise equivalent to that of diagram \ref{fig:migrationDiagram_return_low_5}.}}
\end{figure}

\subsection{Identifying migration status: high-confidence}

\begin{figure}[H]
\renewcommand{\figurename}{Diagram}
\caption{High-confidence migration status: case 1}
\label{fig:migrationDiagram_status_high_1}
\centering
  \includegraphics[width=\textwidth]{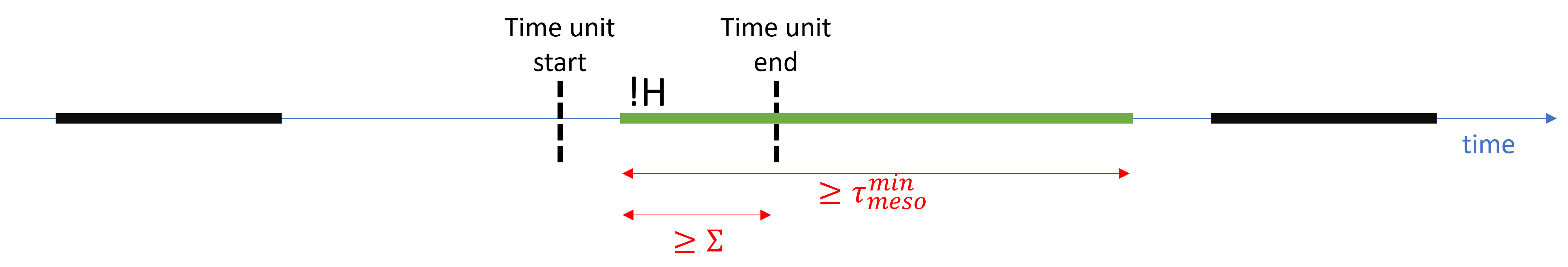}
\caption*{\small{\textit{Note}: The green segment is at a non-home location and has a duration greater than $\tau^{temp}$: it is a migration segment. It overlaps on the right of the time unit for a duration of at least $\Sigma$ days. The user is therefore considered as being in migration during the time unit in the high-confidence estimate.}}
\end{figure}

\begin{figure}[H]
\renewcommand{\figurename}{Diagram}
\caption{High-confidence migration status: case 2}
\label{fig:migrationDiagram_status_high_2}
\centering
  \includegraphics[width=\textwidth]{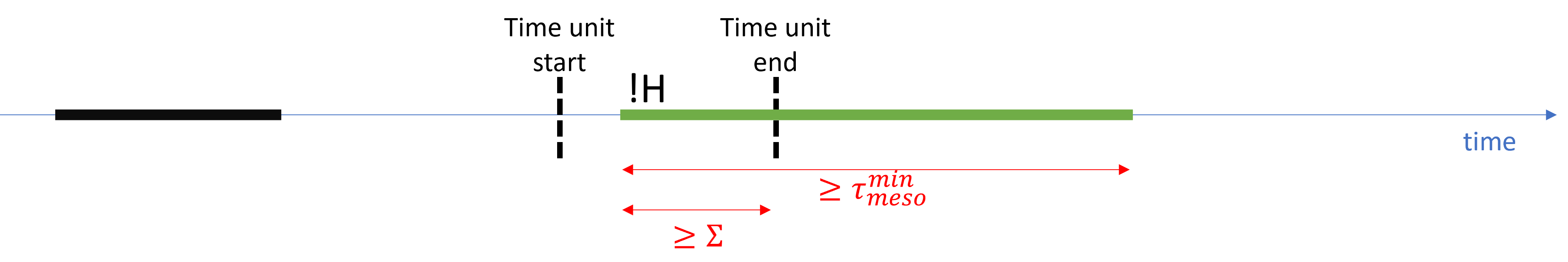}
\caption*{\small{\textit{Note}: This is the same configuration as in diagram \ref{fig:migrationDiagram_status_high_1}, but the user exits the sample at the end of the green segment. The situation is otherwise equivalent to that of diagram \ref{fig:migrationDiagram_status_high_1}.}}
\end{figure}

\begin{figure}[H]
\renewcommand{\figurename}{Diagram}
\caption{High-confidence migration status: case 3}
\label{fig:migrationDiagram_status_high_3}
\centering
  \includegraphics[width=\textwidth]{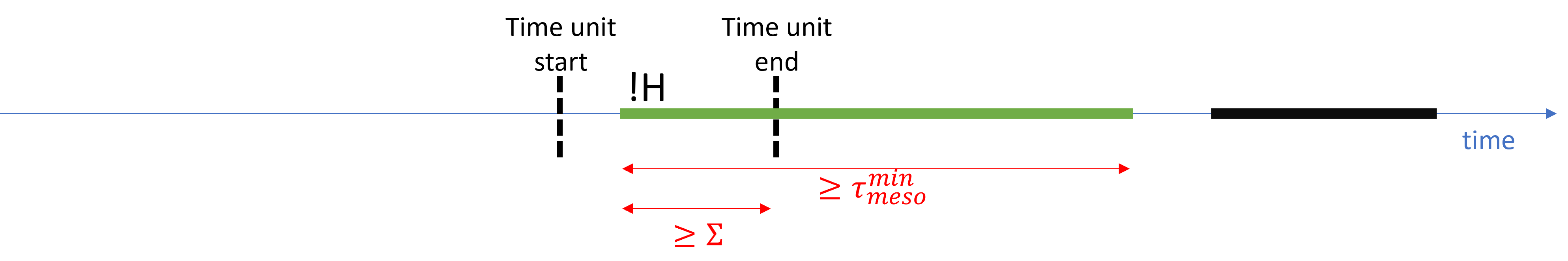}
\caption*{\small{\textit{Note}: This is the same configuration as in diagram \ref{fig:migrationDiagram_status_high_1}, but the user is never observed before the green segment. The situation is otherwise equivalent to that of diagram \ref{fig:migrationDiagram_status_high_1}.}}
\end{figure}

\begin{figure}[H]
\renewcommand{\figurename}{Diagram}
\caption{High-confidence migration status: case 4}
\label{fig:migrationDiagram_status_high_4}
\centering
  \includegraphics[width=\textwidth]{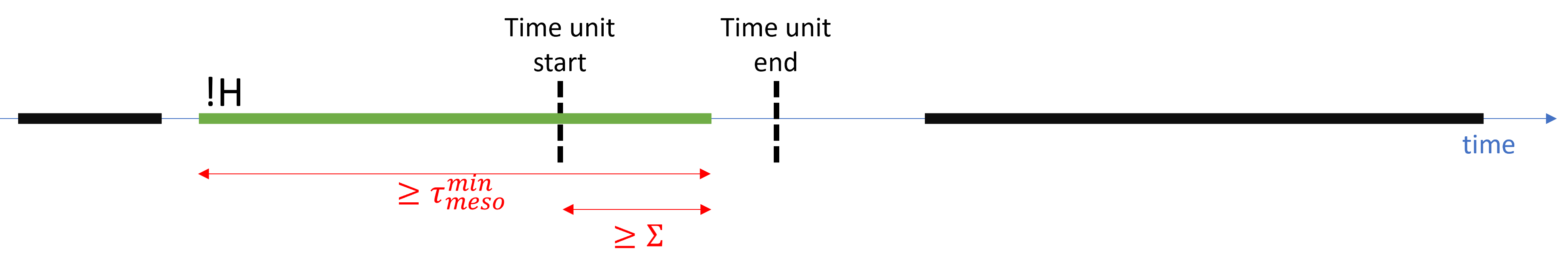}
\caption*{\small{\textit{Note}: The green segment is at a non-home location and has a duration greater than $\tau^{temp}$: it is a migration segment. It overlaps on the left of the time unit for a duration of at least $\Sigma$ days. The user is therefore considered as being in migration during the time unit in the high-confidence estimate.}}
\end{figure}

\begin{figure}[H]
\renewcommand{\figurename}{Diagram}
\caption{High-confidence migration status: case 5}
\label{fig:migrationDiagram_status_high_5}
\centering
  \includegraphics[width=\textwidth]{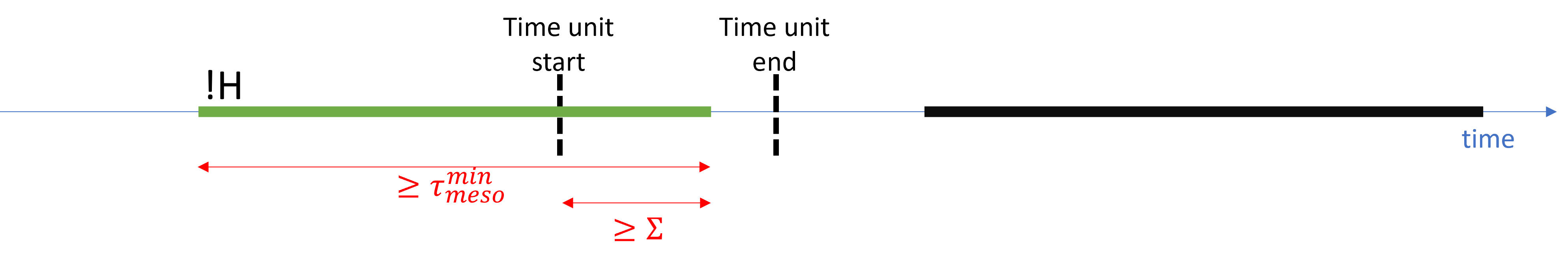}
\caption*{\small{\textit{Note}: This is the same configuration as in diagram \ref{fig:migrationDiagram_status_high_4}, but the user is never observed before the green segment. The situation is otherwise equivalent to that of diagram \ref{fig:migrationDiagram_status_high_4}.}}
\end{figure}

\begin{figure}[H]
\renewcommand{\figurename}{Diagram}
\caption{High-confidence migration status: case 6}
\label{fig:migrationDiagram_status_high_6}
\centering
  \includegraphics[width=\textwidth]{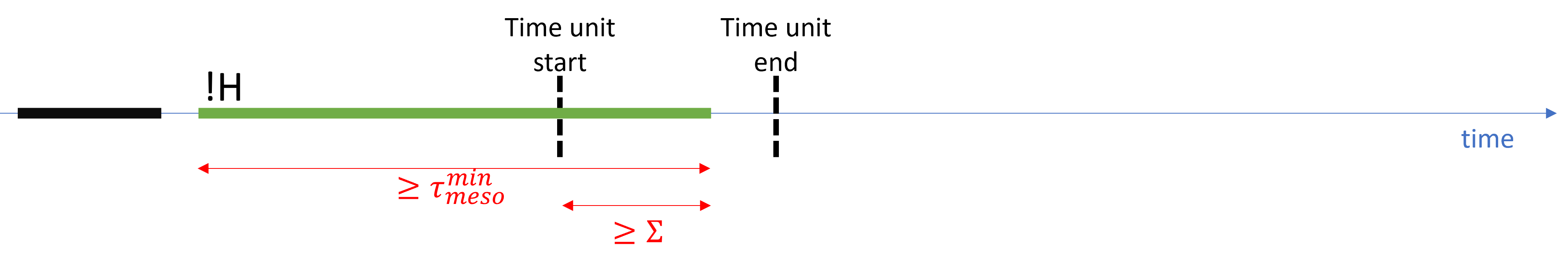}
\caption*{\small{\textit{Note}: This is the same configuration as in diagram \ref{fig:migrationDiagram_status_high_4}, but the user exits the sample at the end of the green segment. The situation is otherwise equivalent to that of diagram \ref{fig:migrationDiagram_status_high_4}.}}
\end{figure}

\begin{figure}[H]
\renewcommand{\figurename}{Diagram}
\caption{High-confidence migration status: case 7}
\label{fig:migrationDiagram_status_high_7}
\centering
  \includegraphics[width=\textwidth]{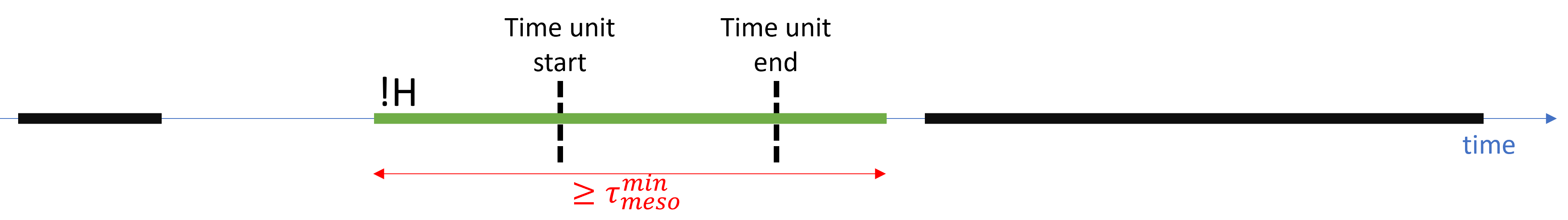}
\caption*{\small{\textit{Note}: The green segment is at a non-home location and has a duration greater than $\tau^{temp}$: it is a migration segment. It covers the entire time unit so the user is considered as being in migration during the time unit in the high-confidence estimate. Note that $\Sigma$ is necessarily set at a value that is lower than the duration of time units considered.}}
\end{figure}

\begin{figure}[H]
\renewcommand{\figurename}{Diagram}
\caption{High-confidence migration status: case 8}
\label{fig:migrationDiagram_status_high_8}
\centering
  \includegraphics[width=\textwidth]{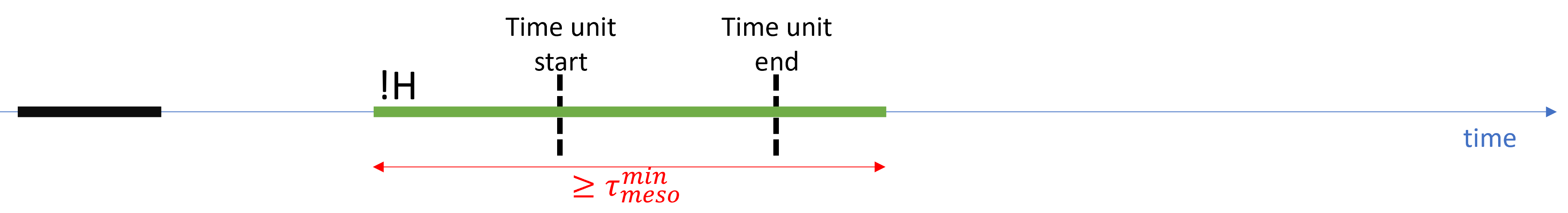}
\caption*{\small{\textit{Note}: This is the same configuration as in diagram \ref{fig:migrationDiagram_status_high_7}, but the user exits the sample at the end of the green segment. The situation is otherwise equivalent to that of diagram \ref{fig:migrationDiagram_status_high_7}.}}
\end{figure}

\begin{figure}[H]
\renewcommand{\figurename}{Diagram}
\caption{High-confidence migration status: case 9}
\label{fig:migrationDiagram_status_high_9}
\centering
  \includegraphics[width=\textwidth]{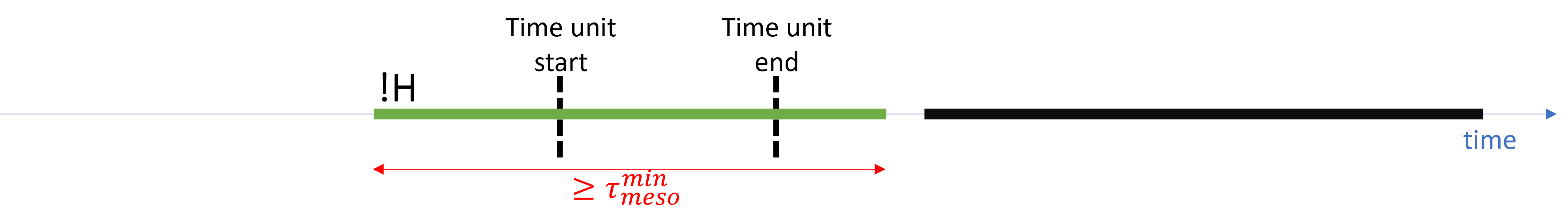}
\caption*{\small{\textit{Note}: This is the same configuration as in diagram \ref{fig:migrationDiagram_status_high_7}, but the user is never observed before the green segment. The situation is otherwise equivalent to that of diagram \ref{fig:migrationDiagram_status_high_7}.}}
\end{figure}

\subsection{Identifying migration status: low-confidence}

\begin{figure}[H]
\renewcommand{\figurename}{Diagram}
\caption{Low-confidence migration status: case 1}
\label{fig:migrationDiagram_status_low_1}
\centering
  \includegraphics[width=\textwidth]{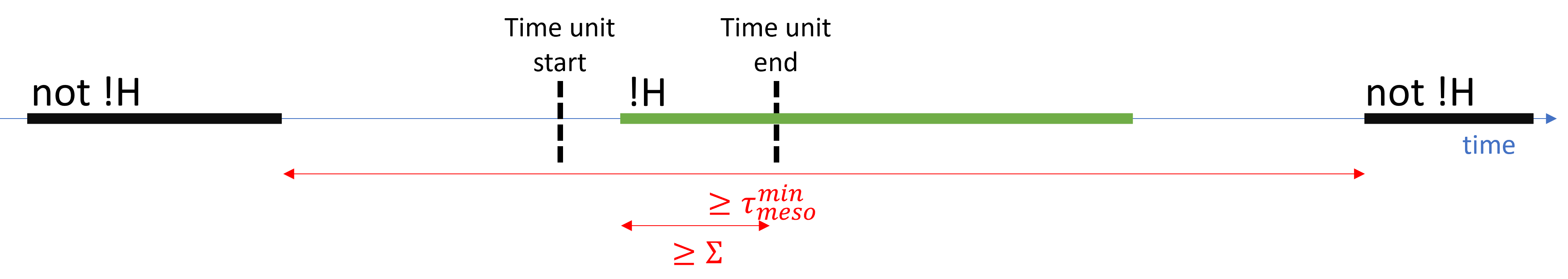}
\caption*{\small{\textit{Note}: The green segment at !H does not necessarily have an observed duration greater than $\tau^{temp}$. The maximum duration is the time elapsed between the day following the previous segment and the date preceding the following segment. When this is greater than $\tau^{temp}$ and the green segment overlaps with the time unit on the right for at least $\Sigma$ days, the user is considered as being in migration during the time unit in the low-confidence estimate.}}
\end{figure}

\begin{figure}[H]
\renewcommand{\figurename}{Diagram}
\caption{Low-confidence migration status: case 2}
\label{fig:migrationDiagram_status_low_2}
\centering
  \includegraphics[width=\textwidth]{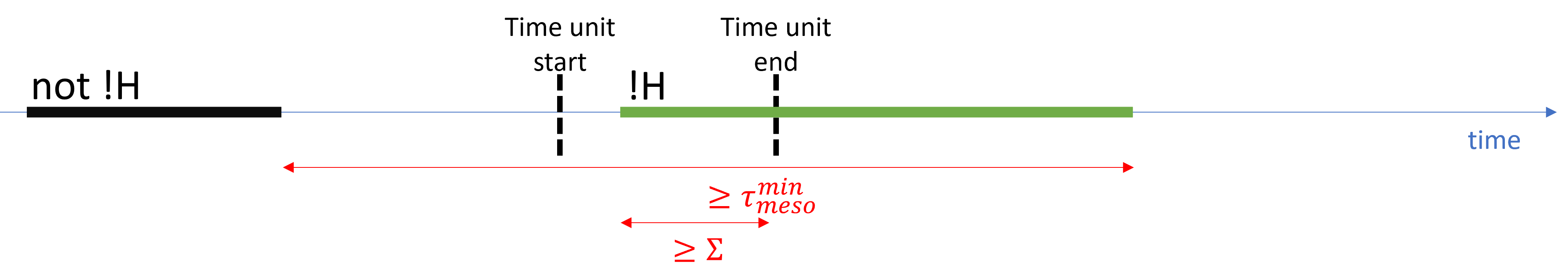}
\caption*{\small{\textit{Note}: This is the same configuration as in diagram \ref{fig:migrationDiagram_status_low_1}, but the user exits the sample at the end of the green segment. In the absence of information about the user's location after the green segment, the maximum duration is limited to the time elapsed between the day following the previous segment and the observed end date. The situation is otherwise equivalent to that of diagram \ref{fig:migrationDiagram_status_low_1}.}}
\end{figure}

\begin{figure}[H]
\renewcommand{\figurename}{Diagram}
\caption{Low-confidence migration status: case 3}
\label{fig:migrationDiagram_status_low_3}
\centering
  \includegraphics[width=\textwidth]{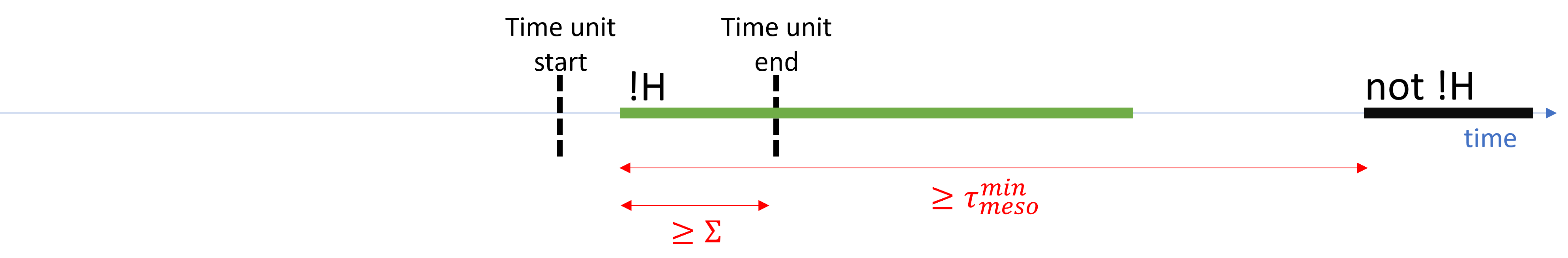}
\caption*{\small{\textit{Note}: This is the same configuration as in diagram \ref{fig:migrationDiagram_status_low_1}, but the user is never observed before the green segment. In the absence of information about the user's location before the green segment, the maximum duration is limited to the time elapsed between the observed start date and the day preceding the following segment. The situation is otherwise equivalent to that of diagram \ref{fig:migrationDiagram_status_low_1}.}}
\end{figure}

\begin{figure}[H]
\renewcommand{\figurename}{Diagram}
\caption{Low-confidence migration status: case 4}
\label{fig:migrationDiagram_status_low_4}
\centering
  \includegraphics[width=\textwidth]{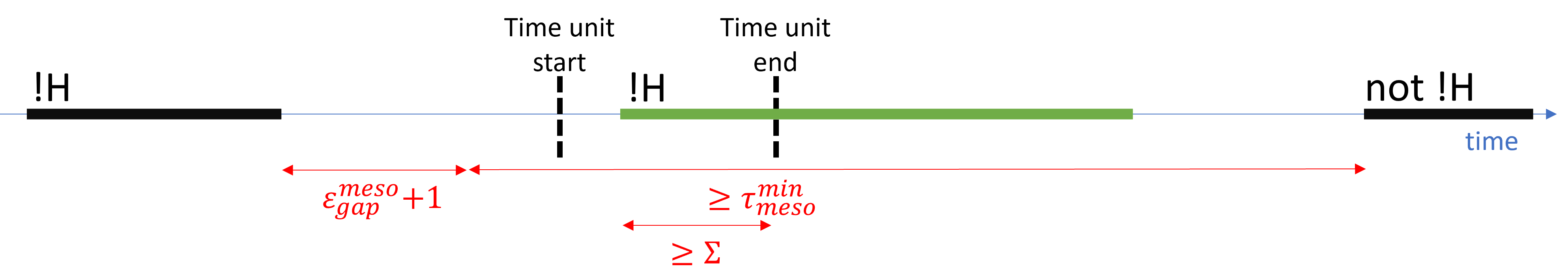}
\caption*{\small{\textit{Note}: This is the same configuration as in diagram \ref{fig:migrationDiagram_status_low_1}, but the segment preceding the green segment is at the same location !H. In this case, the green segment cannot have started less than $\epsilon_{gap}^{meso}+1$ days after that segment. The maximum duration is then the time elapsed between the minimum start date of the green segment (i.e. $\epsilon_{gap}^{meso}+1$ days after the end of the preceding segment) and the day preceding the first day of the following segment. When this is larger than $\tau^{temp}$ and the green segment overlaps with the time unit on the right on at least $\Sigma$ days, the user is considered as being in migration during the time unit in the low-confidence estimate.}}
\end{figure}

\begin{figure}[H]
\renewcommand{\figurename}{Diagram}
\caption{Low-confidence migration status: case 5}
\label{fig:migrationDiagram_status_low_5}
\centering
  \includegraphics[width=\textwidth]{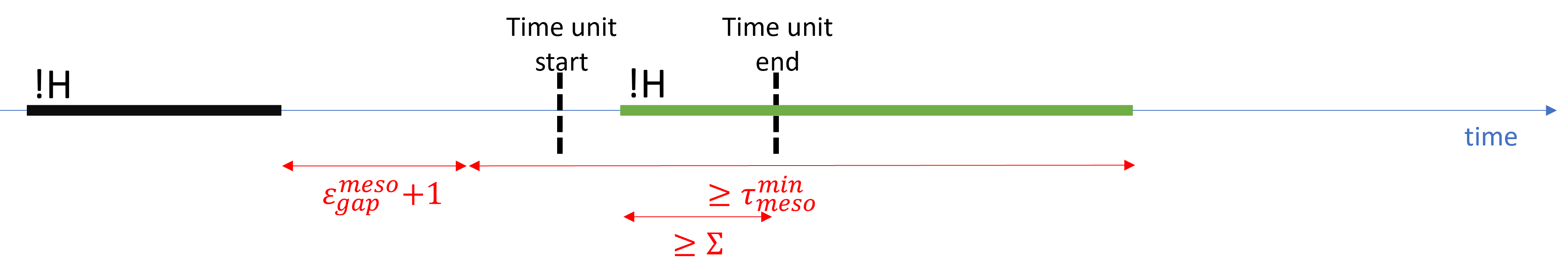}
\caption*{\small{\textit{Note}: This is the same configuration as in diagram \ref{fig:migrationDiagram_status_low_4}, but the user exits the sample after the green segment. In the absence of information about the user's location after the end of the green segment, the maximum end date possible is considered to coincide with the observed end date. The maximum duration is then the time elapsed between the minimum start date of the green segment (i.e. $\epsilon_{gap}^{meso}+1$ days after the end of the preceding segment) and the observed end date. When this is larger than $\tau^{temp}$ and the green segment overlaps with the time unit on the right on at least $\Sigma$ days, the user is considered as being in migration during the time unit in the low-confidence estimate.}}
\end{figure}

\begin{figure}[H]
\renewcommand{\figurename}{Diagram}
\caption{Low-confidence migration status: case 6}
\label{fig:migrationDiagram_status_low_6}
\centering
  \includegraphics[width=\textwidth]{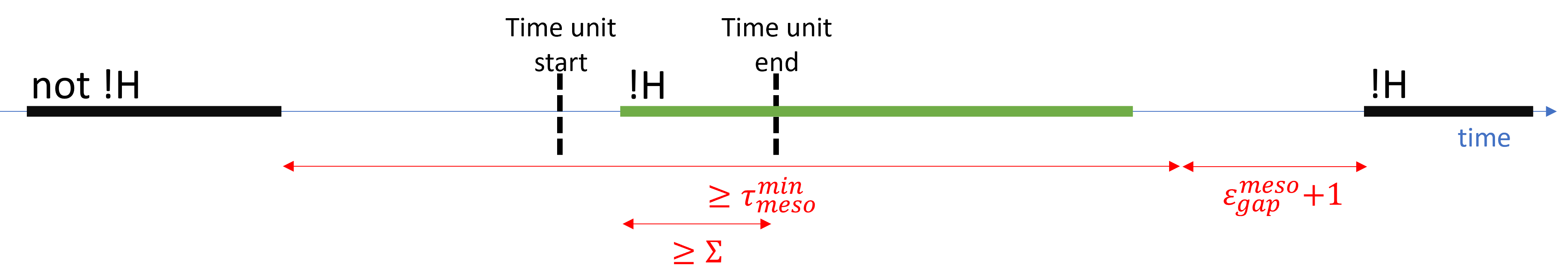}
\caption*{\small{\textit{Note}: This is the same configuration as in diagram \ref{fig:migrationDiagram_status_low_1}, but the segment following the green segment is at the same location !H. In this case, the green segment cannot have ended less than $\epsilon_{gap}^{meso}+1$ days before that segment. The maximum duration is then the time elapsed between the day following the last day of the preceding segment and the maximum end date of the green segment (i.e. $\epsilon_{gap}^{meso}+1$ days before the start of the following segment). When this is larger than $\tau^{temp}$ and the green segment overlaps with the time unit on the right on at least $\Sigma$ days, the user is considered as being in migration during the time unit in the low-confidence estimate.}}
\end{figure}

\begin{figure}[H]
\renewcommand{\figurename}{Diagram}
\caption{Low-confidence migration status: case 7}
\label{fig:migrationDiagram_status_low_7}
\centering
  \includegraphics[width=\textwidth]{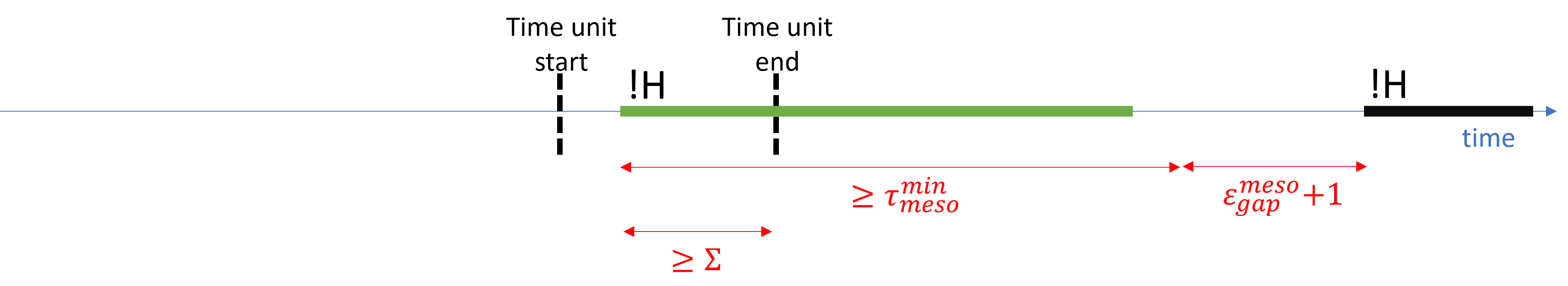}
\caption*{\small{\textit{Note}: This is the same configuration as in diagram \ref{fig:migrationDiagram_status_low_6}, but the user is never seen before the green segment. In the absence of information about the user's location before the green segment started, the minimum start date possible is considered to coincide with the observed start date. The maximum duration is then the time elapsed between the observed start date and the maximum end date of the green segment (i.e. $\epsilon_{gap}^{meso}+1$ days before the start of the following segment). When this is larger than $\tau^{temp}$ and the green segment overlaps with the time unit on the right on at least $\Sigma$ days, the user is considered as being in migration during the time unit in the low-confidence estimate.}}
\end{figure}

\begin{figure}[H]
\renewcommand{\figurename}{Diagram}
\caption{Low-confidence migration status: case 8}
\label{fig:migrationDiagram_status_low_8}
\centering
  \includegraphics[width=\textwidth]{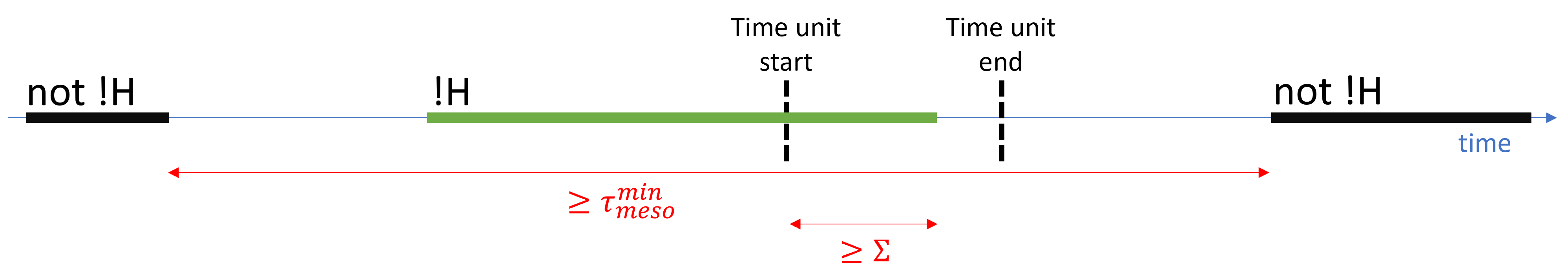}
\caption*{\small{\textit{Note}: This configuration is exactly equivalent to diagram \ref{fig:migrationDiagram_status_low_1} with the green segment overlapping on the left of the time unit.}}
\end{figure}

\begin{figure}[H]
\renewcommand{\figurename}{Diagram}
\caption{Low-confidence migration status: case 9}
\label{fig:migrationDiagram_status_low_9}
\centering
  \includegraphics[width=\textwidth]{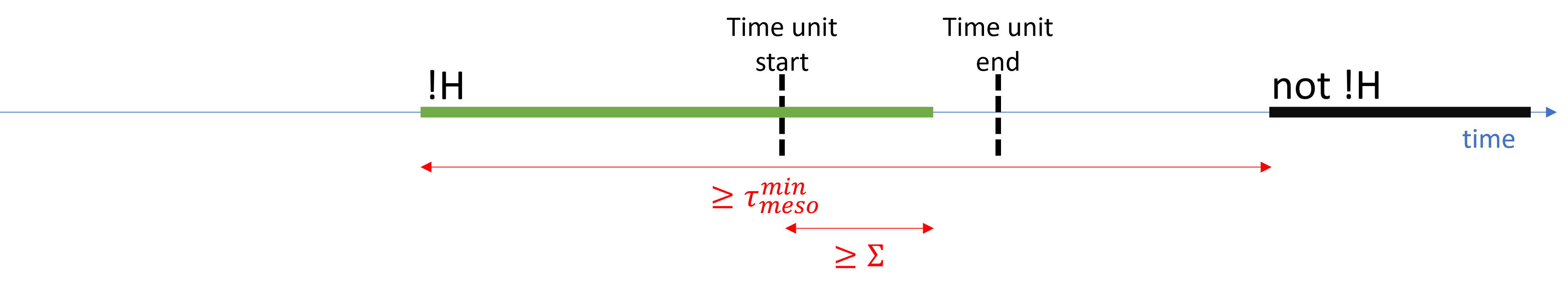}
\caption*{\small{\textit{Note}: This configuration is exactly equivalent to diagram \ref{fig:migrationDiagram_status_low_3} with the green segment overlapping on the left of the time unit.}}
\end{figure}

\begin{figure}[H]
\renewcommand{\figurename}{Diagram}
\caption{Low-confidence migration status: case 10}
\label{fig:migrationDiagram_status_low_10}
\centering
  \includegraphics[width=\textwidth]{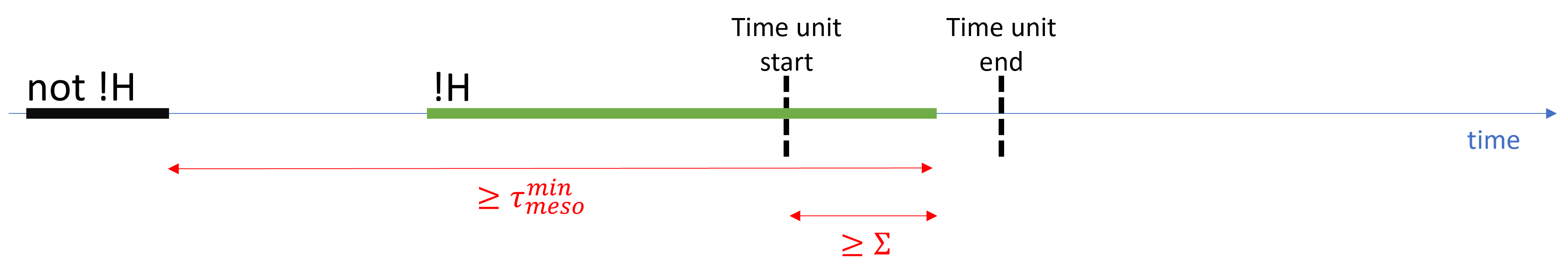}
\caption*{\small{\textit{Note}: This configuration is exactly equivalent to diagram \ref{fig:migrationDiagram_status_low_2} with the green segment overlapping on the left of the time unit.}}
\end{figure}

\begin{figure}[H]
\renewcommand{\figurename}{Diagram}
\caption{Low-confidence migration status: case 11}
\label{fig:migrationDiagram_status_low_11}
\centering
  \includegraphics[width=\textwidth]{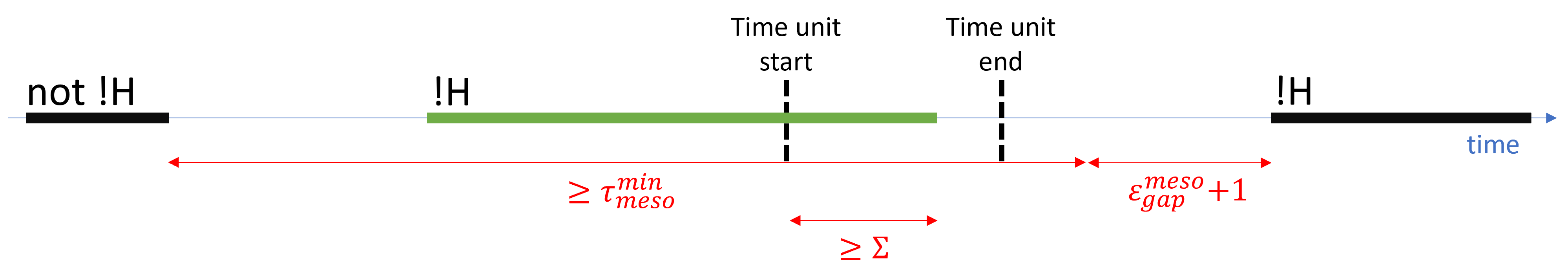}
\caption*{\small{\textit{Note}: This configuration is exactly equivalent to diagram \ref{fig:migrationDiagram_status_low_6} with the green segment overlapping on the left of the time unit.}}
\end{figure}

\begin{figure}[H]
\renewcommand{\figurename}{Diagram}
\caption{Low-confidence migration status: case 12}
\label{fig:migrationDiagram_status_low_12}
\centering
  \includegraphics[width=\textwidth]{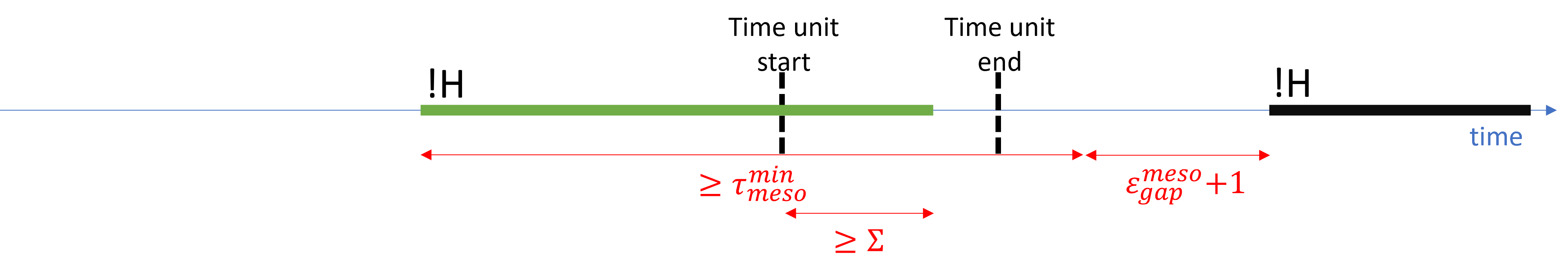}
\caption*{\small{\textit{Note}: This configuration is exactly equivalent to diagram \ref{fig:migrationDiagram_status_low_7} with the green segment overlapping on the left of the time unit.}}
\end{figure}

\begin{figure}[H]
\renewcommand{\figurename}{Diagram}
\caption{Low-confidence migration status: case 13}
\label{fig:migrationDiagram_status_low_13}
\centering
  \includegraphics[width=\textwidth]{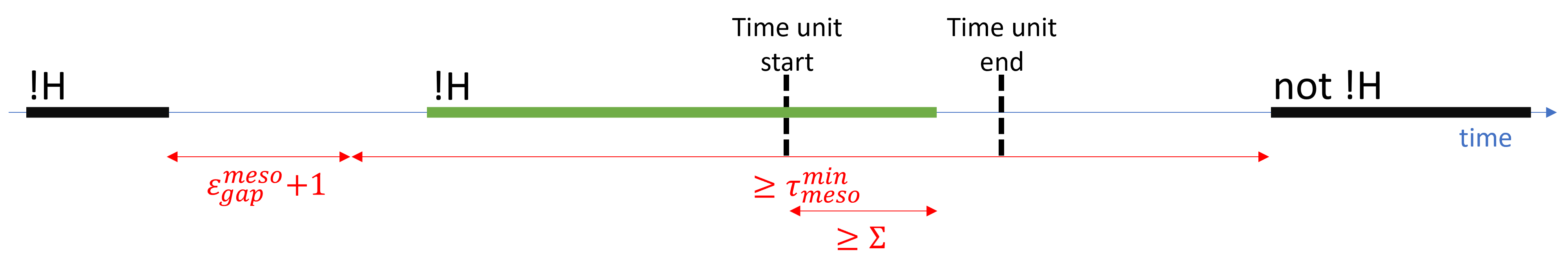}
\caption*{\small{\textit{Note}: This configuration is exactly equivalent to diagram \ref{fig:migrationDiagram_status_low_4} with the green segment overlapping on the left of the time unit.}}
\end{figure}

\begin{figure}[H]
\renewcommand{\figurename}{Diagram}
\caption{Low-confidence migration status: case 14}
\label{fig:migrationDiagram_status_low_14}
\centering
  \includegraphics[width=\textwidth]{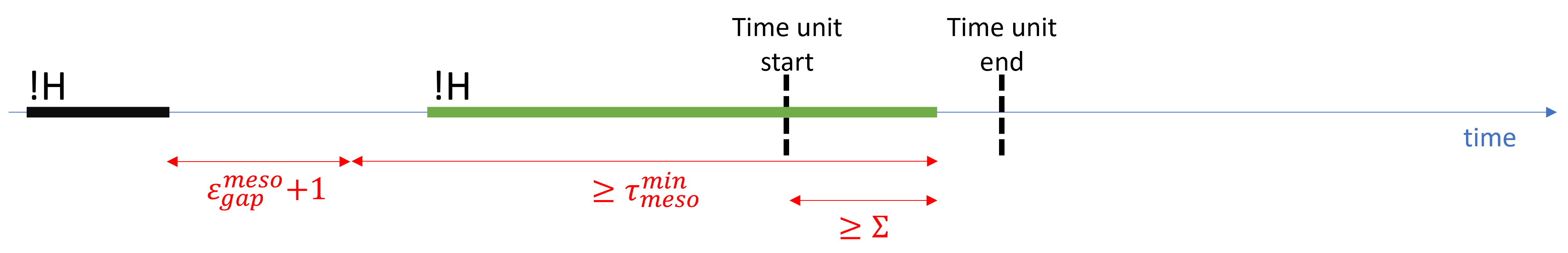}
\caption*{\small{\textit{Note}: This configuration is exactly equivalent to diagram \ref{fig:migrationDiagram_status_low_5} with the green segment overlapping on the left of the time unit.}}
\end{figure}

\section{Algorithmic rules to count the observation status of users by time unit\label{sec:trajectoryToStat_obsStatus_appendix}}

\subsection{Identifying observation status for migration departure}
\begin{figure}[H]
\renewcommand{\figurename}{Diagram}
\caption{Observation status for migration departure: case 1}
\label{fig:obsDiagram_depart_left_1}
\centering
  \includegraphics[width=\textwidth]{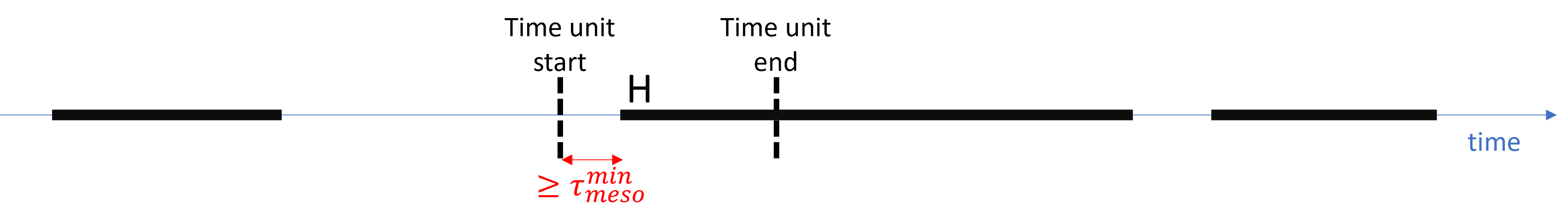}
\caption*{\small{\textit{Note}: An observation gap overlaps with the time unit on the left and is followed by a home segment. If the gap between the time unit start date and the home segment start date is larger than the parameter $\tau^{temp}$, a migration segment may have started during the time unit without being observed. The user is considered as being not observed when calculating the number of migration departures during that time unit.}}
\end{figure}

\begin{figure}[H]
\renewcommand{\figurename}{Diagram}
\caption{Observation status for migration departure: case 2}
\label{fig:obsDiagram_depart_left_2}
\centering
  \includegraphics[width=\textwidth]{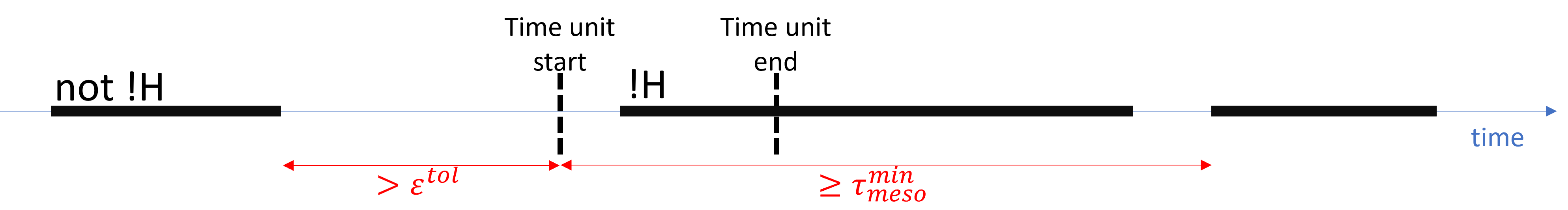}
\caption*{\small{\textit{Note}: An observation gap overlaps with the time unit on the left and is assumed to be strictly smaller than $\tau^{temp}$ -- we are back to case 1 of diagram \ref{fig:obsDiagram_depart_left_1} otherwise. It is followed by a non-home segment at location !H, and preceded by a segment at a location that is not !H. The non-home segment may be a migration segment with a start date within the time unit if the time elapsed between the time unit start date and the day preceding the following segment is greater than $\tau^{temp}$. When the tolerance criterion is exceeded, i.e. the time elapsed between the start of the observation gap and the day preceding the start of the time unit exceeds the parameter $\epsilon^tol$, the uncertainty around the actual start date of the segment at the non-home location is considered as too large and the user is not counted as being observed for the calculation of migration departures during that time unit. Note that if the time elapsed between the time unit start date and the day preceding the following segment is strictly less than $\tau^{temp}$, the non-home segment cannot be a migration segment and the user is then considered as observed and not having departed for migration during the time unit.}}
\end{figure}

\begin{figure}[H]
\renewcommand{\figurename}{Diagram}
\caption{Observation status for migration departure: case 3}
\label{fig:obsDiagram_depart_left_3}
\centering
  \includegraphics[width=\textwidth]{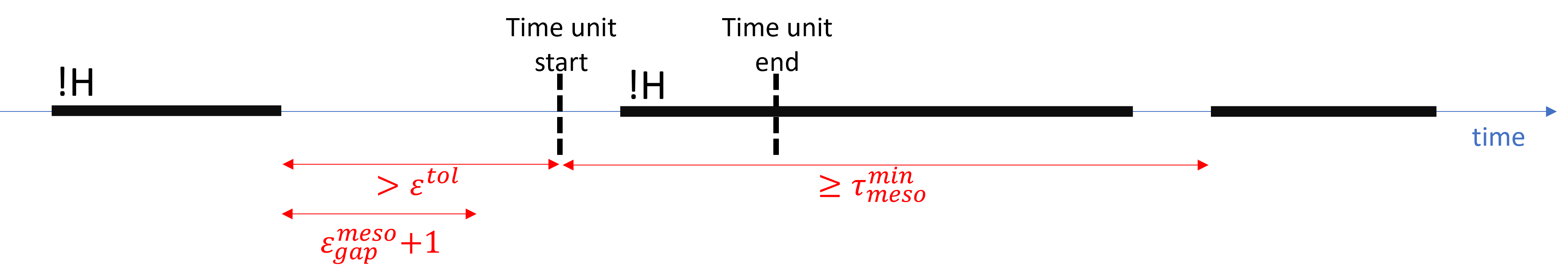}
\caption*{\small{\textit{Note}: This configuration is the same as in diagram \ref{fig:obsDiagram_depart_left_2}, but the segment preceding the observation gap is at the same location !H as the segment following the gap. The observation gap is necessarily strictly larger than $\epsilon_{gap}^{meso}$, otherwise both segments would be have been merged in the migration detection procedure. In this case, the non-home segment overlapping with the time unit cannot have started earlier than $\epsilon_{gap}^{meso}+1$ days after the previous segment. When this minimum start date falls outside the time unit, the maximum possible duration of that non-home segment to still be considered as having potentially started during the time unit is the time elapsed between the time unit start date and the day preceding the first day of the following segment. When this is larger than $\tau^{temp}$, the possibility exists that a migration event started during the time unit. However, if the tolerance criterion is exceeded, i.e. the time elapsed between the start of the observation gap and the day preceding the start of the time unit exceeds the parameter $\epsilon^{tol}$, the uncertainty around the actual start date of the segment at the non-home location is considered as too large and the user is not counted as being observed for the calculation of migration departures during that time unit.}}
\end{figure}

\begin{figure}[H]
\renewcommand{\figurename}{Diagram}
\caption{Observation status for migration departure: case 4}
\label{fig:obsDiagram_depart_left_4}
\centering
  \includegraphics[width=\textwidth]{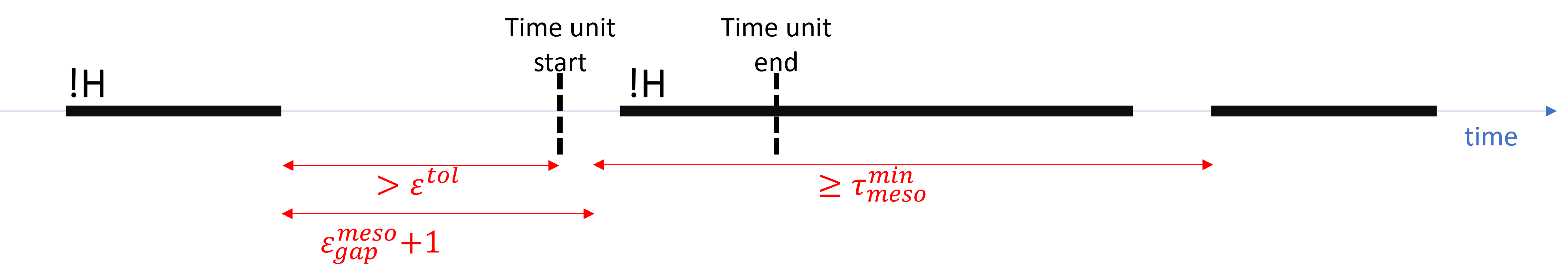}
\caption*{\small{\textit{Note}: This configuration is the same as in diagram \ref{fig:obsDiagram_depart_left_3} for the case when the date corresponding to $\epsilon_{gap}^{meso}+1$ days after the previous segment falls within the time unit. The maximum possible duration of the non-home segment to still be considered as having potentially started during the time unit is now the time elapsed between this date -- instead of the time unit start date-- and the day preceding the first day of the following segment. The situation is then the same as in diagram \ref{fig:obsDiagram_depart_left_3}.}}
\end{figure}

\begin{figure}[H]
\renewcommand{\figurename}{Diagram}
\caption{Observation status for migration departure: case 5}
\label{fig:obsDiagram_depart_left_5}
\centering
  \includegraphics[width=\textwidth]{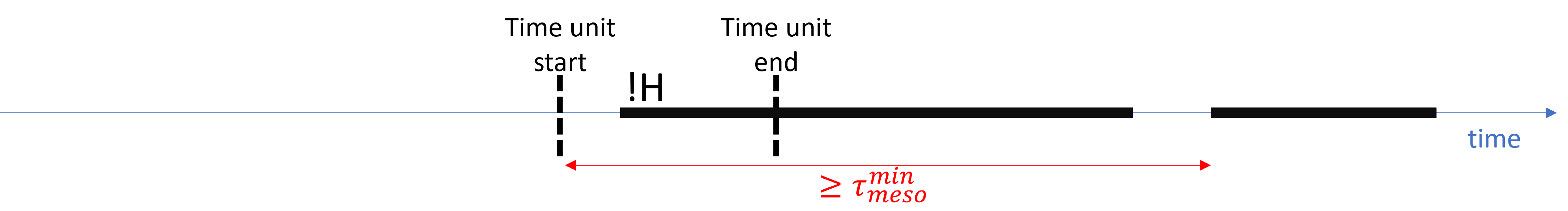}
\caption*{\small{\textit{Note}: This configuration is the same as in diagrams \ref{fig:obsDiagram_depart_left_2}-\ref{fig:obsDiagram_depart_left_4}, but the user is never observed before the non-home segment overlapping with the time unit. In the absence of further information about the user's location, it is assumed that the minimum start date of the non-home segment to consider it started during the time unit coincides with the first day of the time unit. When the maximum duration, i.e the time elapsed between this minimum start date and the day preceding the following segment, is greater than $\tau^{temp}$, the possibility exists that a migration occurred and started during the time unit. Since the user is not observed before the non-home segment, the uncertainty around the actual start date is virtually infinite and the user is not counted as being observed for the calculation of migration departures during that time unit.}}
\end{figure}

\begin{figure}[H]
\renewcommand{\figurename}{Diagram}
\caption{Observation status for migration departure: case 6}
\label{fig:obsDiagram_depart_right_1}
\centering
  \includegraphics[width=\textwidth]{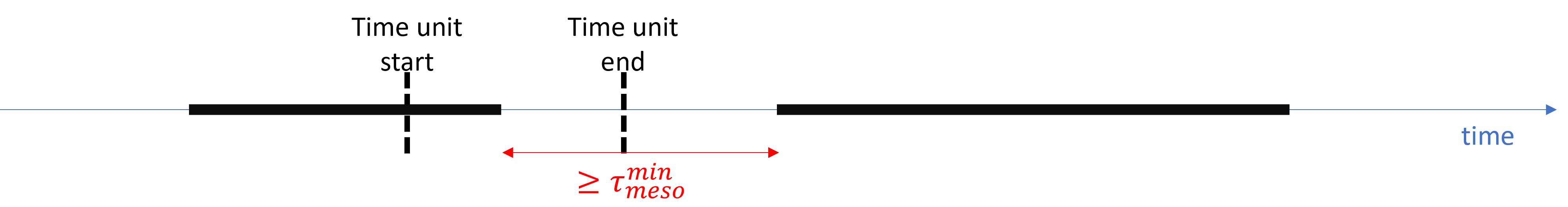}
\caption*{\small{\textit{Note}: An observation gap overlaps with the time unit on the right. Regardless of the locations of the preceding and following segments, if the gap is greater than $\tau^{temp}$ then a migration event could have started during the time unit without being observed. The user is thus considered as being not observed when calculating the number of migration departures during that time unit.}}
\end{figure}

\begin{figure}[H]
\renewcommand{\figurename}{Diagram}
\caption{Observation status for migration departure: case 7}
\label{fig:obsDiagram_depart_right_2}
\centering
  \includegraphics[width=\textwidth]{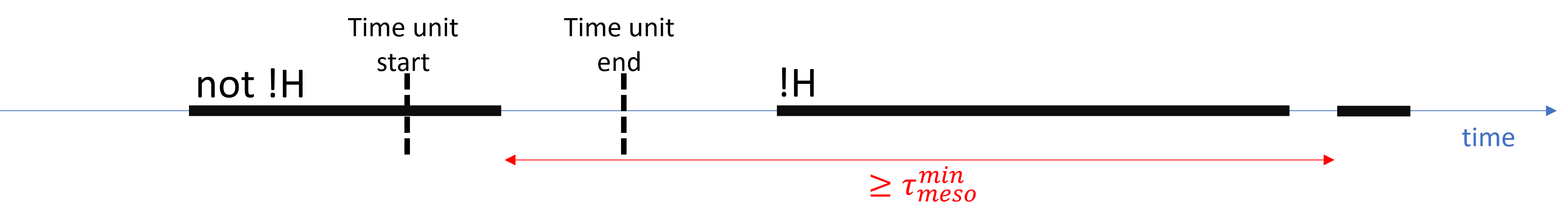}
\caption*{\small{\textit{Note}: An observation gap overlaps with the time unit on the right and is assumed to be strictly smaller than $\tau^{temp}$ -- we are back to case 6 of diagram \ref{fig:obsDiagram_depart_right_1} otherwise. It is followed by a non-home segment at location !H, and preceded by a segment at a location that is not !H. The non-home segment may be a migration segment with a start date within the time unit if the time elapsed between the observation gap start date and the day preceding the following segment is greater than $\tau^{temp}$. In that case, it is impossible to determine whether the user effectively departed for migration during the time unit or not. The user is thus considered as being not observed when calculating the number of migration departures during that time unit. Conversely, if this maximum duration was strictly less than $\tau^{temp}$, the observation gaps could not be concealing a migration event starting during the time unit and the user would be considered as effectively observed and not having departed for migration.}}
\end{figure}

\begin{figure}[H]
\renewcommand{\figurename}{Diagram}
\caption{Observation status for migration departure: case 8}
\label{fig:obsDiagram_depart_right_3}
\centering
  \includegraphics[width=\textwidth]{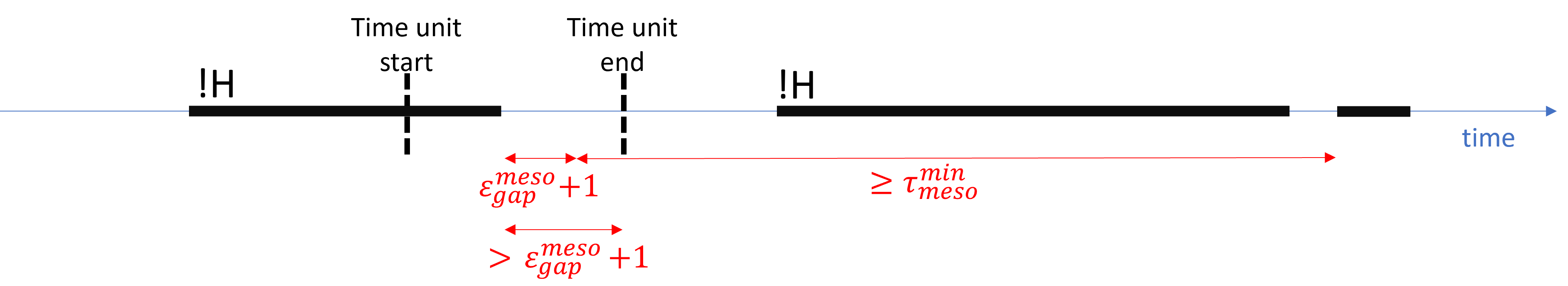}
\caption*{\small{\textit{Note}: This is the same configuration as in diagram \ref{fig:obsDiagram_depart_right_2}, although the segment preceding the observation gap is at the same location !H as the segment following the gap. Two conditions allow for the possibility that the segment following the gap started during the time unit without being observed. First, the minimum start date has to fall within the time unit; the minimum start date is $\epsilon_{gap}^{meso}+2$ days after the preceding segment. Second, the maximum duration, defined as the time elapsed between the minimum start date and the day preceding the following segment, has to be greater than $\tau^{temp}$. When both conditions are met, the degree of uncertainty is such that the user is considered as being not observed when calculating the number of migration departures during that time unit.}}
\end{figure}

\begin{figure}[H]
\renewcommand{\figurename}{Diagram}
\caption{Observation status for migration departure: case 9}
\label{fig:obsDiagram_depart_right_4}
\centering
  \includegraphics[width=\textwidth]{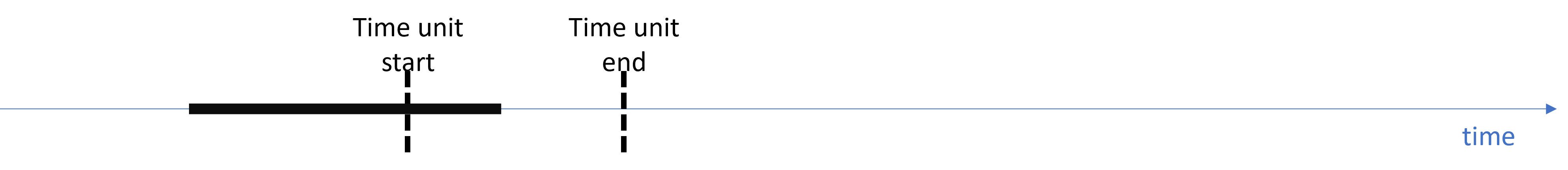}
\caption*{\small{\textit{Note}: An unbounded observation gap overlaps with the time unit on the right; the user exits the sample. The possibility exists that the user departed for migration during the unobserved period on the right of the time unit. Without further information on the user's location after that, the user is considered as being not observed when calculating the number of migration departures during that time unit.}}
\end{figure}

\begin{figure}[H]
\renewcommand{\figurename}{Diagram}
\caption{Observation status for migration departure: case 10}
\label{fig:obsDiagram_depart_full_1}
\centering
  \includegraphics[width=\textwidth]{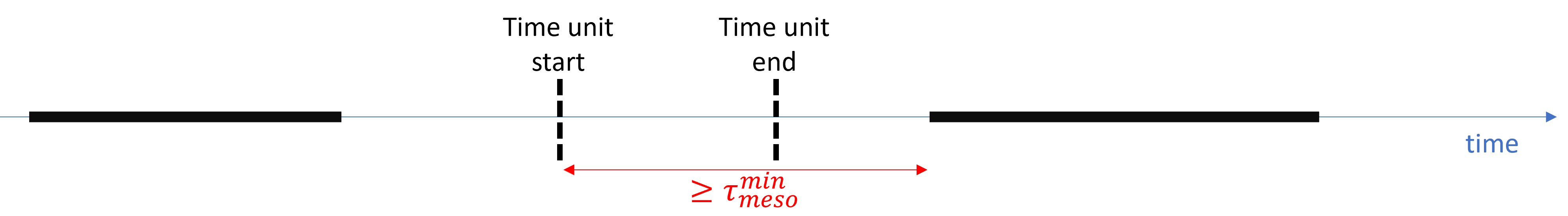}
\caption*{\small{\textit{Note}: An observation gap fully covers the time unit. Regardless of the locations of the preceding and following segments, if the time elapsed between the time unit start date and the end of the observation gap is greater than $\tau^{temp}$ then a migration event could have started during the time unit without being observed. The user is thus considered as being not observed when calculating the number of migration departures during that time unit.}}
\end{figure}

\begin{figure}[H]
\renewcommand{\figurename}{Diagram}
\caption{Observation status for migration departure: case 11}
\label{fig:obsDiagram_depart_full_2}
\centering
  \includegraphics[width=\textwidth]{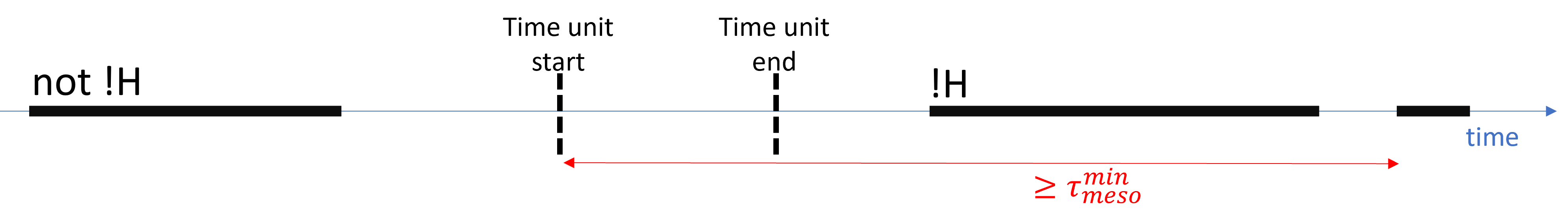}
\caption*{\small{\textit{Note}: An observation gap fully covers the time unit and the fraction of the gap that starts on the first day of the time unit is assumed strictly smaller than $\tau^{temp}$ -- we are back to case 10 of diagram \ref{fig:obsDiagram_depart_full_1} otherwise. It is followed by a non-home segment at location !H, and preceded by a segment at a location that is not !H. The non-home segment may be a migration segment with a start date within the time unit if the time elapsed between the time unit start date and the day preceding the following segment is greater than $\tau^{temp}$. In that case, it is impossible to determine whether the user effectively departed for migration during the time unit or not. The user is thus considered as being not observed when calculating the number of migration departures during that time unit. Conversely, if this maximum duration was strictly less than $\tau^{temp}$, the observation gaps could not be concealing a migration event starting during the time unit and the user would be considered as effectively observed and not having departed for migration.}}
\end{figure}

\begin{figure}[H]
\renewcommand{\figurename}{Diagram}
\caption{Observation status for migration departure: case 12}
\label{fig:obsDiagram_depart_full_3}
\centering
  \includegraphics[width=\textwidth]{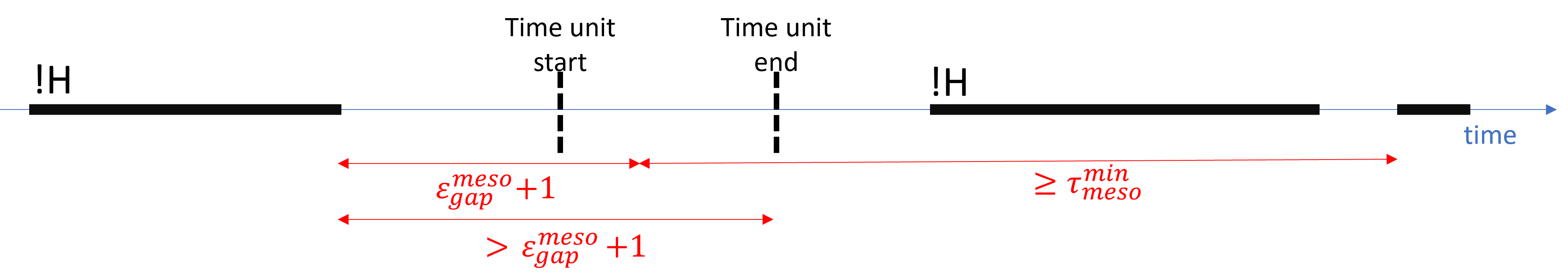}
\caption*{\small{\textit{Note}: This is the same configuration as in diagram \ref{fig:obsDiagram_depart_full_2}, although the segment preceding the observation gap is at the same location !H as the segment following the gap. Two conditions allow for the possibility that the segment following the gap started during the time unit without being observed. First, the minimum start date cannot fall after the end of the time unit; the minimum start date is $\epsilon_{gap}^{meso}+2$ days after the preceding segment. This diagram shows the case when the minimum start date falls within the time unit. Second, the maximum duration, defined as the time elapsed between the minimum start date and the day preceding the following segment, has to be greater than $\tau^{temp}$. When both conditions are met, the degree of uncertainty is such that the user is considered as being not observed when calculating the number of migration departures during that time unit.}}
\end{figure}

\begin{figure}[H]
\renewcommand{\figurename}{Diagram}
\caption{Observation status for migration departure: case 13}
\label{fig:obsDiagram_depart_full_4}
\centering
  \includegraphics[width=\textwidth]{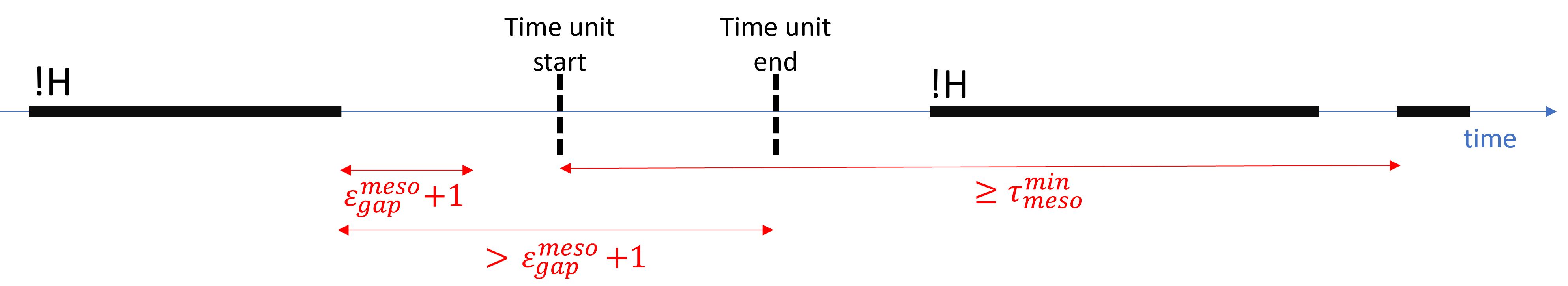}
\caption*{\small{\textit{Note}: This is the same configuration as in diagram \ref{fig:obsDiagram_depart_full_3} above, for the case where the minimum start strictly precedes the start of the time unit. In this case, the maximum duration that allows for the possibility that the non-home segment actually conceals a migration event that started during the time unit is the time elapsed between the time unit start date and the day preceding the following segment. Again, when this is greater than $\tau^{temp}$, the user is considered as being not observed when calculating the number of migration departures during that time unit.}}
\end{figure}

\begin{figure}[H]
\renewcommand{\figurename}{Diagram}
\caption{Observation status for migration departure: case 14}
\label{fig:obsDiagram_depart_full_5}
\centering
  \includegraphics[width=\textwidth]{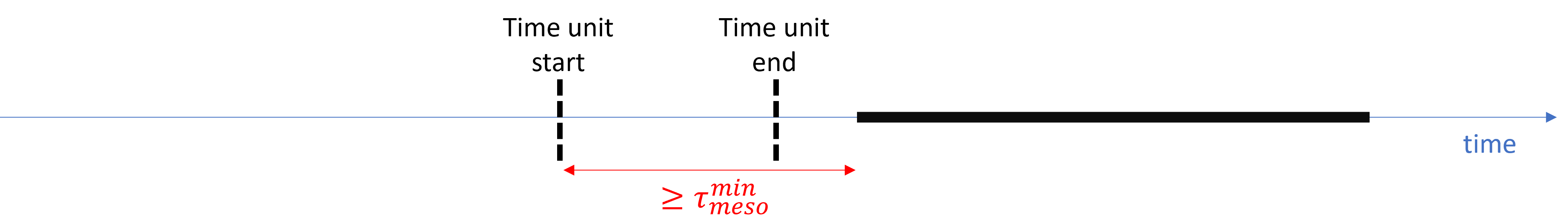}
\caption*{\small{\textit{Note}: This is the same configuration as in case 10 described in diagram \ref{fig:obsDiagram_depart_full_1}, for the case where the user is never observed before the time unit. The criterion to classify the user as not being observed for migration departure for that time unit remains unchanged: the time elapsed between the time unit start date and the day preceding the first segment is greater than $\tau^{temp}$.}}
\end{figure}

\begin{figure}[H]
\renewcommand{\figurename}{Diagram}
\caption{Observation status for migration departure: case 15}
\label{fig:obsDiagram_depart_full_6}
\centering
  \includegraphics[width=\textwidth]{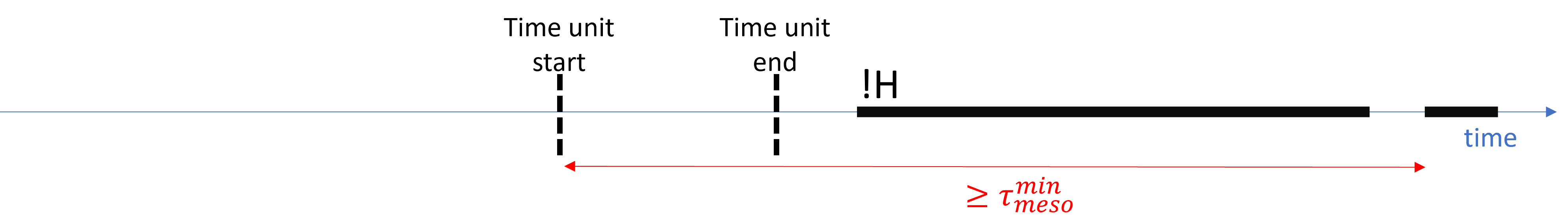}
\caption*{\small{\textit{Note}: This is the same configuration as in case 14 and where it is implicitly assumed that the time elapsed between the time unit start date and the day preceding the first segment is strictly less than $\tau^{temp}$. The first segment is at a non-home location !H. When the time elapsed between the time unit start date and the day preceding the start date of the segment following the non-home segment is greater than $\tau^{temp}$, the possibility exists that a migration departure effectively occurred during the time unit without being observed. The user is thus considered as being not observed when calculating the number of migration departures during that time unit.}}
\end{figure}

\begin{figure}[H]
\renewcommand{\figurename}{Diagram}
\caption{Observation status for migration departure: case 16}
\label{fig:obsDiagram_depart_full_7}
\centering
  \includegraphics[width=\textwidth]{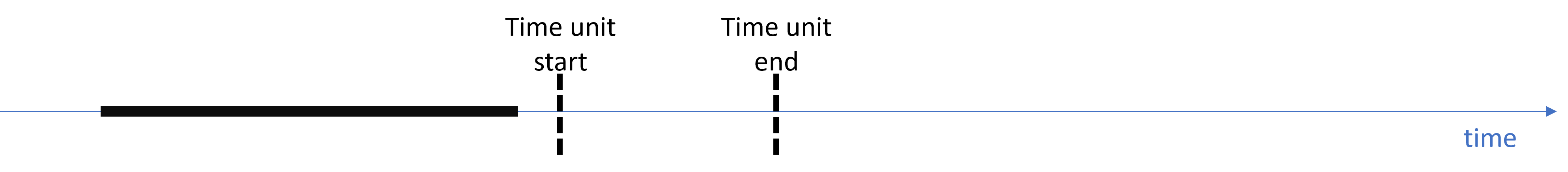}
\caption*{\small{\textit{Note}: The configuration is the same as in case 9 described in diagram \ref{fig:obsDiagram_depart_right_4}, but the observation gap fully covers the time unit. The criterion remains unchanged and in this situation, the user is considered as being not observed when calculating the number of migration departures during that time unit.}}
\end{figure}

\subsection{Identifying observation status for migration return}

\begin{figure}[H]
\renewcommand{\figurename}{Diagram}
\caption{Observation status for migration returns: case 1}
\label{fig:obsDiagram_return_left_1}
\centering
  \includegraphics[width=\textwidth]{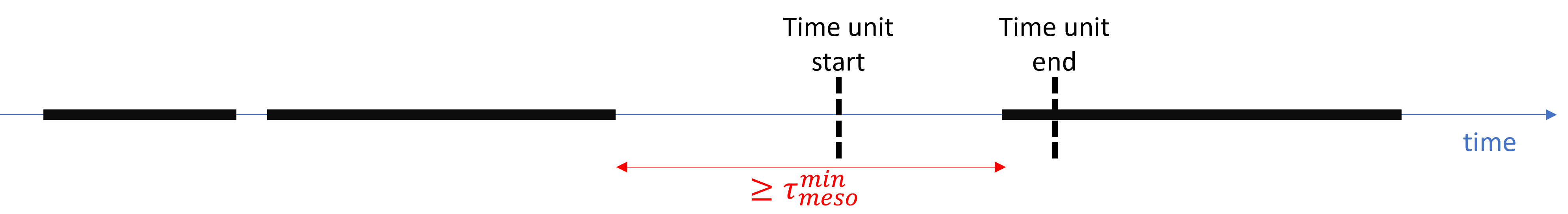}
\caption*{\small{\textit{Note}: An observation gap overlaps with the time unit on the left. Regardless of the locations of the preceding and following segments, if the gap is greater than $\tau^{temp}$ then a migration event could have ended during the time unit without being observed. The user is thus considered as being not observed when calculating the number of migration returns during that time unit.}}
\end{figure}

\begin{figure}[H]
\renewcommand{\figurename}{Diagram}
\caption{Observation status for migration returns: case 2}
\label{fig:obsDiagram_return_left_2}
\centering
  \includegraphics[width=\textwidth]{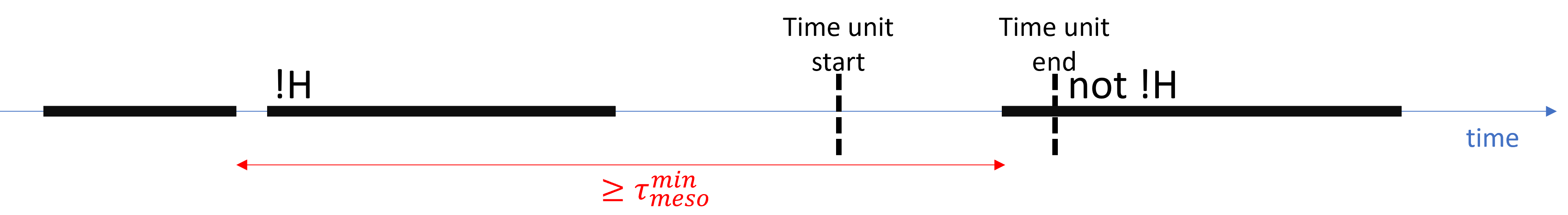}
\caption*{\small{\textit{Note}: An observation gap overlaps with the time unit on the left and is assumed to be strictly smaller than $\tau^{temp}$ -- we are back to case 1 of diagram \ref{fig:obsDiagram_return_left_1} otherwise. It is preceded by a non-home segment at location !H, and followed by a segment at a location that is not !H. The non-home segment may be a migration segment with an end date within the time unit if the time elapsed between the day following the preceding segment and the end of the observation gap is greater than $\tau^{temp}$. In that case, it is impossible to determine whether the user effectively returned from migration during the time unit or not. The user is thus considered as being not observed when calculating the number of migration returns during that time unit. Conversely, if this maximum duration was strictly less than $\tau^{temp}$, the observation gaps could not be concealing a migration event ending during the time unit and the user would be considered as effectively observed and not having returned from migration.}}
\end{figure}

\begin{figure}[H]
\renewcommand{\figurename}{Diagram}
\caption{Observation status for migration returns: case 3}
\label{fig:obsDiagram_return_left_3}
\centering
  \includegraphics[width=\textwidth]{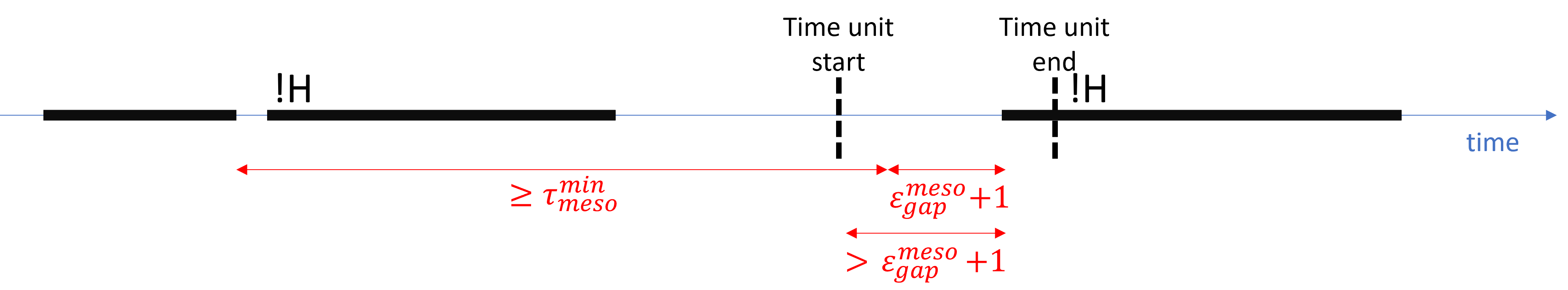}
\caption*{\small{\textit{Note}: This is the same configuration as in diagram \ref{fig:obsDiagram_return_left_2}, although the segment following the observation gap is at the same location !H as the segment preceding the gap. Two conditions allow for the possibility that the non-home segment preceding the gap reflects a migration event that ended during the time unit without being observed. First, the maximum end date has to fall within the time unit; the maximum end date is $\epsilon_{gap}^{meso}+2$ days before the following segment. Second, the maximum duration, defined as the time elapsed between the day following the preceding segment and the maximum end date, has to be greater than $\tau^{temp}$. When both conditions are met, the degree of uncertainty is such that the user is considered as being not observed when calculating the number of migration returns during that time unit.}}
\end{figure}

\begin{figure}[H]
\renewcommand{\figurename}{Diagram}
\caption{Observation status for migration returns: case 4}
\label{fig:obsDiagram_return_left_4}
\centering
  \includegraphics[width=\textwidth]{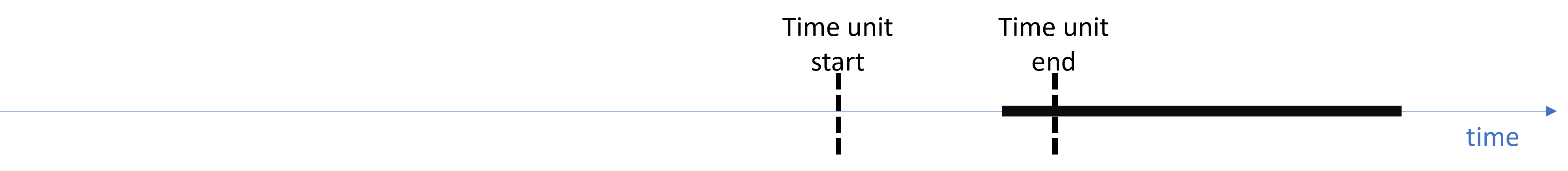}
\caption*{\small{\textit{Note}: An unbounded observation gap overlaps with the time unit on the left; the user enters the sample during the time unit. The possibility exists that the user returned from migration during the unobserved period on the left of the time unit. Without further information on the user's location before the first segment observed, the user is considered as being not observed when calculating the number of migration returns during that time unit.}}
\end{figure}

\begin{figure}[H]
\renewcommand{\figurename}{Diagram}
\caption{Observation status for migration returns: case 5}
\label{fig:obsDiagram_return_right_1}
\centering
  \includegraphics[width=\textwidth]{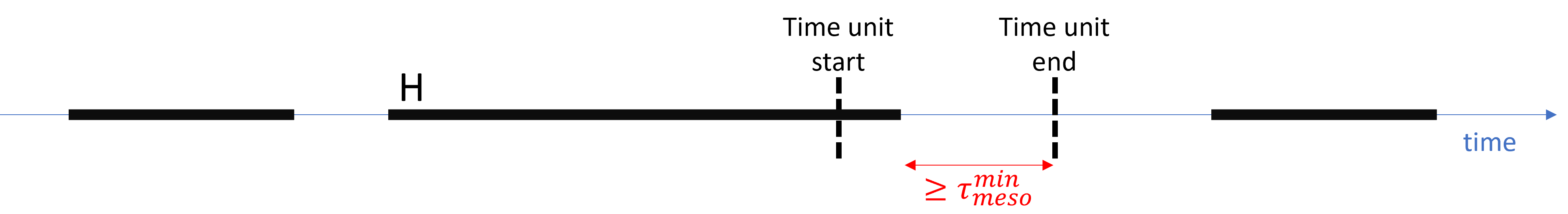}
\caption*{\small{\textit{Note}: An observation gap overlaps with the time unit on the right and is preceded by a home segment. If the gap between the observation gap start date and the time unit end date is larger than the parameter $\tau^{temp}$, a migration segment may have occurred and ended during this portion of the time unit, without being observed. The user is considered as being not observed when calculating the number of migration returns during that time unit.}}
\end{figure}

\begin{figure}[H]
\renewcommand{\figurename}{Diagram}
\caption{Observation status for migration returns: case 6}
\label{fig:obsDiagram_return_right_2}
\centering
  \includegraphics[width=\textwidth]{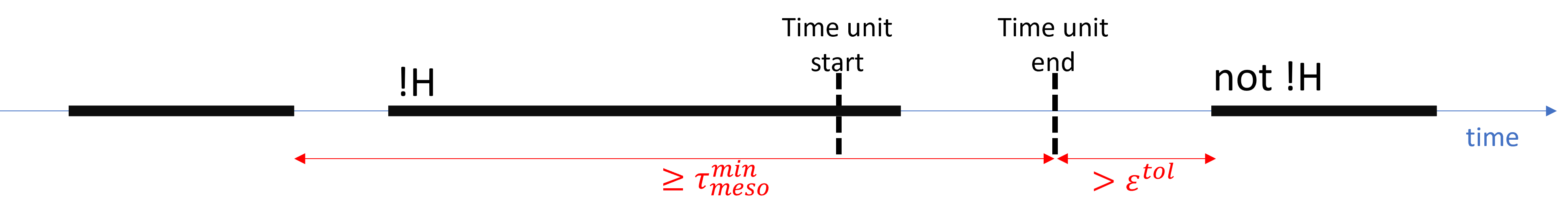}
\caption*{\small{\textit{Note}: An observation gap overlaps with the time unit on the right and the fraction of the gap ending at the end of the time unit is assumed strictly smaller than $\tau^{temp}$ -- we are back to case 5 of diagram \ref{fig:obsDiagram_return_right_1} otherwise. It is preceded by a non-home segment at location !H, and followed by a segment at a location that is not !H. The non-home segment may be a migration segment ending within the time unit if the time elapsed between the day following the preceding segment and the time unit end date is greater than $\tau^{temp}$. When the tolerance criterion is exceeded, i.e. the time elapsed between the day following the time unit end date and the day preceding the first day of the following segment exceeds the parameter $\epsilon^tol$, the uncertainty around the actual end date of the segment at the non-home location is considered as too large and the user is not counted as being observed for the calculation of migration returns during that time unit. Note that if the maximum duration considered is strictly less than $\tau^{temp}$, the non-home segment cannot be a migration segment and the user is then considered as observed and not having returned from migration during the time unit.}}
\end{figure}

\begin{figure}[H]
\renewcommand{\figurename}{Diagram}
\caption{Observation status for migration returns: case 7}
\label{fig:obsDiagram_return_right_3}
\centering
  \includegraphics[width=\textwidth]{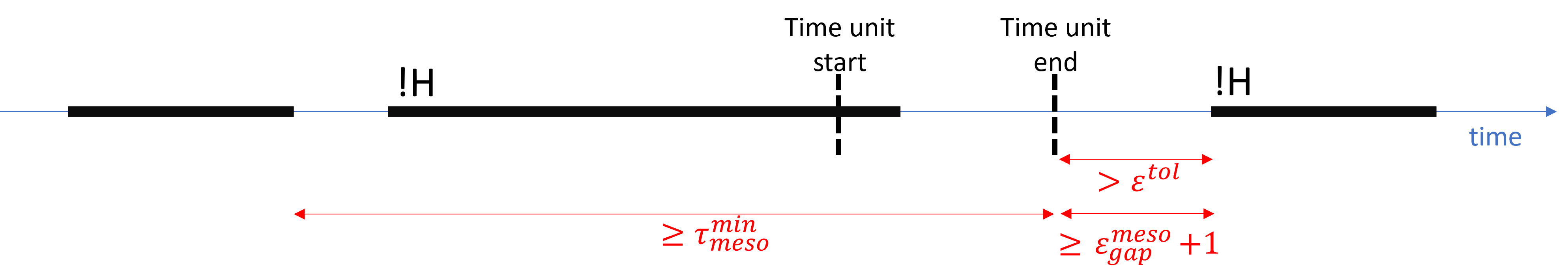}
\caption*{\small{\textit{Note}: This is the same configuration as in case 6 described in diagram \ref{fig:obsDiagram_return_right_2}, although the segment following the observation gap is at the same location !H as the segment preceding the gap. The non-home segment preceding the gap cannot have ended later than $\epsilon_{gap}^{meso}+2$ days before the following non-home segment started -- they would have been merged by the detection algorithm otherwise. This maximum end date falls either within or after the time unit. Case 7 deals with the latter configuration whereas case 8 in the following diagram (\ref{fig:obsDiagram_return_right_4}) considers the former. In this case, the maximum end date possible for the first non-home segment to consider it ended during the time unit coincides with the time unit end date. Two conditions allow for the possibility that the non-home segment preceding the gap reflects a migration event that ended during the time unit without being observed. First, the maximum duration, i.e. the time elapsed between the day following the last day of the previous segment and the maximum end date, is greater than $\tau^{temp}$. Second, the tolerance criterion is exceeded: the time elapsed between the maximum end date and the day preceding the first day of the following segment is strictly greater than $\epsilon^{tol}$. When both conditions are met, the degree of uncertainty is such that the user is considered as being not observed when calculating the number of migration returns during that time unit.}}
\end{figure}

\begin{figure}[H]
\renewcommand{\figurename}{Diagram}
\caption{Observation status for migration returns: case 8}
\label{fig:obsDiagram_return_right_4}
\centering
  \includegraphics[width=\textwidth]{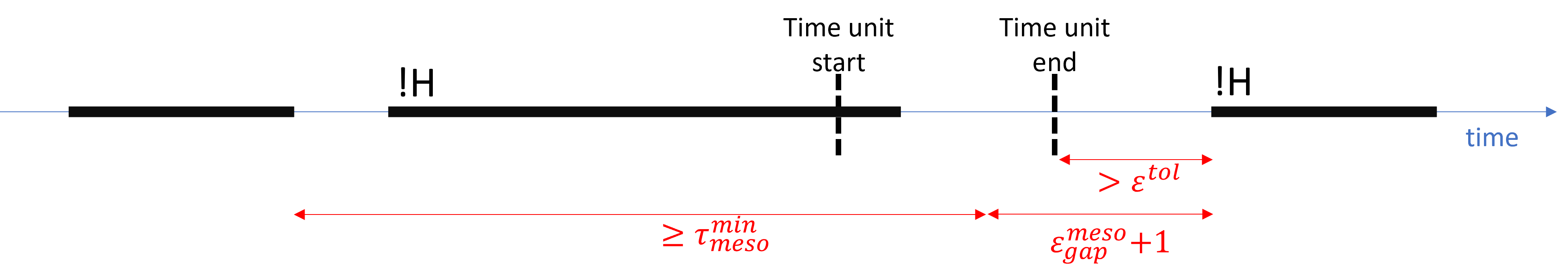}
\caption*{\small{\textit{Note}: This is the same configuration as in case 7 above, for the case where the maximum end date falls within the time unit. In this case the maximum duration is the time elapsed between the day following the last day of the previous segment and the maximum end date, which is strictly lower than the time unit end date. The criteria to define the user as not being observed are then equivalent.}}
\end{figure}

\begin{figure}[H]
\renewcommand{\figurename}{Diagram}
\caption{Observation status for migration returns: case 9}
\label{fig:obsDiagram_return_right_5}
\centering
  \includegraphics[width=\textwidth]{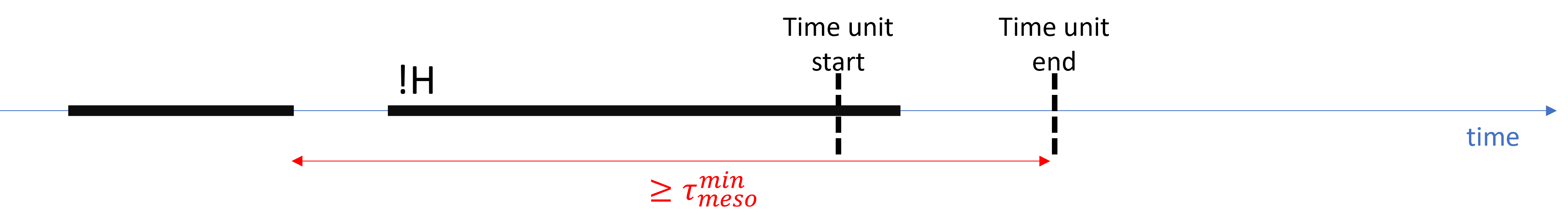}
\caption*{\small{\textit{Note}:  This configuration is the same as in diagrams \ref{fig:obsDiagram_return_right_2}-\ref{fig:obsDiagram_return_right_4}, but the user exits the sample after the non-home segment overlapping with the time unit. In the absence of further information about the user's location, it is assumed that the maximum end date of the non-home segment to consider it ended during the time unit coincides with the last day of the time unit. When the maximum duration, i.e the time elapsed between the day following the preceding segment and this maximum end date, is greater than $\tau^{temp}$, the possibility exists that the observation gaps conceal a migration event that ended during the time unit. Since the user is not observed after the non-home segment, the uncertainty around the actual end date is virtually infinite (i.e. the tolerance criterion is exceeded) and the user is not counted as being observed for the calculation of migration returns during that time unit.}}
\end{figure}

\begin{figure}[H]
\renewcommand{\figurename}{Diagram}
\caption{Observation status for migration returns: case 10}
\label{fig:obsDiagram_return_full_1}
\centering
  \includegraphics[width=\textwidth]{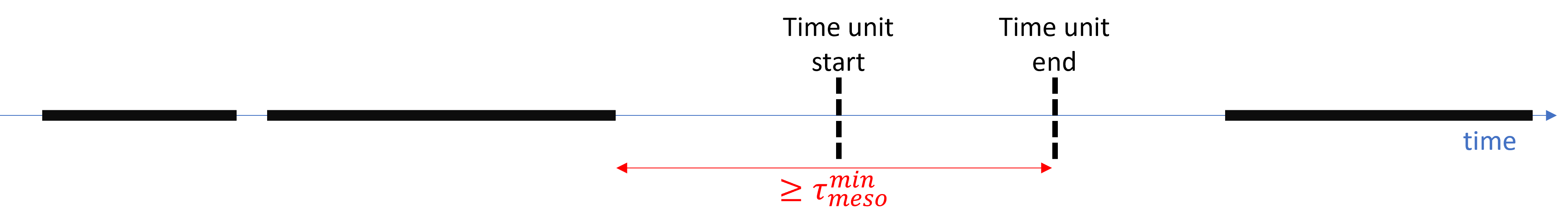}
\caption*{\small{\textit{Note}: An observation gap fully covers the time unit. Regardless of the locations of the preceding and following segments, if the time elapsed between the observation gap start date and the time unit end date is greater than $\tau^{temp}$, then a migration event could have occurred and ended during the time unit without being observed. The user is thus considered as being not observed when calculating the number of migration returns during that time unit.}}
\end{figure}

\begin{figure}[H]
\renewcommand{\figurename}{Diagram}
\caption{Observation status for migration returns: case 11}
\label{fig:obsDiagram_return_full_2}
\centering
  \includegraphics[width=\textwidth]{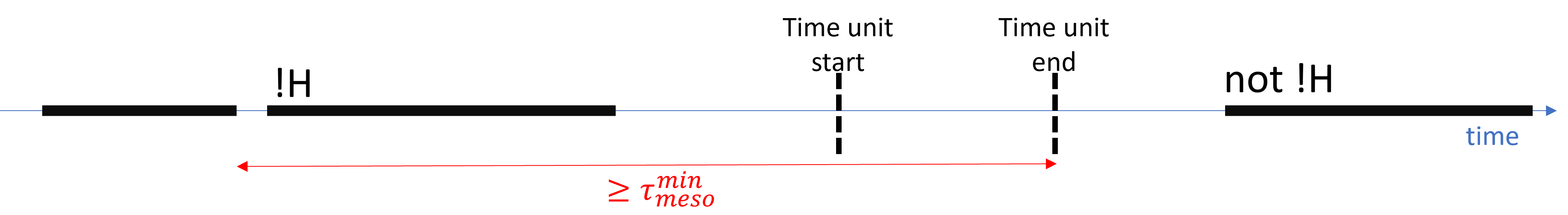}
\caption*{\small{\textit{Note}: An observation gap fully covers the time unit and the left-hand side portion of the gap that ends on the last day of the time unit is assumed strictly smaller than $\tau^{temp}$ -- we are back to case 10 of diagram \ref{fig:obsDiagram_return_full_1} otherwise. It is preceded by a non-home segment at location !H, and followed by a segment at a location that is not !H. The non-home segment may be a migration segment with an end date within the time unit if the time elapsed between the day following the preceding segment and the time unit end date is greater than $\tau^{temp}$. In that case, it is impossible to determine whether the user effectively returned from migration during the time unit or not. The user is thus considered as being not observed when calculating the number of migration returns during that time unit. Conversely, if this maximum duration was strictly less than $\tau^{temp}$, the observation gaps could not be concealing a migration event ending during the time unit and the user would be considered as effectively observed and not having returned from migration.}}
\end{figure}

\begin{figure}[H]
\renewcommand{\figurename}{Diagram}
\caption{Observation status for migration returns: case 12}
\label{fig:obsDiagram_return_full_3}
\centering
  \includegraphics[width=\textwidth]{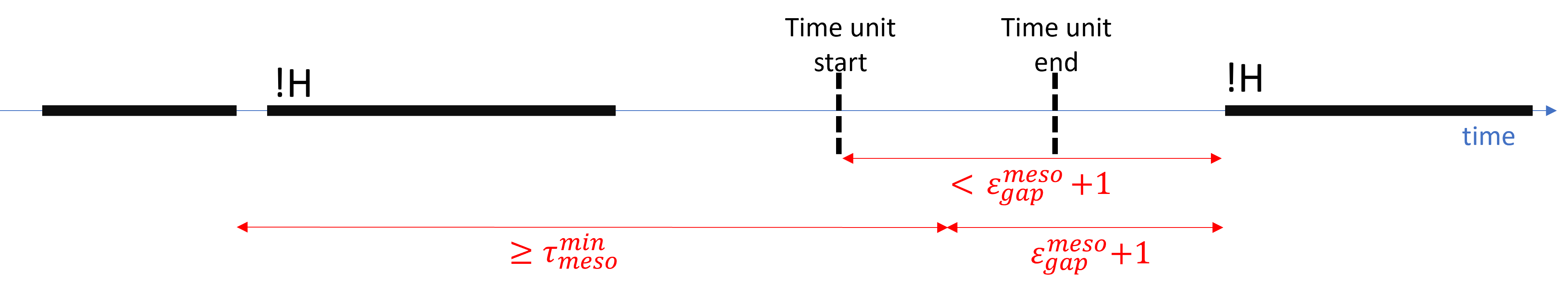}
\caption*{\small{\textit{Note}: This is the same configuration as in diagram \ref{fig:obsDiagram_return_full_2}, although the segment following the observation gap is at the same location !H as the segment preceding the gap. Two conditions allow for the possibility that the segment preceding the gap ended during the time unit without being observed. First, the maximum end date cannot fall before the start of the time unit; the maximum end date is $\epsilon_{gap}^{meso}+2$ days before the following segment. In other words, and as showed on the diagram, the gap between the time unit start date and the end of the observation gap has to be strictly lower than $\epsilon_{gap}^{meso}+1$ days. Then, the maximum end date falls either within or strictly after the time unit. The present case treats the former while case 13 below considers the latter. In this case, the maximum duration is defined as the time elapsed between the day following the preceding segment and the maximum end date. The second condition then implies that this duration has to be greater than $\tau^{temp}$. When both conditions are met, the degree of uncertainty is such that the user is considered as being not observed when calculating the number of migration returns during that time unit.}}
\end{figure}

\begin{figure}[H]
\renewcommand{\figurename}{Diagram}
\caption{Observation status for migration returns: case 13}
\label{fig:obsDiagram_return_full_4}
\centering
  \includegraphics[width=\textwidth]{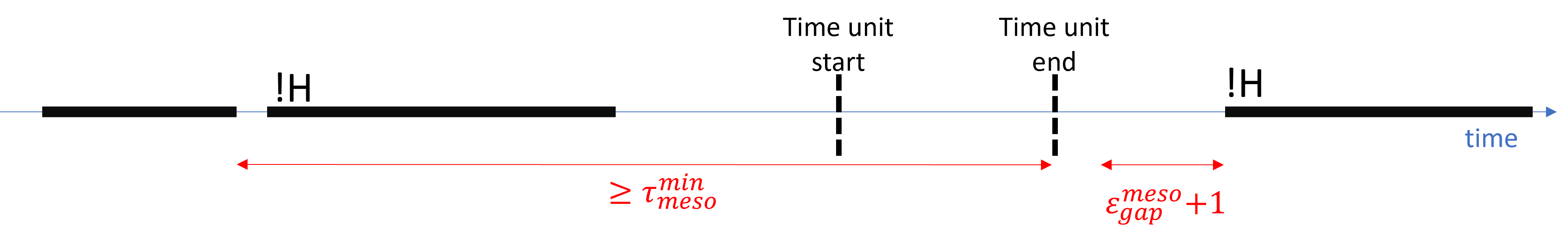}
\caption*{\small{\textit{Note}: This is the same configuration as in case 12 above, for the case where the maximum end date falls after the time unit. In this case the maximum duration considered is the time elapsed between the day following the last day of the previous segment and the last day of the time unit. The criteria to define the user as not being observed are then equivalent.}}
\end{figure}

\begin{figure}[H]
\renewcommand{\figurename}{Diagram}
\caption{Observation status for migration returns: case 14}
\label{fig:obsDiagram_return_full_5}
\centering
  \includegraphics[width=\textwidth]{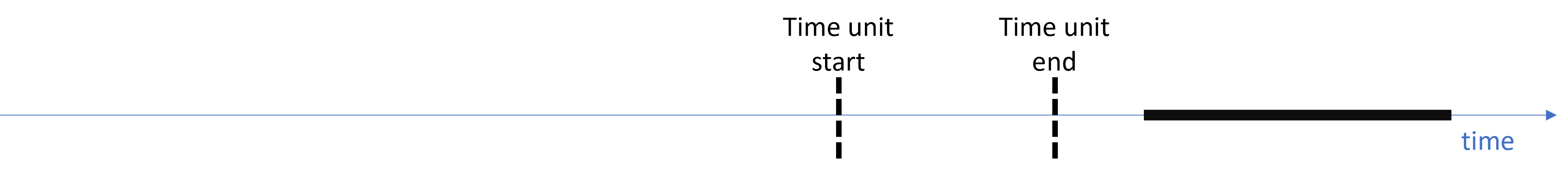}
\caption*{\small{\textit{Note}: An unbounded observation gap fully covers the time unit; the user is never observed before and during the time unit. This is the most straightforward case where a user is considered as being not observed for the calculation of migration returns during that time unit.}}
\end{figure}

\begin{figure}[H]
\renewcommand{\figurename}{Diagram}
\caption{Observation status for migration returns: case 15}
\label{fig:obsDiagram_return_full_6}
\centering
  \includegraphics[width=\textwidth]{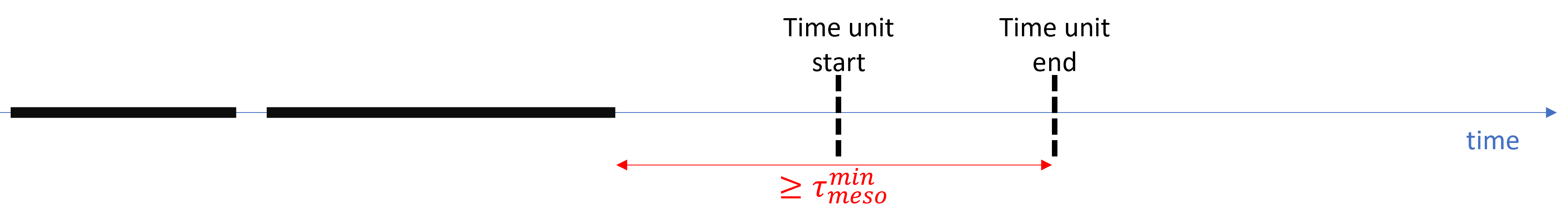}
\caption*{\small{\textit{Note}: The user exits the sample before the time unit start date. Regardless of the last segment location, when the time elapsed between the day following the last day observed and the end of time unit is greater than $\tau^{temp}$, the possibility exists that a migration event ending within the time unit occurred without being observed. The user is thus considered as being not observed for the calculation of migration returns during that time unit.}}
\end{figure}

\begin{figure}[H]
\renewcommand{\figurename}{Diagram}
\caption{Observation status for migration returns: case 16}
\label{fig:obsDiagram_return_full_7}
\centering
  \includegraphics[width=\textwidth]{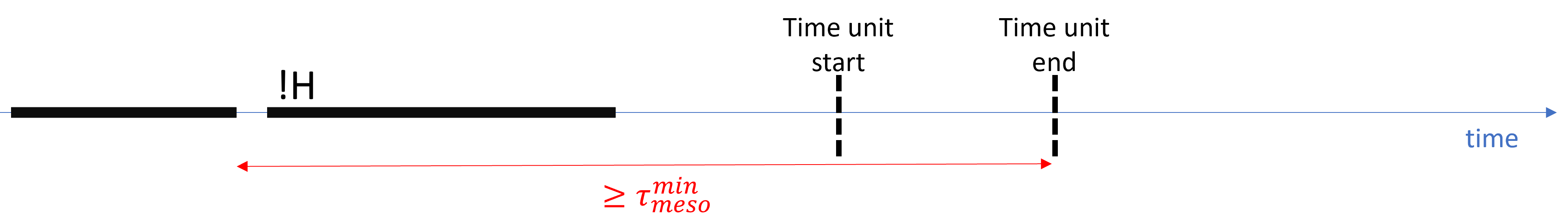}
\caption*{\small{\textit{Note}: This is the same configuration as in case 15, but it is implicitly assumed that the time elapsed between the day following the last day observed and the end of time unit is strictly less than $\tau^{temp}$ -- we are back to case 15 of diagram \ref{fig:obsDiagram_return_full_6} otherwise. The last segment is at a non-home location !H. It may have ended during the time unit if the time elapsed between the day following the last day of the preceding segment and the end of the time unit is greater than $\tau^{temp}$. Since the user is not observed after the non-home segment, the uncertainty around the actual end date is virtually infinite (i.e. the tolerance criterion is exceeded) and the user is not counted as being observed for the calculation of migration returns during that time unit.}}
\end{figure}

\subsection{Identifying observation status for migration status}

\begin{figure}[H]
\renewcommand{\figurename}{Diagram}
\caption{Observation status for migration status: case 1}
\label{fig:obsDiagram_status_left_1}
\centering
  \includegraphics[width=\textwidth]{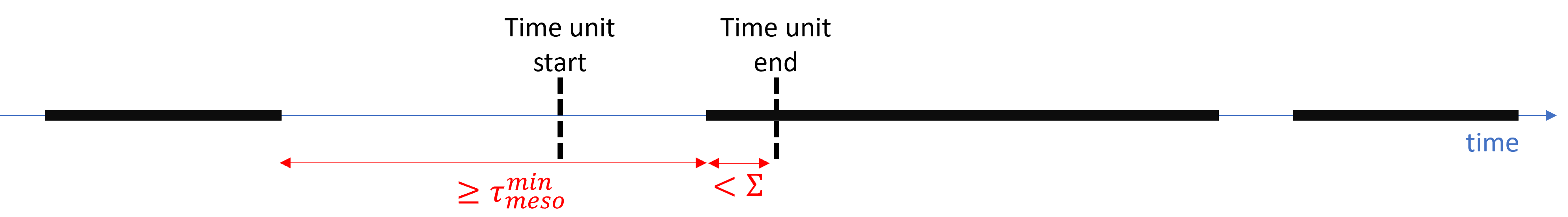}
\caption*{\small{\textit{Note}: An observation gap overlaps with the time unit on the left. The segment following the gap overlaps with the gap for a duration that is strictly lower than $\Sigma$ days: the certainty criterion is not satisfied with respect to the only segment overlapping with the time unit. If the observation gap is greater than $\tau^{temp}$, the possibility exists that a migration event overlapping with the time unit on at least $\Sigma$ days occurred during the observation gap. The user is thus considered as being not observed for the calculation of the migration stock during that time unit. Note that when the observation gap is strictly less than $\tau^{temp}$, the non-observation conditions depend on the characteristics of the preceding and following segments. The corresponding configurations are considered in the following diagrams.}}
\end{figure}

\begin{figure}[H]
\renewcommand{\figurename}{Diagram}
\caption{Observation status for migration status: case 2}
\label{fig:obsDiagram_status_left_2}
\centering
  \includegraphics[width=\textwidth]{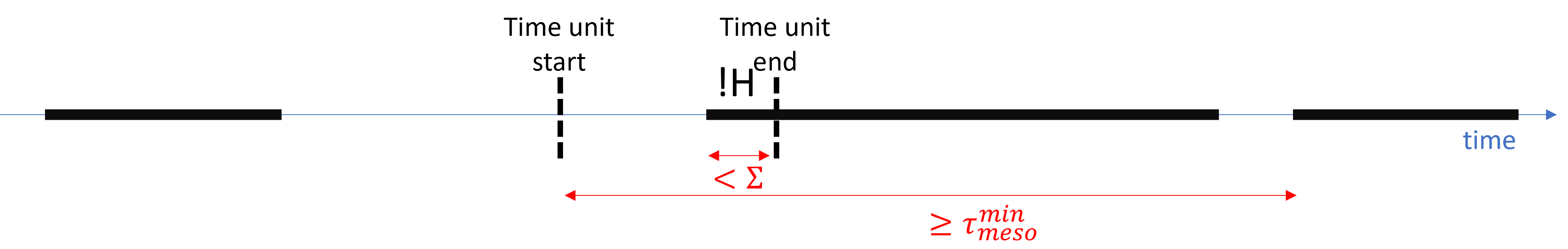}
\caption*{\small{\textit{Note}: This case is equivalent to case 1 described in diagram \ref{fig:obsDiagram_status_left_1} but implicitly assumes that the observation gap is strictly less than $\tau^{temp}$ -- otherwise we are back to case 1. The segment following the gap is at a non-home location !H and overlaps with the time unit for a duration strictly lower than $\Sigma$ days. If the time elapsed between the first day of the time unit and the day preceding the first day of the following segment is greater than $\tau^{temp}$, the possibility exists that the observed segment conceals a migration episode overlapping with the time unit on at least $\Sigma$ days. Given this uncertainty, the user is considered as being not observed for the calculation of the migration stock during that time unit. Note that, conversely, if the duration considered is strictly lower than $\tau^{temp}$, then we are certain that no migration event overlapping with the time unit occurred, and the user is considered as observed and not in migration.}}
\end{figure}

\begin{figure}[H]
\renewcommand{\figurename}{Diagram}
\caption{Observation status for migration status: case 3}
\label{fig:obsDiagram_status_left_3}
\centering
  \includegraphics[width=\textwidth]{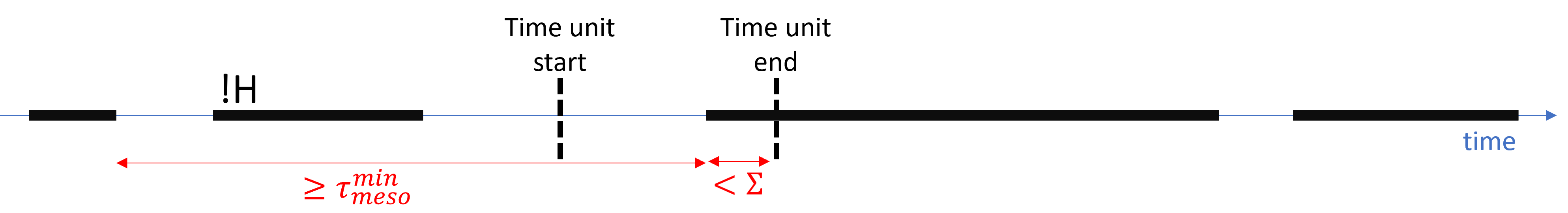}
\caption*{\small{\textit{Note}: This is the same configuration as in case 2 above, where the non-observation conditions with respect to the characteristics of the segment preceding the gap are considered. Similarly, the preceding segment is at a non-home location !H and the following segment still overlaps with the time unit for less than $\Sigma$ days. If the time elapsed between the day following the last day of the preceding segment and the observation gap end date is greater than $\tau_temp$, the possibility exists that the observed non-home segments conceals a migration episode overlapping with the time unit on at least $\Sigma$ days. Given this uncertainty, the user is considered as being not observed for the calculation of the migration stock during that time unit.}}
\end{figure}

\begin{figure}[H]
\renewcommand{\figurename}{Diagram}
\caption{Observation status for migration status: case 4}
\label{fig:obsDiagram_status_left_4}
\centering
  \includegraphics[width=\textwidth]{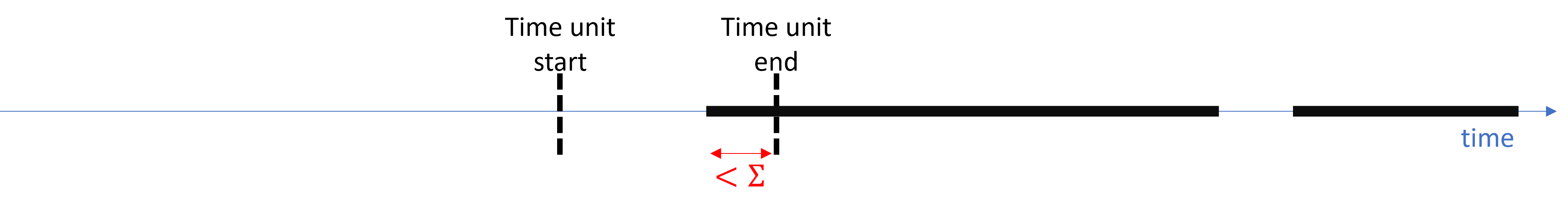}
\caption*{\small{\textit{Note}: The user enters the sample during the time unit. If the entry date is such that the first segment overlaps with the time unit on strictly less than $\Sigma$ days, the possibility exists that a migration event overlapping with at least the first $\Sigma$ days of the time unit occurred without being observed. The user is thus considered as being not observed for the calculation of the migration stock during that time unit.}}
\end{figure}

\begin{figure}[H]
\renewcommand{\figurename}{Diagram}
\caption{Observation status for migration status: case 5}
\label{fig:obsDiagram_status_right_1}
\centering
  \includegraphics[width=\textwidth]{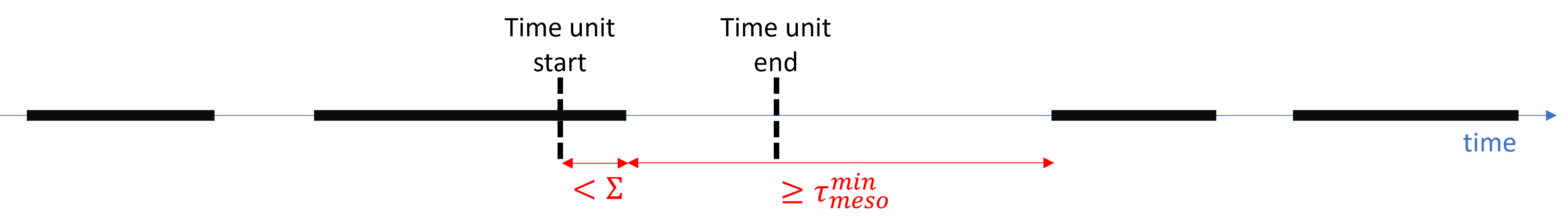}
\caption*{\small{\textit{Note}: An observation gap overlaps with the time unit on the right. The segment preceding the gap overlaps with the gap for a duration that is strictly lower than $\Sigma$ days: the certainty criterion is not satisfied with respect to the only segment overlapping with the time unit. If the observation gap is greater than $\tau^{temp}$, the possibility exists that a migration event overlapping with at least the last $\Sigma$ days of the time unit occurred during the observation gap. The user is thus considered as being not observed for the calculation of the migration stock during that time unit. Note that when the observation gap is strictly less than $\tau^{temp}$, the non-observation conditions depend on the characteristics of the preceding and following segments. The corresponding configurations are considered in the following diagrams.}}
\end{figure}

\begin{figure}[H]
\renewcommand{\figurename}{Diagram}
\caption{Observation status for migration status: case 6}
\label{fig:obsDiagram_status_right_2}
\centering
  \includegraphics[width=\textwidth]{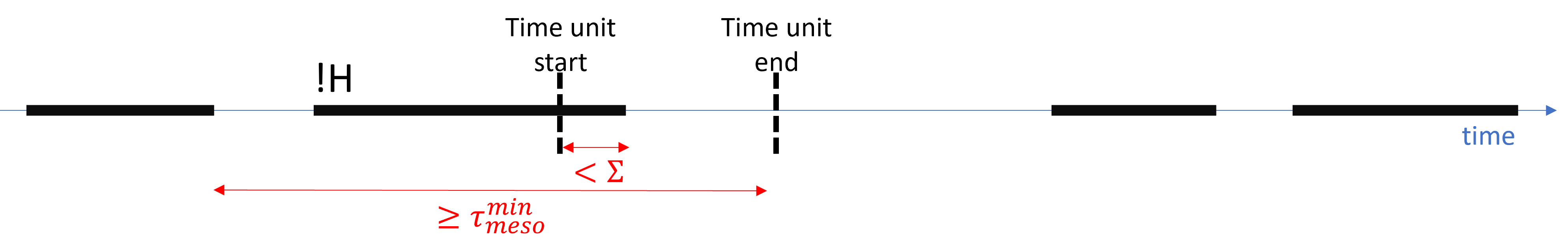}
\caption*{\small{\textit{Note}: This case is equivalent to case 5 described in diagram \ref{fig:obsDiagram_status_right_1} but implicitly assumes that the observation gap is strictly less than $\tau^{temp}$ -- otherwise we are back to case 5. The segment preceding the gap is at a non-home location !H and overlaps with the time unit for a duration strictly lower than $\Sigma$ days. If the time elapsed between the day following the last day of the preceding segment and the time unit end date is greater than $\tau^{temp}$, the possibility exists that the observed segment conceals a migration episode overlapping with the time unit on at least $\Sigma$ days. Given this uncertainty, the user is considered as being not observed for the calculation of the migration stock during that time unit. Note that, conversely, if the duration considered is strictly lower than $\tau^{temp}$, then we are certain that no migration event overlapping with the time unit occurred, and the user is considered as observed and not in migration.}}
\end{figure}

\begin{figure}[H]
\renewcommand{\figurename}{Diagram}
\caption{Observation status for migration status: case 7}
\label{fig:obsDiagram_status_right_3}
\centering
  \includegraphics[width=\textwidth]{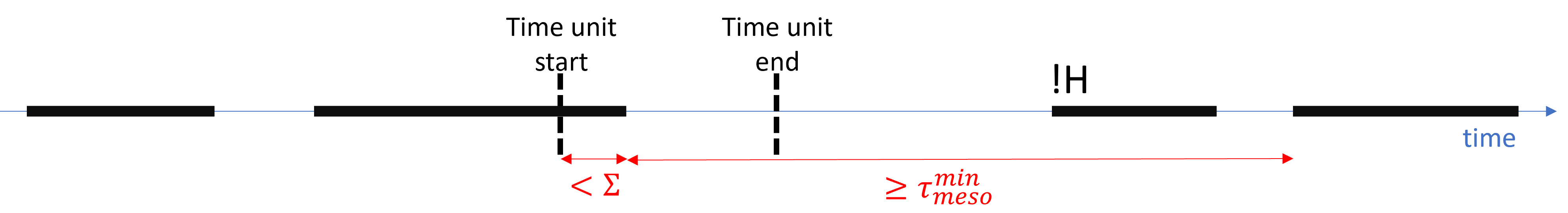}
\caption*{\small{\textit{Note}: This is the same configuration as in case 6 above, where the non-observation conditions with respect to the characteristics of the segment following the gap are considered. Similarly, the following segment is at a non-home location !H and the preceding segment still overlaps with the time unit for less than $\Sigma$ days. If the time elapsed between the observation gap start date and the day preceding the first day of the following segment is greater than $\tau_temp$, the possibility exists that the observed non-home segments conceals a migration episode overlapping with the time unit on at least $\Sigma$ days. Given this uncertainty, the user is considered as being not observed for the calculation of the migration stock during that time unit.}}
\end{figure}

\begin{figure}[H]
\renewcommand{\figurename}{Diagram}
\caption{Observation status for migration status: case 8}
\label{fig:obsDiagram_status_right_4}
\centering
  \includegraphics[width=\textwidth]{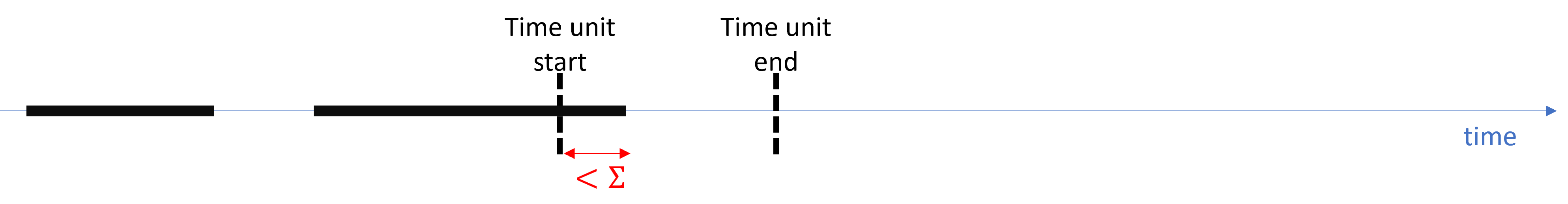}
\caption*{\small{\textit{Note}: The user exits the sample during the time unit. If the exit date is such that the last segment overlaps with the time unit on strictly less than $\Sigma$ days, the possibility exists that a migration event overlapping with at least the first $\Sigma$ days of the time unit occurred without being observed. The user is thus considered as being not observed for the calculation of the migration stock during that time unit.}}
\end{figure}

\begin{figure}[H]
\renewcommand{\figurename}{Diagram}
\caption{Observation status for migration status: case 9}
\label{fig:obsDiagram_status_full_1}
\centering
  \includegraphics[width=\textwidth]{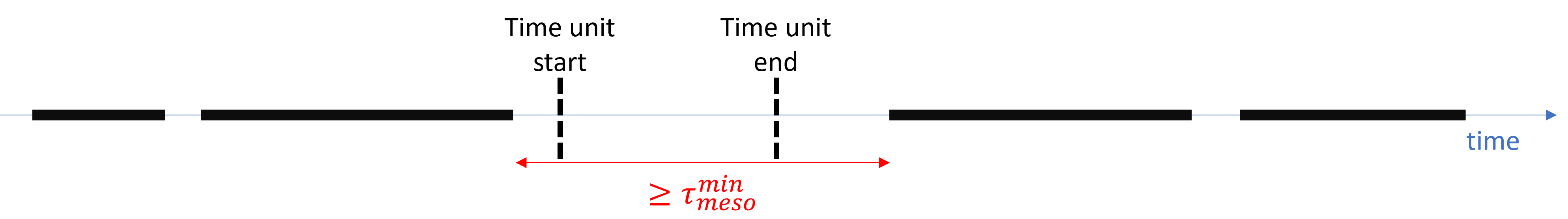}
\caption*{\small{\textit{Note}: An observation gap fully covers the time unit. When this gap is larger than $\tau^{temp}$ days, the possibility exists that a migration event overlapping with the time unit on at least $\Sigma$ days occurred during the observation gap. The user is thus considered as being not observed for the calculation of the migration stock during that time unit. Note that when the observation gap is strictly less than $\tau^{temp}$, the non-observation conditions depend on the characteristics of the preceding and following segments. The corresponding configurations are considered in the following diagrams.}}
\end{figure}

\begin{figure}[H]
\renewcommand{\figurename}{Diagram}
\caption{Observation status for migration status: case 10}
\label{fig:obsDiagram_status_full_2}
\centering
  \includegraphics[width=\textwidth]{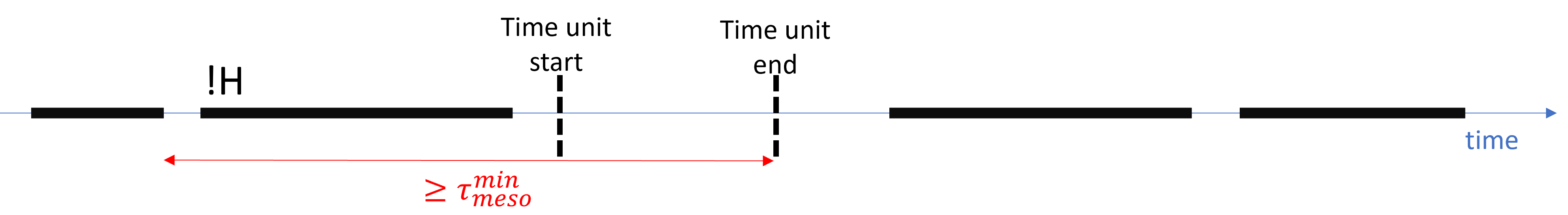}
\caption*{\small{\textit{Note}: This case is equivalent to case 9 described in diagram \ref{fig:obsDiagram_status_full_1} but implicitly assumes that the observation gap is strictly less than $\tau^{temp}$ -- otherwise we are back to case 9. The segment preceding the gap is at a non-home location !H -- and does not overlap with the time unit. If the time elapsed between the day following the last day of the preceding segment and the time unit end date is greater than $\tau^{temp}$, the possibility exists that the observed segment conceals a migration episode overlapping with the time unit on at least $\Sigma$ days. Given this uncertainty, the user is considered as being not observed for the calculation of the migration stock during that time unit. Note that, conversely, if the duration considered is strictly lower than $\tau^{temp}$, then we are certain that no migration event overlapping with the time unit occurred, and the user is considered as observed and not in migration.}}
\end{figure}

\begin{figure}[H]
\renewcommand{\figurename}{Diagram}
\caption{Observation status for migration status: case 11}
\label{fig:obsDiagram_status_full_3}
\centering
  \includegraphics[width=\textwidth]{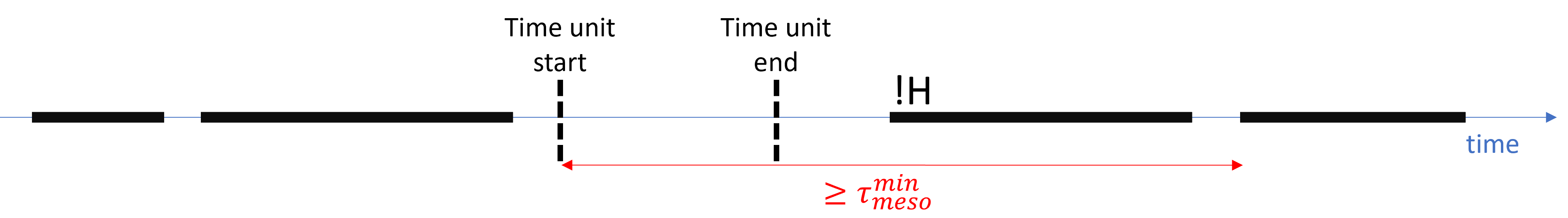}
\caption*{\small{\textit{Note}: This is the same configuration as in case 10 above, where the non-observation conditions with respect to the characteristics of the segment following the gap are considered. Similarly, the following segment is at a non-home location !H and the preceding segment does not overlap with the time unit. If the time elapsed between the time unit start date and the day preceding the first day of the following segment is greater than $\tau_temp$, the possibility exists that the observed non-home segments conceals a migration episode overlapping with the time unit on at least $\Sigma$ days. Given this uncertainty, the user is considered as being not observed for the calculation of the migration stock during that time unit.}}
\end{figure}

\begin{figure}[H]
\renewcommand{\figurename}{Diagram}
\caption{Observation status for migration status: case 12}
\label{fig:obsDiagram_status_full_4}
\centering
  \includegraphics[width=\textwidth]{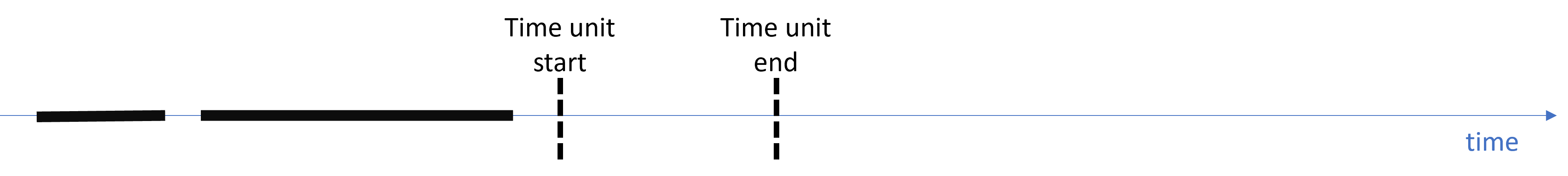}
\caption*{\small{\textit{Note}: The user exits the sample before the time unit. There is complete uncertainty about the user's migration status for that time unit and he is considered as no observed.}}
\end{figure}

\begin{figure}[H]
\renewcommand{\figurename}{Diagram}
\caption{Observation status for migration status: case 13}
\label{fig:obsDiagram_status_full_5}
\centering
  \includegraphics[width=\textwidth]{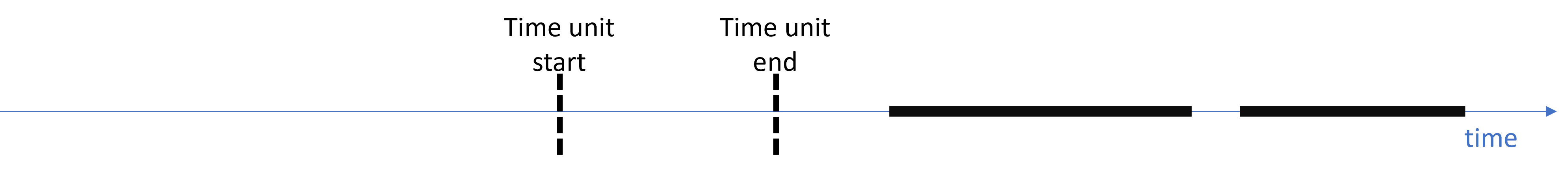}
\caption*{\small{\textit{Note}: The user enters the sample only after the time unit. There is complete uncertainty about the user's migration status for that time unit and he is considered as no observed.}}
\end{figure}

\begin{figure}[H]
\renewcommand{\figurename}{Diagram}
\caption{Observation status for migration status: case 14}
\label{fig:obsDiagram_status_within_1}
\centering
  \includegraphics[width=\textwidth]{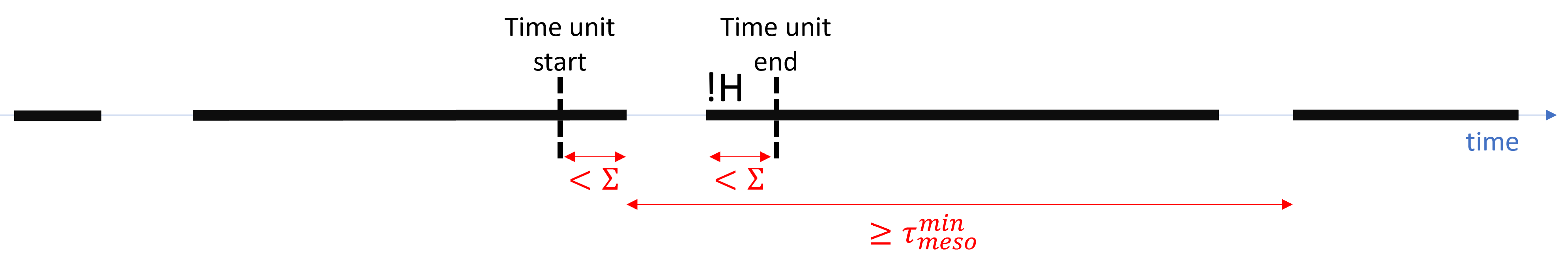}
\caption*{\small{\textit{Note}: The observation gap is strictly within the time unit. Both the preceding and following segments overlap with the time unit on strictly less than $\Sigma$ days. In this case, the following segment is at a non-home location !H. If the maximum duration of that segment is greater than $\tau^{temp}$, the possibility exists that it conceals a migration event overlapping with the time unit on at least $\Sigma$ days. Given this uncertainty, the user is considered as being not observed for the calculation of the migration stock during that time unit.}}
\end{figure}

\begin{figure}[H]
\renewcommand{\figurename}{Diagram}
\caption{Observation status for migration status: case 15}
\label{fig:obsDiagram_status_within_2}
\centering
  \includegraphics[width=\textwidth]{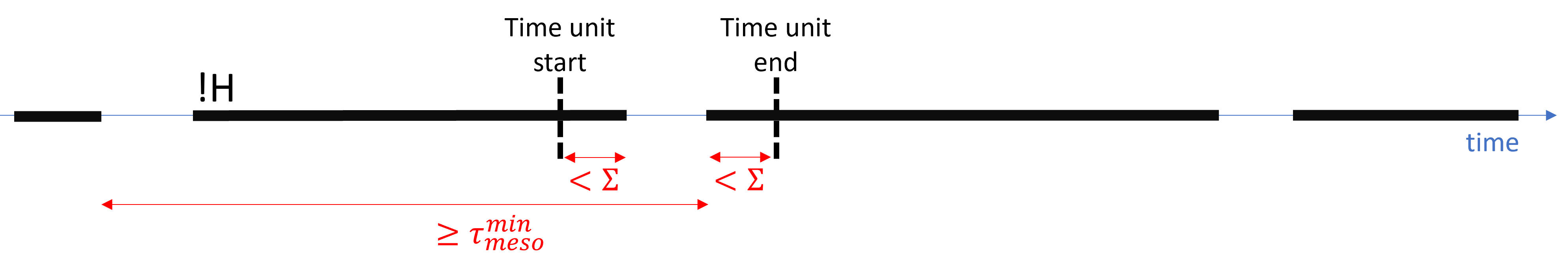}
\caption*{\small{\textit{Note}: The observation gap is strictly within the time unit. Both the preceding and following segments overlap with the time unit on strictly less than $\Sigma$ days. In this case, the preceding segment is at a non-home location !H. If the maximum duration of that segment is greater than $\tau^{temp}$, the possibility exists that it conceals a migration event overlapping with the time unit on at least $\Sigma$ days. Given this uncertainty, the user is considered as being not observed for the calculation of the migration stock during that time unit..}}
\end{figure}

\begin{figure}[H]
\renewcommand{\figurename}{Diagram}
\caption{Observation status for migration status: case 16}
\label{fig:obsDiagram_status_within_3}
\centering
  \includegraphics[width=\textwidth]{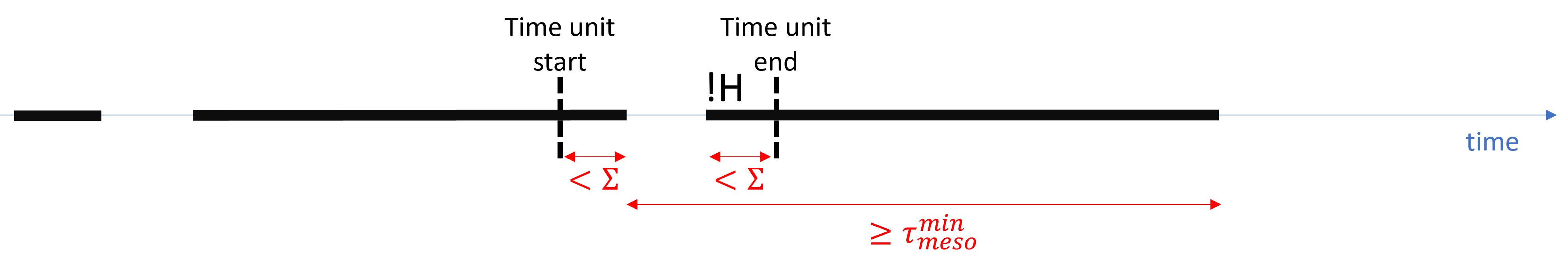}
\caption*{\small{\textit{Note}: The observation gap is strictly within the time unit. Both the preceding and following segments overlap with the time unit on strictly less than $\Sigma$ days. In this case, the following segment is at a non-home location !H and the user exits the sample at the end of this segment. Compared with case 14, the maximum end date is therefore considered to coincide with the segment end date and the maximum duration is shorter. Similarly, if it is greater than $\tau^{temp}$, the possibility exists that it conceals a migration event overlapping with the time unit on at least $\Sigma$ days. Given this uncertainty, the user is considered as being not observed for the calculation of the migration stock during that time unit.}}
\end{figure}

\begin{figure}[H]
\renewcommand{\figurename}{Diagram}
\caption{Observation status for migration status: case 17}
\label{fig:obsDiagram_status_within_4}
\centering
  \includegraphics[width=\textwidth]{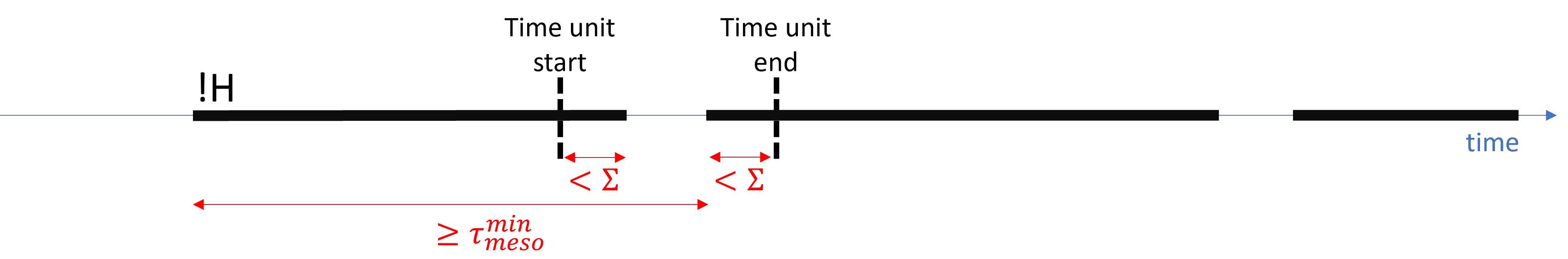}
\caption*{\small{\textit{Note}: The observation gap is strictly within the time unit. Both the preceding and following segments overlap with the time unit on strictly less than $\Sigma$ days. In this case, the preceding segment is at a non-home location !H and the user actually enters the sample on the start date of this segment. Compared with case 15, the minimum start date is therefore considered to coincide with the segment start date and the maximum duration is shorter. If the maximum duration of that segment is greater than $\tau^{temp}$, the possibility exists that it conceals a migration event overlapping with the time unit on at least $\Sigma$ days. Given this uncertainty, the user is considered as being not observed for the calculation of the migration stock during that time unit..}}
\end{figure}

\section{Statistical differences in individual characteristics between phone users and the population}

\begin{table}[H]
\centering
\setlength{\aboverulesep}{0pt}
\setlength{\belowrulesep}{0pt}
\resizebox{\textwidth}{!}{
\begin{tabular}{|l|>{\centering}m{0.7in}|>{\centering}m{0.7in}|>{\centering}m{0.7in}|>{\centering}m{0.7in}|>{\centering}m{0.7in}|>{\centering\arraybackslash}m{0.7in}|}
  \toprule 
  & \multicolumn{3}{c|}{\textbf{\underline{Urban}}} & \multicolumn{3}{c|}{\textbf{\underline{Rural}}} \\ 
 & \textbf{Phone users} & \textbf{All} & \textbf{Diff.} & \textbf{Phone users} & \textbf{All} & \textbf{Diff.} \\ 
   & (1) & (2) & (1)-(2) & (3) & (4) & (3)-(4) \\ 
 \midrule
wealth group: richest & 0.352 & 0.344 & 0.008$^{}$ & 0.866 & 0.862 & 0.004$^{}$ \\ 
  wealth group: richer & 0.282 & 0.274 & 0.008$^{}$ & 0.134 & 0.138 & -0.004$^{}$ \\ 
  wealth group: middle & 0.193 & 0.202 & -0.008$^{}$ & 0.000 & 0.000 & 0$^{NA}$ \\ 
  wealth group: poorer & 0.152 & 0.157 & -0.006$^{}$ & 0.000 & 0.000 & 0$^{NA}$ \\ 
  wealth group: poorest & 0.021 & 0.024 & -0.002$^{}$ & 0.000 & 0.000 & 0$^{NA}$ \\ 
  Years of education & 7.297 & 7.101 & 0.196$^{}$ & 6.293 & 6.328 & -0.035$^{}$ \\ 
  Age & 31.472 & 30.424 & 1.048$^{**}$ & 33.517 & 32.139 & 1.378$^{}$ \\ 
  Male & 0.520 & 0.506 & 0.014$^{}$ & 0.428 & 0.388 & 0.041$^{}$ \\ 
  Married & 0.486 & 0.454 & 0.031$^{}$ & 0.633 & 0.590 & 0.044$^{}$ \\ 
  Has a bank account & 0.273 & 0.248 & 0.025$^{}$ & 0.194 & 0.175 & 0.018$^{}$ \\ 
  occupation: not working & 0.270 & 0.296 & -0.026$^{}$ & 0.224 & 0.258 & -0.034$^{}$ \\ 
  occupation: agriculture & 0.009 & 0.009 & 0$^{}$ & 0.007 & 0.007 & 0.001$^{}$ \\ 
  occupation: sales & 0.207 & 0.203 & 0.004$^{}$ & 0.293 & 0.282 & 0.011$^{}$ \\ 
  occupation: household/domestic & 0.048 & 0.050 & -0.002$^{}$ & 0.028 & 0.034 & -0.006$^{}$ \\ 
  occupation: unskilled & 0.168 & 0.166 & 0.002$^{}$ & 0.132 & 0.120 & 0.013$^{}$ \\ 
  Household size & 11.177 & 11.365 & -0.189$^{}$ & 10.461 & 10.547 & -0.086$^{}$ \\ 
  Water access & 0.962 & 0.960 & 0.003$^{}$ & 0.606 & 0.577 & 0.03$^{}$ \\ 
  Access to sanitation & 0.989 & 0.989 & 0$^{}$ & 0.803 & 0.813 & -0.01$^{}$ \\ 
  Electricity & 0.982 & 0.981 & 0.001$^{}$ & 0.916 & 0.916 & 0$^{}$ \\ 
   \bottomrule
\end{tabular}}
\caption{Differences in characteristics between phone users and the population, Dakar. Statistics were derived from the Senegal 2017 DHS men and women individual datasets.} 
\label{}
\end{table}

\begin{table}[H]
\centering
\setlength{\aboverulesep}{0pt}
\setlength{\belowrulesep}{0pt}
\resizebox{\textwidth}{!}{
\begin{tabular}{|l|>{\centering}m{0.7in}|>{\centering}m{0.7in}|>{\centering}m{0.7in}|>{\centering}m{0.7in}|>{\centering}m{0.7in}|>{\centering\arraybackslash}m{0.7in}|}
  \toprule 
  & \multicolumn{3}{c|}{\textbf{\underline{Urban}}} & \multicolumn{3}{c|}{\textbf{\underline{Rural}}} \\ 
 & \textbf{Phone users} & \textbf{All} & \textbf{Diff.} & \textbf{Phone users} & \textbf{All} & \textbf{Diff.} \\ 
   & (1) & (2) & (1)-(2) & (3) & (4) & (3)-(4) \\ 
 \midrule
wealth group: richest & 0.088 & 0.074 & 0.014$^{}$ & 0.548 & 0.519 & 0.029$^{}$ \\ 
  wealth group: richer & 0.145 & 0.129 & 0.016$^{}$ & 0.140 & 0.139 & 0.002$^{}$ \\ 
  wealth group: middle & 0.132 & 0.132 & 0.001$^{}$ & 0.124 & 0.134 & -0.01$^{}$ \\ 
  wealth group: poorer & 0.391 & 0.399 & -0.008$^{}$ & 0.109 & 0.111 & -0.002$^{}$ \\ 
  wealth group: poorest & 0.244 & 0.266 & -0.022$^{}$ & 0.079 & 0.098 & -0.019$^{}$ \\ 
  Years of education & 5.427 & 5.097 & 0.33$^{}$ & 1.852 & 1.740 & 0.112$^{}$ \\ 
  Age & 30.304 & 28.550 & 1.755$^{}$ & 30.175 & 28.524 & 1.651$^{**}$ \\ 
  Male & 0.501 & 0.446 & 0.055$^{**}$ & 0.465 & 0.418 & 0.046$^{}$ \\ 
  Married & 0.544 & 0.511 & 0.032$^{}$ & 0.689 & 0.636 & 0.053$^{}$ \\ 
  Has a bank account & 0.199 & 0.159 & 0.04$^{}$ & 0.114 & 0.083 & 0.031$^{}$ \\ 
  occupation: not working & 0.310 & 0.364 & -0.054$^{}$ & 0.243 & 0.271 & -0.028$^{}$ \\ 
  occupation: agriculture & 0.033 & 0.027 & 0.006$^{}$ & 0.174 & 0.198 & -0.024$^{}$ \\ 
  occupation: sales & 0.205 & 0.197 & 0.008$^{}$ & 0.208 & 0.189 & 0.019$^{}$ \\ 
  occupation: household/domestic & 0.033 & 0.041 & -0.008$^{}$ & 0.034 & 0.044 & -0.009$^{}$ \\ 
  occupation: unskilled & 0.219 & 0.204 & 0.015$^{}$ & 0.211 & 0.195 & 0.016$^{}$ \\ 
  Household size & 14.394 & 14.557 & -0.163$^{}$ & 14.187 & 14.381 & -0.194$^{}$ \\ 
  Water access & 0.352 & 0.339 & 0.013$^{}$ & 0.528 & 0.532 & -0.004$^{}$ \\ 
  Access to sanitation & 0.754 & 0.750 & 0.004$^{}$ & 0.606 & 0.583 & 0.023$^{}$ \\ 
  Electricity & 0.916 & 0.896 & 0.02$^{}$ & 0.602 & 0.579 & 0.023$^{}$ \\ 
   \bottomrule
\end{tabular}}
\caption{Differences in characteristics between phone users and the population, Diourbel. Statistics were derived from the Senegal 2017 DHS men and women individual datasets.} 
\label{}
\end{table}

\begin{table}[H]
\centering
\setlength{\aboverulesep}{0pt}
\setlength{\belowrulesep}{0pt}
\resizebox{\textwidth}{!}{
\begin{tabular}{|l|>{\centering}m{0.7in}|>{\centering}m{0.7in}|>{\centering}m{0.7in}|>{\centering}m{0.7in}|>{\centering}m{0.7in}|>{\centering\arraybackslash}m{0.7in}|}
  \toprule 
  & \multicolumn{3}{c|}{\textbf{\underline{Urban}}} & \multicolumn{3}{c|}{\textbf{\underline{Rural}}} \\ 
 & \textbf{Phone users} & \textbf{All} & \textbf{Diff.} & \textbf{Phone users} & \textbf{All} & \textbf{Diff.} \\ 
   & (1) & (2) & (1)-(2) & (3) & (4) & (3)-(4) \\ 
 \midrule
wealth group: richest & 0.069 & 0.059 & 0.01$^{}$ & 0.202 & 0.175 & 0.027$^{}$ \\ 
  wealth group: richer & 0.214 & 0.177 & 0.036$^{}$ & 0.366 & 0.342 & 0.025$^{}$ \\ 
  wealth group: middle & 0.144 & 0.130 & 0.014$^{}$ & 0.203 & 0.210 & -0.007$^{}$ \\ 
  wealth group: poorer & 0.188 & 0.179 & 0.009$^{}$ & 0.149 & 0.172 & -0.023$^{}$ \\ 
  wealth group: poorest & 0.385 & 0.455 & -0.07$^{}$ & 0.080 & 0.102 & -0.022$^{}$ \\ 
  Years of education & 6.166 & 5.895 & 0.271$^{}$ & 4.448 & 4.080 & 0.368$^{}$ \\ 
  Age & 31.417 & 29.736 & 1.681$^{*}$ & 30.937 & 29.140 & 1.797$^{**}$ \\ 
  Male & 0.496 & 0.449 & 0.047$^{*}$ & 0.571 & 0.501 & 0.07$^{***}$ \\ 
  Married & 0.603 & 0.547 & 0.056$^{}$ & 0.596 & 0.552 & 0.044$^{}$ \\ 
  Has a bank account & 0.270 & 0.218 & 0.051$^{}$ & 0.121 & 0.084 & 0.037$^{}$ \\ 
  occupation: not working & 0.230 & 0.277 & -0.046$^{}$ & 0.202 & 0.250 & -0.048$^{*}$ \\ 
  occupation: agriculture & 0.110 & 0.113 & -0.003$^{}$ & 0.388 & 0.391 & -0.003$^{}$ \\ 
  occupation: sales & 0.212 & 0.203 & 0.009$^{}$ & 0.120 & 0.104 & 0.016$^{}$ \\ 
  occupation: household/domestic & 0.022 & 0.037 & -0.015$^{}$ & 0.034 & 0.040 & -0.006$^{}$ \\ 
  occupation: unskilled & 0.150 & 0.142 & 0.007$^{}$ & 0.064 & 0.066 & -0.003$^{}$ \\ 
  Household size & 12.426 & 12.451 & -0.026$^{}$ & 13.714 & 13.678 & 0.037$^{}$ \\ 
  Water access & 0.425 & 0.444 & -0.019$^{}$ & 0.383 & 0.359 & 0.024$^{}$ \\ 
  Access to sanitation & 0.588 & 0.546 & 0.041$^{}$ & 0.244 & 0.230 & 0.014$^{}$ \\ 
  Electricity & 0.930 & 0.908 & 0.023$^{}$ & 0.493 & 0.461 & 0.032$^{}$ \\ 
   \bottomrule
\end{tabular}}
\caption{Differences in characteristics between phone users and the population, Fatick. Statistics were derived from the Senegal 2017 DHS men and women individual datasets.} 
\label{}
\end{table}

\begin{table}[H]
\centering
\setlength{\aboverulesep}{0pt}
\setlength{\belowrulesep}{0pt}
\resizebox{\textwidth}{!}{
\begin{tabular}{|l|>{\centering}m{0.7in}|>{\centering}m{0.7in}|>{\centering}m{0.7in}|>{\centering}m{0.7in}|>{\centering}m{0.7in}|>{\centering\arraybackslash}m{0.7in}|}
  \toprule 
  & \multicolumn{3}{c|}{\textbf{\underline{Urban}}} & \multicolumn{3}{c|}{\textbf{\underline{Rural}}} \\ 
 & \textbf{Phone users} & \textbf{All} & \textbf{Diff.} & \textbf{Phone users} & \textbf{All} & \textbf{Diff.} \\ 
   & (1) & (2) & (1)-(2) & (3) & (4) & (3)-(4) \\ 
 \midrule
wealth group: richest & 0.015 & 0.010 & 0.005$^{}$ & 0.072 & 0.045 & 0.026$^{}$ \\ 
  wealth group: richer & 0.028 & 0.023 & 0.005$^{}$ & 0.113 & 0.105 & 0.008$^{}$ \\ 
  wealth group: middle & 0.065 & 0.056 & 0.008$^{}$ & 0.134 & 0.112 & 0.022$^{}$ \\ 
  wealth group: poorer & 0.131 & 0.113 & 0.017$^{}$ & 0.292 & 0.299 & -0.006$^{}$ \\ 
  wealth group: poorest & 0.762 & 0.797 & -0.035$^{}$ & 0.389 & 0.439 & -0.05$^{}$ \\ 
  Years of education & 3.453 & 3.083 & 0.371$^{}$ & 1.313 & 1.165 & 0.148$^{}$ \\ 
  Age & 29.723 & 28.088 & 1.634$^{}$ & 30.905 & 29.108 & 1.796$^{**}$ \\ 
  Male & 0.524 & 0.459 & 0.065$^{*}$ & 0.603 & 0.442 & 0.16$^{***}$ \\ 
  Married & 0.642 & 0.617 & 0.025$^{}$ & 0.775 & 0.748 & 0.027$^{}$ \\ 
  Has a bank account & 0.091 & 0.064 & 0.027$^{}$ & 0.056 & 0.032 & 0.024$^{**}$ \\ 
  occupation: not working & 0.327 & 0.348 & -0.022$^{}$ & 0.145 & 0.174 & -0.029$^{}$ \\ 
  occupation: agriculture & 0.119 & 0.173 & -0.055$^{}$ & 0.495 & 0.528 & -0.033$^{}$ \\ 
  occupation: sales & 0.213 & 0.173 & 0.04$^{}$ & 0.107 & 0.087 & 0.021$^{}$ \\ 
  occupation: household/domestic & 0.008 & 0.022 & -0.014$^{}$ & 0.013 & 0.010 & 0.003$^{}$ \\ 
  occupation: unskilled & 0.142 & 0.148 & -0.006$^{}$ & 0.136 & 0.130 & 0.006$^{}$ \\ 
  Household size & 13.436 & 14.029 & -0.593$^{}$ & 14.175 & 14.197 & -0.022$^{}$ \\ 
  Water access & 0.622 & 0.628 & -0.005$^{}$ & 0.563 & 0.509 & 0.054$^{}$ \\ 
  Access to sanitation & 0.226 & 0.192 & 0.034$^{}$ & 0.176 & 0.159 & 0.018$^{}$ \\ 
  Electricity & 0.539 & 0.485 & 0.054$^{}$ & 0.191 & 0.176 & 0.015$^{}$ \\ 
   \bottomrule
\end{tabular}}
\caption{Differences in characteristics between phone users and the population, Kaffrine. Statistics were derived from the Senegal 2017 DHS men and women individual datasets.} 
\label{}
\end{table}

\begin{table}[H]
\centering
\setlength{\aboverulesep}{0pt}
\setlength{\belowrulesep}{0pt}
\resizebox{\textwidth}{!}{
\begin{tabular}{|l|>{\centering}m{0.7in}|>{\centering}m{0.7in}|>{\centering}m{0.7in}|>{\centering}m{0.7in}|>{\centering}m{0.7in}|>{\centering\arraybackslash}m{0.7in}|}
  \toprule 
  & \multicolumn{3}{c|}{\textbf{\underline{Urban}}} & \multicolumn{3}{c|}{\textbf{\underline{Rural}}} \\ 
 & \textbf{Phone users} & \textbf{All} & \textbf{Diff.} & \textbf{Phone users} & \textbf{All} & \textbf{Diff.} \\ 
   & (1) & (2) & (1)-(2) & (3) & (4) & (3)-(4) \\ 
 \midrule
wealth group: richest & 0.095 & 0.085 & 0.01$^{}$ & 0.096 & 0.087 & 0.01$^{}$ \\ 
  wealth group: richer & 0.200 & 0.184 & 0.016$^{}$ & 0.263 & 0.223 & 0.04$^{}$ \\ 
  wealth group: middle & 0.189 & 0.192 & -0.003$^{}$ & 0.290 & 0.291 & -0.001$^{}$ \\ 
  wealth group: poorer & 0.255 & 0.262 & -0.007$^{}$ & 0.280 & 0.295 & -0.015$^{}$ \\ 
  wealth group: poorest & 0.261 & 0.277 & -0.016$^{}$ & 0.071 & 0.104 & -0.033$^{}$ \\ 
  Years of education & 5.439 & 5.272 & 0.167$^{}$ & 3.074 & 2.812 & 0.262$^{}$ \\ 
  Age & 30.371 & 28.446 & 1.925$^{**}$ & 31.408 & 29.337 & 2.071$^{*}$ \\ 
  Male & 0.468 & 0.451 & 0.017$^{}$ & 0.564 & 0.476 & 0.088$^{**}$ \\ 
  Married & 0.525 & 0.461 & 0.064$^{*}$ & 0.693 & 0.621 & 0.072$^{}$ \\ 
  Has a bank account & 0.226 & 0.179 & 0.047$^{}$ & 0.150 & 0.090 & 0.06$^{}$ \\ 
  occupation: not working & 0.302 & 0.343 & -0.042$^{}$ & 0.200 & 0.213 & -0.013$^{}$ \\ 
  occupation: agriculture & 0.037 & 0.036 & 0.001$^{}$ & 0.530 & 0.548 & -0.019$^{}$ \\ 
  occupation: sales & 0.253 & 0.237 & 0.016$^{}$ & 0.087 & 0.071 & 0.016$^{}$ \\ 
  occupation: household/domestic & 0.033 & 0.044 & -0.01$^{}$ & 0.013 & 0.012 & 0.001$^{}$ \\ 
  occupation: unskilled & 0.138 & 0.148 & -0.01$^{}$ & 0.099 & 0.097 & 0.002$^{}$ \\ 
  Household size & 10.462 & 10.474 & -0.012$^{}$ & 14.009 & 13.704 & 0.305$^{}$ \\ 
  Water access & 0.697 & 0.688 & 0.009$^{}$ & 0.518 & 0.489 & 0.03$^{}$ \\ 
  Access to sanitation & 0.845 & 0.825 & 0.02$^{}$ & 0.364 & 0.323 & 0.042$^{}$ \\ 
  Electricity & 0.844 & 0.836 & 0.008$^{}$ & 0.320 & 0.282 & 0.037$^{}$ \\ 
   \bottomrule
\end{tabular}}
\caption{Differences in characteristics between phone users and the population, Kaolack. Statistics were derived from the Senegal 2017 DHS men and women individual datasets.} 
\label{}
\end{table}

\begin{table}[H]
\centering
\setlength{\aboverulesep}{0pt}
\setlength{\belowrulesep}{0pt}
\resizebox{\textwidth}{!}{
\begin{tabular}{|l|>{\centering}m{0.7in}|>{\centering}m{0.7in}|>{\centering}m{0.7in}|>{\centering}m{0.7in}|>{\centering}m{0.7in}|>{\centering\arraybackslash}m{0.7in}|}
  \toprule 
  & \multicolumn{3}{c|}{\textbf{\underline{Urban}}} & \multicolumn{3}{c|}{\textbf{\underline{Rural}}} \\ 
 & \textbf{Phone users} & \textbf{All} & \textbf{Diff.} & \textbf{Phone users} & \textbf{All} & \textbf{Diff.} \\ 
   & (1) & (2) & (1)-(2) & (3) & (4) & (3)-(4) \\ 
 \midrule
wealth group: richest & 0.048 & 0.041 & 0.007$^{}$ & 0.022 & 0.016 & 0.006$^{}$ \\ 
  wealth group: richer & 0.015 & 0.013 & 0.002$^{}$ & 0.191 & 0.181 & 0.011$^{}$ \\ 
  wealth group: middle & 0.039 & 0.045 & -0.006$^{}$ & 0.253 & 0.217 & 0.037$^{}$ \\ 
  wealth group: poorer & 0.186 & 0.164 & 0.021$^{}$ & 0.172 & 0.205 & -0.033$^{}$ \\ 
  wealth group: poorest & 0.713 & 0.737 & -0.024$^{}$ & 0.361 & 0.381 & -0.02$^{}$ \\ 
  Years of education & 5.938 & 5.499 & 0.439$^{}$ & 3.640 & 3.091 & 0.549$^{}$ \\ 
  Age & 30.023 & 28.412 & 1.611$^{*}$ & 29.931 & 29.104 & 0.827$^{}$ \\ 
  Male & 0.488 & 0.477 & 0.011$^{}$ & 0.563 & 0.466 & 0.097$^{***}$ \\ 
  Married & 0.635 & 0.576 & 0.058$^{}$ & 0.715 & 0.706 & 0.009$^{}$ \\ 
  Has a bank account & 0.210 & 0.168 & 0.042$^{}$ & 0.088 & 0.055 & 0.033$^{}$ \\ 
  occupation: not working & 0.308 & 0.318 & -0.011$^{}$ & 0.172 & 0.188 & -0.016$^{}$ \\ 
  occupation: agriculture & 0.146 & 0.155 & -0.009$^{}$ & 0.429 & 0.457 & -0.028$^{}$ \\ 
  occupation: sales & 0.162 & 0.158 & 0.004$^{}$ & 0.038 & 0.028 & 0.01$^{}$ \\ 
  occupation: household/domestic & 0.016 & 0.015 & 0.001$^{}$ & 0.000 & 0.003 & -0.003$^{}$ \\ 
  occupation: unskilled & 0.091 & 0.110 & -0.019$^{}$ & 0.213 & 0.183 & 0.03$^{}$ \\ 
  Household size & 12.865 & 12.781 & 0.085$^{}$ & 13.778 & 14.028 & -0.249$^{}$ \\ 
  Water access & 0.498 & 0.479 & 0.019$^{}$ & 0.009 & 0.007 & 0.002$^{}$ \\ 
  Access to sanitation & 0.299 & 0.290 & 0.009$^{}$ & 0.005 & 0.003 & 0.002$^{}$ \\ 
  Electricity & 0.752 & 0.732 & 0.02$^{}$ & 0.245 & 0.231 & 0.015$^{}$ \\ 
   \bottomrule
\end{tabular}}
\caption{Differences in characteristics between phone users and the population, Kedougou. Statistics were derived from the Senegal 2017 DHS men and women individual datasets.} 
\label{}
\end{table}

\begin{table}[H]
\centering
\setlength{\aboverulesep}{0pt}
\setlength{\belowrulesep}{0pt}
\resizebox{\textwidth}{!}{
\begin{tabular}{|l|>{\centering}m{0.7in}|>{\centering}m{0.7in}|>{\centering}m{0.7in}|>{\centering}m{0.7in}|>{\centering}m{0.7in}|>{\centering\arraybackslash}m{0.7in}|}
  \toprule 
  & \multicolumn{3}{c|}{\textbf{\underline{Urban}}} & \multicolumn{3}{c|}{\textbf{\underline{Rural}}} \\ 
 & \textbf{Phone users} & \textbf{All} & \textbf{Diff.} & \textbf{Phone users} & \textbf{All} & \textbf{Diff.} \\ 
   & (1) & (2) & (1)-(2) & (3) & (4) & (3)-(4) \\ 
 \midrule
wealth group: richest & 0.029 & 0.026 & 0.003$^{}$ & 0.015 & 0.017 & -0.002$^{}$ \\ 
  wealth group: richer & 0.072 & 0.057 & 0.014$^{}$ & 0.036 & 0.027 & 0.009$^{}$ \\ 
  wealth group: middle & 0.097 & 0.082 & 0.015$^{}$ & 0.064 & 0.070 & -0.005$^{}$ \\ 
  wealth group: poorer & 0.215 & 0.179 & 0.037$^{}$ & 0.231 & 0.211 & 0.02$^{}$ \\ 
  wealth group: poorest & 0.587 & 0.656 & -0.069$^{}$ & 0.654 & 0.676 & -0.021$^{}$ \\ 
  Years of education & 6.141 & 5.575 & 0.566$^{}$ & 2.181 & 1.971 & 0.209$^{}$ \\ 
  Age & 31.318 & 29.182 & 2.136$^{**}$ & 30.964 & 29.314 & 1.65$^{**}$ \\ 
  Male & 0.540 & 0.495 & 0.045$^{}$ & 0.659 & 0.478 & 0.181$^{***}$ \\ 
  Married & 0.632 & 0.581 & 0.051$^{}$ & 0.737 & 0.724 & 0.013$^{}$ \\ 
  Has a bank account & 0.244 & 0.183 & 0.061$^{}$ & 0.070 & 0.037 & 0.033$^{**}$ \\ 
  occupation: not working & 0.335 & 0.399 & -0.063$^{**}$ & 0.144 & 0.169 & -0.025$^{}$ \\ 
  occupation: agriculture & 0.056 & 0.057 & -0.001$^{}$ & 0.598 & 0.660 & -0.062$^{}$ \\ 
  occupation: sales & 0.205 & 0.183 & 0.023$^{}$ & 0.119 & 0.078 & 0.041$^{}$ \\ 
  occupation: household/domestic & 0.017 & 0.025 & -0.008$^{}$ & 0.004 & 0.004 & 0.001$^{}$ \\ 
  occupation: unskilled & 0.145 & 0.141 & 0.004$^{}$ & 0.045 & 0.036 & 0.009$^{}$ \\ 
  Household size & 11.808 & 11.895 & -0.087$^{}$ & 12.792 & 13.007 & -0.214$^{}$ \\ 
  Water access & 0.329 & 0.304 & 0.025$^{}$ & 0.073 & 0.063 & 0.01$^{}$ \\ 
  Access to sanitation & 0.293 & 0.259 & 0.034$^{}$ & 0.004 & 0.002 & 0.002$^{}$ \\ 
  Electricity & 0.716 & 0.665 & 0.051$^{}$ & 0.081 & 0.066 & 0.015$^{}$ \\ 
   \bottomrule
\end{tabular}}
\caption{Differences in characteristics between phone users and the population, Kolda. Statistics were derived from the Senegal 2017 DHS men and women individual datasets.} 
\label{}
\end{table}

\begin{table}[H]
\centering
\setlength{\aboverulesep}{0pt}
\setlength{\belowrulesep}{0pt}
\resizebox{\textwidth}{!}{
\begin{tabular}{|l|>{\centering}m{0.7in}|>{\centering}m{0.7in}|>{\centering}m{0.7in}|>{\centering}m{0.7in}|>{\centering}m{0.7in}|>{\centering\arraybackslash}m{0.7in}|}
  \toprule 
  & \multicolumn{3}{c|}{\textbf{\underline{Urban}}} & \multicolumn{3}{c|}{\textbf{\underline{Rural}}} \\ 
 & \textbf{Phone users} & \textbf{All} & \textbf{Diff.} & \textbf{Phone users} & \textbf{All} & \textbf{Diff.} \\ 
   & (1) & (2) & (1)-(2) & (3) & (4) & (3)-(4) \\ 
 \midrule
wealth group: richest & 0.112 & 0.092 & 0.02$^{}$ & 0.210 & 0.186 & 0.025$^{}$ \\ 
  wealth group: richer & 0.123 & 0.113 & 0.01$^{}$ & 0.227 & 0.220 & 0.006$^{}$ \\ 
  wealth group: middle & 0.208 & 0.200 & 0.008$^{}$ & 0.280 & 0.275 & 0.005$^{}$ \\ 
  wealth group: poorer & 0.242 & 0.252 & -0.01$^{}$ & 0.135 & 0.157 & -0.021$^{}$ \\ 
  wealth group: poorest & 0.314 & 0.342 & -0.028$^{}$ & 0.147 & 0.162 & -0.014$^{}$ \\ 
  Years of education & 5.157 & 4.954 & 0.203$^{}$ & 2.198 & 2.019 & 0.179$^{}$ \\ 
  Age & 29.232 & 27.858 & 1.373$^{**}$ & 30.143 & 28.344 & 1.799$^{***}$ \\ 
  Male & 0.516 & 0.481 & 0.035$^{}$ & 0.482 & 0.413 & 0.068$^{*}$ \\ 
  Married & 0.513 & 0.474 & 0.039$^{}$ & 0.712 & 0.659 & 0.053$^{}$ \\ 
  Has a bank account & 0.160 & 0.126 & 0.034$^{*}$ & 0.052 & 0.037 & 0.015$^{}$ \\ 
  occupation: not working & 0.272 & 0.320 & -0.049$^{}$ & 0.250 & 0.290 & -0.04$^{}$ \\ 
  occupation: agriculture & 0.067 & 0.074 & -0.007$^{}$ & 0.387 & 0.388 & -0.001$^{}$ \\ 
  occupation: sales & 0.213 & 0.193 & 0.02$^{}$ & 0.167 & 0.147 & 0.019$^{}$ \\ 
  occupation: household/domestic & 0.042 & 0.047 & -0.005$^{}$ & 0.014 & 0.013 & 0$^{}$ \\ 
  occupation: unskilled & 0.162 & 0.157 & 0.004$^{}$ & 0.093 & 0.093 & 0$^{}$ \\ 
  Household size & 13.658 & 13.974 & -0.316$^{}$ & 11.747 & 11.799 & -0.052$^{}$ \\ 
  Water access & 0.942 & 0.944 & -0.002$^{}$ & 0.682 & 0.657 & 0.025$^{}$ \\ 
  Access to sanitation & 0.946 & 0.947 & -0.001$^{}$ & 0.556 & 0.532 & 0.024$^{}$ \\ 
  Electricity & 0.894 & 0.894 & 0$^{}$ & 0.432 & 0.406 & 0.026$^{}$ \\ 
   \bottomrule
\end{tabular}}
\caption{Differences in characteristics between phone users and the population, Louga. Statistics were derived from the Senegal 2017 DHS men and women individual datasets.} 
\label{}
\end{table}

\begin{table}[H]
\centering
\setlength{\aboverulesep}{0pt}
\setlength{\belowrulesep}{0pt}
\resizebox{\textwidth}{!}{
\begin{tabular}{|l|>{\centering}m{0.7in}|>{\centering}m{0.7in}|>{\centering}m{0.7in}|>{\centering}m{0.7in}|>{\centering}m{0.7in}|>{\centering\arraybackslash}m{0.7in}|}
  \toprule 
  & \multicolumn{3}{c|}{\textbf{\underline{Urban}}} & \multicolumn{3}{c|}{\textbf{\underline{Rural}}} \\ 
 & \textbf{Phone users} & \textbf{All} & \textbf{Diff.} & \textbf{Phone users} & \textbf{All} & \textbf{Diff.} \\ 
   & (1) & (2) & (1)-(2) & (3) & (4) & (3)-(4) \\ 
 \midrule
wealth group: richest & 0.063 & 0.051 & 0.012$^{}$ & 0.308 & 0.241 & 0.066$^{}$ \\ 
  wealth group: richer & 0.014 & 0.010 & 0.003$^{}$ & 0.278 & 0.281 & -0.003$^{}$ \\ 
  wealth group: middle & 0.058 & 0.051 & 0.008$^{}$ & 0.171 & 0.168 & 0.003$^{}$ \\ 
  wealth group: poorer & 0.179 & 0.172 & 0.007$^{}$ & 0.162 & 0.195 & -0.033$^{}$ \\ 
  wealth group: poorest & 0.686 & 0.717 & -0.031$^{}$ & 0.082 & 0.115 & -0.033$^{}$ \\ 
  Years of education & 4.157 & 3.824 & 0.333$^{}$ & 2.790 & 2.366 & 0.424$^{}$ \\ 
  Age & 30.932 & 29.384 & 1.548$^{}$ & 31.133 & 29.171 & 1.962$^{***}$ \\ 
  Male & 0.481 & 0.442 & 0.039$^{}$ & 0.530 & 0.458 & 0.072$^{**}$ \\ 
  Married & 0.590 & 0.539 & 0.051$^{}$ & 0.656 & 0.605 & 0.051$^{}$ \\ 
  Has a bank account & 0.152 & 0.117 & 0.035$^{}$ & 0.071 & 0.047 & 0.024$^{}$ \\ 
  occupation: not working & 0.364 & 0.393 & -0.028$^{}$ & 0.343 & 0.405 & -0.062$^{*}$ \\ 
  occupation: agriculture & 0.103 & 0.106 & -0.003$^{}$ & 0.226 & 0.236 & -0.01$^{}$ \\ 
  occupation: sales & 0.189 & 0.174 & 0.014$^{}$ & 0.171 & 0.141 & 0.03$^{}$ \\ 
  occupation: household/domestic & 0.007 & 0.020 & -0.014$^{*}$ & 0.004 & 0.004 & 0$^{}$ \\ 
  occupation: unskilled & 0.123 & 0.122 & 0.001$^{}$ & 0.098 & 0.090 & 0.008$^{}$ \\ 
  Household size & 13.208 & 13.950 & -0.742$^{}$ & 15.376 & 15.615 & -0.239$^{}$ \\ 
  Water access & 0.871 & 0.832 & 0.039$^{}$ & 0.739 & 0.641 & 0.099$^{}$ \\ 
  Access to sanitation & 0.495 & 0.445 & 0.05$^{}$ & 0.359 & 0.308 & 0.051$^{}$ \\ 
  Electricity & 0.737 & 0.707 & 0.03$^{}$ & 0.563 & 0.500 & 0.063$^{}$ \\ 
   \bottomrule
\end{tabular}}
\caption{Differences in characteristics between phone users and the population, Matam. Statistics were derived from the Senegal 2017 DHS men and women individual datasets.} 
\label{}
\end{table}

\begin{table}[H]
\centering
\setlength{\aboverulesep}{0pt}
\setlength{\belowrulesep}{0pt}
\resizebox{\textwidth}{!}{
\begin{tabular}{|l|>{\centering}m{0.7in}|>{\centering}m{0.7in}|>{\centering}m{0.7in}|>{\centering}m{0.7in}|>{\centering}m{0.7in}|>{\centering\arraybackslash}m{0.7in}|}
  \toprule 
  & \multicolumn{3}{c|}{\textbf{\underline{Urban}}} & \multicolumn{3}{c|}{\textbf{\underline{Rural}}} \\ 
 & \textbf{Phone users} & \textbf{All} & \textbf{Diff.} & \textbf{Phone users} & \textbf{All} & \textbf{Diff.} \\ 
   & (1) & (2) & (1)-(2) & (3) & (4) & (3)-(4) \\ 
 \midrule
wealth group: richest & 0.088 & 0.075 & 0.013$^{}$ & 0.219 & 0.178 & 0.041$^{}$ \\ 
  wealth group: richer & 0.139 & 0.124 & 0.015$^{}$ & 0.171 & 0.153 & 0.018$^{}$ \\ 
  wealth group: middle & 0.253 & 0.244 & 0.009$^{}$ & 0.265 & 0.265 & -0.001$^{}$ \\ 
  wealth group: poorer & 0.314 & 0.330 & -0.016$^{}$ & 0.198 & 0.233 & -0.035$^{}$ \\ 
  wealth group: poorest & 0.206 & 0.227 & -0.021$^{}$ & 0.148 & 0.171 & -0.024$^{}$ \\ 
  Years of education & 5.656 & 5.391 & 0.265$^{}$ & 3.435 & 3.408 & 0.027$^{}$ \\ 
  Age & 31.160 & 29.509 & 1.651$^{***}$ & 30.574 & 28.830 & 1.744$^{***}$ \\ 
  Male & 0.503 & 0.475 & 0.028$^{}$ & 0.509 & 0.454 & 0.055$^{}$ \\ 
  Married & 0.546 & 0.502 & 0.044$^{}$ & 0.631 & 0.598 & 0.033$^{}$ \\ 
  Has a bank account & 0.159 & 0.127 & 0.032$^{}$ & 0.072 & 0.051 & 0.021$^{}$ \\ 
  occupation: not working & 0.288 & 0.342 & -0.054$^{**}$ & 0.309 & 0.349 & -0.04$^{}$ \\ 
  occupation: agriculture & 0.043 & 0.046 & -0.003$^{}$ & 0.283 & 0.310 & -0.027$^{}$ \\ 
  occupation: sales & 0.228 & 0.208 & 0.02$^{}$ & 0.123 & 0.112 & 0.011$^{}$ \\ 
  occupation: household/domestic & 0.018 & 0.025 & -0.008$^{}$ & 0.006 & 0.004 & 0.002$^{}$ \\ 
  occupation: unskilled & 0.157 & 0.145 & 0.013$^{}$ & 0.095 & 0.087 & 0.008$^{}$ \\ 
  Household size & 11.199 & 11.362 & -0.163$^{}$ & 10.973 & 10.974 & 0$^{}$ \\ 
  Water access & 0.976 & 0.973 & 0.002$^{}$ & 0.492 & 0.428 & 0.064$^{}$ \\ 
  Access to sanitation & 0.916 & 0.913 & 0.003$^{}$ & 0.477 & 0.437 & 0.039$^{}$ \\ 
  Electricity & 0.943 & 0.944 & -0.001$^{}$ & 0.340 & 0.294 & 0.046$^{}$ \\ 
   \bottomrule
\end{tabular}}
\caption{Differences in characteristics between phone users and the population, Saint-Louis. Statistics were derived from the Senegal 2017 DHS men and women individual datasets.} 
\label{}
\end{table}

\begin{table}[H]
\centering
\setlength{\aboverulesep}{0pt}
\setlength{\belowrulesep}{0pt}
\resizebox{\textwidth}{!}{
\begin{tabular}{|l|>{\centering}m{0.7in}|>{\centering}m{0.7in}|>{\centering}m{0.7in}|>{\centering}m{0.7in}|>{\centering}m{0.7in}|>{\centering\arraybackslash}m{0.7in}|}
  \toprule 
  & \multicolumn{3}{c|}{\textbf{\underline{Urban}}} & \multicolumn{3}{c|}{\textbf{\underline{Rural}}} \\ 
 & \textbf{Phone users} & \textbf{All} & \textbf{Diff.} & \textbf{Phone users} & \textbf{All} & \textbf{Diff.} \\ 
   & (1) & (2) & (1)-(2) & (3) & (4) & (3)-(4) \\ 
 \midrule
wealth group: richest & 0.017 & 0.016 & 0.001$^{}$ & 0.027 & 0.025 & 0.002$^{}$ \\ 
  wealth group: richer & 0.017 & 0.013 & 0.004$^{}$ & 0.122 & 0.114 & 0.008$^{}$ \\ 
  wealth group: middle & 0.092 & 0.081 & 0.012$^{}$ & 0.298 & 0.279 & 0.019$^{}$ \\ 
  wealth group: poorer & 0.245 & 0.231 & 0.014$^{}$ & 0.420 & 0.436 & -0.015$^{}$ \\ 
  wealth group: poorest & 0.629 & 0.660 & -0.03$^{}$ & 0.133 & 0.146 & -0.014$^{}$ \\ 
  Years of education & 6.420 & 6.190 & 0.23$^{}$ & 3.616 & 3.252 & 0.364$^{}$ \\ 
  Age & 28.466 & 26.785 & 1.681$^{**}$ & 30.727 & 28.508 & 2.219$^{***}$ \\ 
  Male & 0.583 & 0.557 & 0.025$^{}$ & 0.667 & 0.530 & 0.137$^{***}$ \\ 
  Married & 0.438 & 0.386 & 0.052$^{}$ & 0.613 & 0.589 & 0.025$^{}$ \\ 
  Has a bank account & 0.180 & 0.142 & 0.038$^{}$ & 0.088 & 0.061 & 0.027$^{}$ \\ 
  occupation: not working & 0.313 & 0.349 & -0.036$^{}$ & 0.111 & 0.146 & -0.035$^{}$ \\ 
  occupation: agriculture & 0.146 & 0.161 & -0.014$^{}$ & 0.618 & 0.633 & -0.015$^{}$ \\ 
  occupation: sales & 0.183 & 0.174 & 0.009$^{}$ & 0.077 & 0.071 & 0.006$^{}$ \\ 
  occupation: household/domestic & 0.012 & 0.015 & -0.003$^{}$ & 0.013 & 0.020 & -0.008$^{}$ \\ 
  occupation: unskilled & 0.130 & 0.119 & 0.011$^{}$ & 0.047 & 0.037 & 0.01$^{}$ \\ 
  Household size & 14.502 & 14.485 & 0.016$^{}$ & 16.229 & 16.436 & -0.207$^{}$ \\ 
  Water access & 0.584 & 0.576 & 0.008$^{}$ & 0.017 & 0.019 & -0.002$^{}$ \\ 
  Access to sanitation & 0.462 & 0.433 & 0.029$^{}$ & 0.170 & 0.160 & 0.01$^{}$ \\ 
  Electricity & 0.837 & 0.814 & 0.023$^{}$ & 0.201 & 0.195 & 0.006$^{}$ \\ 
   \bottomrule
\end{tabular}}
\caption{Differences in characteristics between phone users and the population, Sedhiou. Statistics were derived from the Senegal 2017 DHS men and women individual datasets.} 
\label{}
\end{table}

\begin{table}[H]
\centering
\setlength{\aboverulesep}{0pt}
\setlength{\belowrulesep}{0pt}
\resizebox{\textwidth}{!}{
\begin{tabular}{|l|>{\centering}m{0.7in}|>{\centering}m{0.7in}|>{\centering}m{0.7in}|>{\centering}m{0.7in}|>{\centering}m{0.7in}|>{\centering\arraybackslash}m{0.7in}|}
  \toprule 
  & \multicolumn{3}{c|}{\textbf{\underline{Urban}}} & \multicolumn{3}{c|}{\textbf{\underline{Rural}}} \\ 
 & \textbf{Phone users} & \textbf{All} & \textbf{Diff.} & \textbf{Phone users} & \textbf{All} & \textbf{Diff.} \\ 
   & (1) & (2) & (1)-(2) & (3) & (4) & (3)-(4) \\ 
 \midrule
wealth group: richest & 0.070 & 0.059 & 0.011$^{}$ & 0.041 & 0.030 & 0.011$^{}$ \\ 
  wealth group: richer & 0.136 & 0.116 & 0.019$^{}$ & 0.149 & 0.130 & 0.018$^{}$ \\ 
  wealth group: middle & 0.175 & 0.159 & 0.016$^{}$ & 0.210 & 0.185 & 0.025$^{}$ \\ 
  wealth group: poorer & 0.305 & 0.293 & 0.012$^{}$ & 0.176 & 0.188 & -0.012$^{}$ \\ 
  wealth group: poorest & 0.313 & 0.372 & -0.059$^{}$ & 0.424 & 0.466 & -0.042$^{}$ \\ 
  Years of education & 6.188 & 5.714 & 0.474$^{}$ & 2.233 & 1.839 & 0.393$^{}$ \\ 
  Age & 29.117 & 27.873 & 1.245$^{}$ & 30.068 & 28.463 & 1.606$^{*}$ \\ 
  Male & 0.497 & 0.474 & 0.023$^{}$ & 0.677 & 0.457 & 0.22$^{***}$ \\ 
  Married & 0.443 & 0.426 & 0.017$^{}$ & 0.719 & 0.720 & -0.001$^{}$ \\ 
  Has a bank account & 0.252 & 0.204 & 0.047$^{}$ & 0.078 & 0.041 & 0.037$^{}$ \\ 
  occupation: not working & 0.338 & 0.349 & -0.012$^{}$ & 0.208 & 0.298 & -0.09$^{}$ \\ 
  occupation: agriculture & 0.092 & 0.101 & -0.01$^{}$ & 0.485 & 0.496 & -0.012$^{}$ \\ 
  occupation: sales & 0.191 & 0.190 & 0.001$^{}$ & 0.099 & 0.062 & 0.037$^{}$ \\ 
  occupation: household/domestic & 0.021 & 0.032 & -0.011$^{}$ & 0.004 & 0.003 & 0.001$^{}$ \\ 
  occupation: unskilled & 0.123 & 0.127 & -0.004$^{}$ & 0.107 & 0.077 & 0.03$^{}$ \\ 
  Household size & 12.763 & 12.876 & -0.113$^{}$ & 20.155 & 19.910 & 0.245$^{}$ \\ 
  Water access & 0.821 & 0.796 & 0.025$^{}$ & 0.279 & 0.246 & 0.033$^{}$ \\ 
  Access to sanitation & 0.736 & 0.690 & 0.046$^{}$ & 0.038 & 0.031 & 0.007$^{}$ \\ 
  Electricity & 0.894 & 0.859 & 0.035$^{}$ & 0.180 & 0.159 & 0.02$^{}$ \\ 
   \bottomrule
\end{tabular}}
\caption{Differences in characteristics between phone users and the population, Tambacounda. Statistics were derived from the Senegal 2017 DHS men and women individual datasets.} 
\label{}
\end{table}

\begin{table}[H]
\centering
\setlength{\aboverulesep}{0pt}
\setlength{\belowrulesep}{0pt}
\resizebox{\textwidth}{!}{
\begin{tabular}{|l|>{\centering}m{0.7in}|>{\centering}m{0.7in}|>{\centering}m{0.7in}|>{\centering}m{0.7in}|>{\centering}m{0.7in}|>{\centering\arraybackslash}m{0.7in}|}
  \toprule 
  & \multicolumn{3}{c|}{\textbf{\underline{Urban}}} & \multicolumn{3}{c|}{\textbf{\underline{Rural}}} \\ 
 & \textbf{Phone users} & \textbf{All} & \textbf{Diff.} & \textbf{Phone users} & \textbf{All} & \textbf{Diff.} \\ 
   & (1) & (2) & (1)-(2) & (3) & (4) & (3)-(4) \\ 
 \midrule
wealth group: richest & 0.158 & 0.137 & 0.021$^{}$ & 0.279 & 0.261 & 0.018$^{}$ \\ 
  wealth group: richer & 0.190 & 0.177 & 0.013$^{}$ & 0.337 & 0.329 & 0.008$^{}$ \\ 
  wealth group: middle & 0.221 & 0.222 & 0$^{}$ & 0.239 & 0.245 & -0.006$^{}$ \\ 
  wealth group: poorer & 0.236 & 0.244 & -0.008$^{}$ & 0.109 & 0.114 & -0.005$^{}$ \\ 
  wealth group: poorest & 0.195 & 0.220 & -0.025$^{}$ & 0.037 & 0.051 & -0.015$^{}$ \\ 
  Years of education & 5.668 & 5.444 & 0.223$^{}$ & 3.543 & 3.342 & 0.201$^{}$ \\ 
  Age & 30.771 & 29.316 & 1.455$^{**}$ & 30.653 & 29.555 & 1.098$^{}$ \\ 
  Male & 0.558 & 0.520 & 0.038$^{}$ & 0.594 & 0.502 & 0.092$^{**}$ \\ 
  Married & 0.481 & 0.444 & 0.036$^{}$ & 0.603 & 0.591 & 0.012$^{}$ \\ 
  Has a bank account & 0.224 & 0.186 & 0.038$^{}$ & 0.110 & 0.083 & 0.027$^{}$ \\ 
  occupation: not working & 0.249 & 0.287 & -0.038$^{}$ & 0.149 & 0.194 & -0.045$^{}$ \\ 
  occupation: agriculture & 0.035 & 0.037 & -0.002$^{}$ & 0.371 & 0.369 & 0.002$^{}$ \\ 
  occupation: sales & 0.192 & 0.192 & -0.001$^{}$ & 0.134 & 0.140 & -0.006$^{}$ \\ 
  occupation: household/domestic & 0.032 & 0.039 & -0.007$^{}$ & 0.035 & 0.042 & -0.007$^{}$ \\ 
  occupation: unskilled & 0.233 & 0.217 & 0.017$^{}$ & 0.127 & 0.114 & 0.013$^{}$ \\ 
  Household size & 12.440 & 12.696 & -0.256$^{}$ & 13.749 & 13.827 & -0.077$^{}$ \\ 
  Water access & 0.830 & 0.823 & 0.007$^{}$ & 0.652 & 0.635 & 0.017$^{}$ \\ 
  Access to sanitation & 0.674 & 0.658 & 0.016$^{}$ & 0.243 & 0.243 & 0$^{}$ \\ 
  Electricity & 0.919 & 0.911 & 0.007$^{}$ & 0.532 & 0.515 & 0.017$^{}$ \\ 
   \bottomrule
\end{tabular}}
\caption{Differences in characteristics between phone users and the population, Thies. Statistics were derived from the Senegal 2017 DHS men and women individual datasets.} 
\label{}
\end{table}

\begin{table}[H]
\centering
\setlength{\aboverulesep}{0pt}
\setlength{\belowrulesep}{0pt}
\resizebox{\textwidth}{!}{
\begin{tabular}{|l|>{\centering}m{0.7in}|>{\centering}m{0.7in}|>{\centering}m{0.7in}|>{\centering}m{0.7in}|>{\centering}m{0.7in}|>{\centering\arraybackslash}m{0.7in}|}
  \toprule 
  & \multicolumn{3}{c|}{\textbf{\underline{Urban}}} & \multicolumn{3}{c|}{\textbf{\underline{Rural}}} \\ 
 \toprule
 & \textbf{Phone users} & \textbf{All} & \textbf{Diff.} & \textbf{Phone users} & \textbf{All} & \textbf{Diff.} \\ 
   & (1) & (2) & (1)-(2) & (3) & (4) & (3)-(4) \\ 
 \midrule
wealth group: richest & 0.029 & 0.026 & 0.003$^{}$ & 0.189 & 0.170 & 0.019$^{}$ \\ 
  wealth group: richer & 0.101 & 0.090 & 0.011$^{}$ & 0.332 & 0.321 & 0.011$^{}$ \\ 
  wealth group: middle & 0.151 & 0.146 & 0.005$^{}$ & 0.262 & 0.277 & -0.015$^{}$ \\ 
  wealth group: poorer & 0.207 & 0.196 & 0.01$^{}$ & 0.205 & 0.222 & -0.017$^{}$ \\ 
  wealth group: poorest & 0.512 & 0.542 & -0.03$^{}$ & 0.012 & 0.011 & 0.001$^{}$ \\ 
  Years of education & 7.841 & 7.498 & 0.343$^{}$ & 6.923 & 6.502 & 0.421$^{}$ \\ 
  Age & 31.180 & 29.372 & 1.808$^{*}$ & 31.247 & 28.998 & 2.25$^{**}$ \\ 
  Male & 0.594 & 0.561 & 0.033$^{}$ & 0.606 & 0.564 & 0.042$^{}$ \\ 
  Married & 0.440 & 0.399 & 0.041$^{}$ & 0.490 & 0.431 & 0.058$^{}$ \\ 
  Has a bank account & 0.262 & 0.216 & 0.046$^{}$ & 0.149 & 0.113 & 0.036$^{}$ \\ 
  occupation: not working & 0.207 & 0.244 & -0.037$^{}$ & 0.182 & 0.214 & -0.032$^{}$ \\ 
  occupation: agriculture & 0.124 & 0.135 & -0.012$^{}$ & 0.325 & 0.329 & -0.004$^{}$ \\ 
  occupation: sales & 0.161 & 0.154 & 0.006$^{}$ & 0.148 & 0.168 & -0.019$^{}$ \\ 
  occupation: household/domestic & 0.025 & 0.028 & -0.003$^{}$ & 0.022 & 0.031 & -0.009$^{}$ \\ 
  occupation: unskilled & 0.201 & 0.195 & 0.006$^{}$ & 0.108 & 0.088 & 0.019$^{}$ \\ 
  Household size & 12.961 & 13.067 & -0.106$^{}$ & 13.490 & 13.777 & -0.288$^{}$ \\ 
  Water access & 0.432 & 0.416 & 0.015$^{}$ & 0.149 & 0.144 & 0.005$^{}$ \\ 
  Access to sanitation & 0.723 & 0.679 & 0.044$^{}$ & 0.193 & 0.178 & 0.015$^{}$ \\ 
  Electricity & 0.920 & 0.905 & 0.015$^{}$ & 0.570 & 0.555 & 0.014$^{}$ \\ 
   \bottomrule
\end{tabular}}
\caption{Differences in characteristics between phone users and the population, Ziguinchor. Statistics were derived from the Senegal 2017 DHS men and women individual datasets.} 
\label{}
\end{table}

\end{document}